\documentclass[11pt, oneside]{article}
\usepackage[margin=2.5cm]{geometry}
\usepackage{setspace}
\usepackage{comment}
\usepackage{natbib}
\usepackage{algorithm}
\usepackage{algpseudocode} %psuedo code
\usepackage{amsmath} 
\usepackage{amsfonts}
\usepackage{amssymb}
\usepackage{mathtools}
\usepackage{graphicx}
\usepackage{bbm}

\usepackage{tabularx}   % for the X column type
\usepackage{booktabs}   % for \toprule, \midrule, \bottomrule
\usepackage{subfig}
\usepackage{diagbox}
\usepackage{multirow}
\usepackage{pgfplots}
\usepackage{caption}
\usepackage{subcaption}

\usepackage{pdfpages}
\usepackage{nomencl}  %nomenclature
\pgfplotsset{compat = newest}				
\usepackage{soul}

\usepackage[utf8]{inputenc} % allow utf-8 input
\usepackage[T1]{fontenc}    % use 8-bit T1 fonts
\usepackage{hyperref}       % hyperlinks
\usepackage{url}            % simple URL typesetting
\hypersetup{hidelinks}
\usepackage{booktabs}       % professional-quality tables

\title{Adaptive Conformal Inference Under Delayed Feedback: Coverage Guarantees and a Delay-to-Memory Diagnostic}

\author{
  \begin{tabular}[t]{c}
    Lama El Halabi \\
    Stanford University \\
    \texttt{lie08@stanford.edu}
  \end{tabular}
  \hspace{2cm}
  \begin{tabular}[t]{c}
    Adam Brandt \\
    Stanford University \\
    \texttt{abrandt@stanford.edu}
  \end{tabular}
}

\begin{document}
\maketitle
\begin{abstract}
Adaptive Conformal Inference (ACI) extends conformal prediction to non-exchangeable settings by adjusting the nominal miscoverage level online in response to recent coverage errors. When forecasts are issued with horizon $\tau$, however, the outcome needed to evaluate a prediction is observed only $\tau$ steps later, so these adaptive updates must rely on delayed feedback. We study this setting and ask how the effect of delay depends on the persistence of the residual process. First, we show that the $\tau$-delayed ACI recursion can be decomposed into $\tau$ interleaved ACI-like sequences. This representation yields a finite-sample bound on long-run empirical coverage with explicit dependence on $\tau$. We also derive an approximate marginal coverage bound that relates coverage deviation to changes in the underlying environment across the forecast horizon and to the adaptation rate $\gamma$. We then introduce the delay-to-memory ratio $r=\tau/L$, where $L$ is the time scale over which the temporal signal driving non-exchangeability decays. Simulation results show that the usefulness of this ratio depends on the form of temporal dependence: it strongly organizes performance under AR(1) dependence, is less predictive of overall performance under GARCH(1,1) and Markov switching, but more clearly characterizes when scale normalization remains useful in those settings. Abrupt mean- and variance-shift experiments further show that the preferred adaptation rate depends on the residual dynamics. Overall, the results show that the effect of forecast delay is best understood relative to the time scale over which past residual information remains relevant.
\end{abstract}

\pagenumbering{arabic}
\section{Introduction}

Conformal prediction provides a principled framework for uncertainty quantification that can be applied to any black-box prediction model \citep{shafer2007tutorialconformalprediction}. It converts point predictions into prediction sets with valid coverage guarantees without requiring strong distributional assumptions. Classical conformal methods rely on exchangeability, but recent work has extended conformal prediction to settings with distribution shifts through reweighting \citep{barber2023conformal}, distributionally robust methods \citep{cauchois2021knowing}, and online learning \citep{gibbs2021adaptive,zaffran2022adaptive,gibbs2024conformal,bhatnagar2023improved}. \citet{gibbs2021adaptive} introduced Adaptive Conformal Inference (ACI), which responds to distribution shifts by updating the miscoverage level online according to whether recently issued prediction sets cover their targets. In its standard form, however, ACI assumes that the outcome needed to evaluate a prediction is available before the next update is made.\\

Forecasting with a fixed lead time changes this setting. If a prediction issued at time $t$ targets the outcome at some future time $t+\tau$, its coverage outcome is not observed until $\tau$ steps later. The adaptive update must therefore operate using delayed feedback.  In extreme cases, there may be significant delay before the effectiveness of a coverage shift can be determined. \citet{gibbs2021adaptive} identify this as an open direction. The importance of this delay depends on the dynamics of the residual process. A delay of ten steps may be substantial for a process whose dependence disappears within a few observations, but relatively short for one whose residual structure persists over much longer horizons. Thus, the relevant question is not simply whether feedback is delayed, but whether the information contained in that feedback remains useful at the forecast target. Understanding \emph{when} delay becomes consequential is the central question of this paper.\\

\paragraph{Related work:}
Recent work on multi-step conformal forecasting also encounters delayed feedback: a forecast issued at time \(t\) for the outcome at time \(t+\tau\) cannot be evaluated until that outcome is observed \(\tau\) steps later. \citet{szabadvary2024adaptive} extend ACI to this setting by maintaining a separate adaptive sequence for each forecast horizon, with horizon-specific target miscoverage levels and learning rates. Their construction inherits ACI-style finite-sample coverage-frequency guarantees at each horizon and for the overall error rate. Their numerical examples illustrate these choices on electricity-demand forecasts, but do not systematically study how delayed feedback affects adaptation or how its effect depends on the persistence of the underlying process.\\

\citet{wang2024online} introduce Multi-step Adaptive Conformal Prediction (MACP) as a multi-step extension of ACI and use it as a baseline for their proposed Autocorrelated Multi-step Conformal Prediction method (AcMCP). AcMCP explicitly exploits temporal dependence in multi-step forecast errors. They show that optimal $h$-step-ahead forecast errors, where $h$ denotes the forecast horizon, follow an approximate moving-average process of order $h-1$, denoted MA$(h-1)$, under a general non-stationary autoregressive data-generating process. AcMCP uses this structure to predict part of the upcoming forecast error from historical and preceding-horizon errors. This predicted error component is incorporated directly into the conformal quantile update, alongside feedback terms designed to control long-run coverage. Their simulation studies consider forecast horizons of up to three time steps.\\

A broader line of work uses temporal or multi-step information to improve online conformal adaptation. \citet{yang2024bellman} propose Bellman Conformal Inference (BCI), which uses available multi-step prediction intervals within a stochastic-control problem to select the nominal miscoverage level while explicitly trading off coverage and interval length. Rather than updating the ACI level directly, BCI optimizes the current decision using information about predicted future intervals while retaining a long-run coverage guarantee. More recently, \citet{li2026optimization} propose Optimization-Based Online Conformal Prediction (O$^2$CP), a general framework that augments online conformal methods such as ACI and conformal PID with cross-horizon optimization. O$^2$CP constructs admissible sets around the control variables produced by the underlying online update and then jointly selects values across forecast horizons using an estimated cross-horizon error distribution, preserving the long-run marginal coverage guarantee of the base procedure. These approaches illustrate complementary ways of using temporal information beyond the most recent coverage error: BCI uses multi-step forecasts to improve interval efficiency, while O$^2$CP exploits dependence across forecast horizons.\\

Related work also considers missing or intermittent rather than delayed-but-guaranteed feedback \citep{wang2025mirror,skalse2026online}. Separately, \citet{gibbs2024conformal}, \citet{angelopoulos2023conformal,angelopoulos2024online}, and \cite{zaffran2022adaptive} extend ACI through step-size adaptation, expert aggregation, and control-theoretic formulations under distribution shift, but do not focus on delayed feedback.\\

Taken together, these works address several consequences of multi-step prediction and imperfect feedback, including horizon-specific adaptation, dependence among multi-step forecast errors, cross-horizon optimization, and limited feedback. Our focus is complementary: we study the delayed ACI recursion itself and ask how the consequences of a forecast delay depend on the persistence of the temporal signal driving non-exchangeability. Rather than exploiting a particular temporal dependence structure to improve interval construction, we use persistence as an explanatory time scale and examine when feedback remains informative relative to the delay.\\

\paragraph{Contributions:}
This paper makes two main contributions.

\begin{enumerate}

\item \textbf{A delay-to-memory diagnostic for adaptive conformal inference with delayed feedback:}
We characterize the feedback delay relative to the persistence of the temporal signal driving non-exchangeability through the ratio
$r=\tau/L$, where $L$ denotes the time scale over which that signal decays. For each persistent residual process, $L$ is defined from the decay rate of the relevant dependence structure: residual autocorrelation for AR(1), dependence in squared residuals for GARCH(1,1), and latent-state dependence for Markov switching. We introduce a curve-collapse diagnostic to assess whether $r$ explains performance more effectively than the raw delay $\tau$. The experiments show that the information captured by $r$ depends on the mechanism of non-exchangeability. Under AR(1), $r$ strongly explains overall interval performance. Under GARCH(1,1), it provides a more moderate organization of the raw interval score but more clearly identifies when scale normalization is beneficial or detrimental. Under Markov switching, $r$ does not produce a strong collapse in overall interval performance, but it does strongly characterize how the benefit of state-dependent normalization changes as the forecast delay grows relative to regime persistence. Complementary mean- and variance-shift experiments examine abrupt changes for which no stationary memory scale is available. Overall, the delay-to-memory ratio allows delays across processes with different time scales to be compared on a common axis. The experiments show that its explanatory power depends on the form of non-exchangeability, but across persistent processes it provides information that the raw horizon $\tau$ alone cannot.

\item \textbf{Coverage analysis of the $\tau$-delayed ACI recursion:}
We introduce a phase-based representation that decomposes the delayed recursion into $\tau$ interleaved standard-ACI-like sequences (Section~\ref{sec:alternative-form}). Each phase is updated once every $\tau$ calendar steps, and recombining the phase-wise bounds yields boundedness and a finite-sample long-run coverage bound with explicit dependence on $\tau$ (Sections~\ref{sec:boundedness}--\ref{sec:long-term-coverage}). We additionally derive an approximate marginal coverage bound under a hidden Markov model (Section~\ref{sec:approx-marginal}), relating coverage deviation to distributional change between prediction and target time and to the adaptation rate $\gamma$. To our knowledge, no analogous marginal coverage bound or phase-based representation has been established for $\tau$-delayed ACI.

\end{enumerate}

The theoretical analysis shows how delay affects both long-run and marginal coverage, while the simulations show that its practical effect depends on the persistence of temporal dependence and where that persistence appears in the residual process.

\section{Forecasting Setup}
\label{sec:forecasting-setup}
In our forecasting framework, the objective is to predict the target output at a future time \( t + \tau \), where, for example, \( \tau \in \{5, 10, 15, 20, 30, 40\} \) minutes denotes the forecast horizon. At each time step \( t \), the model receives a history of recent sequential observations:
\[
\{(X_{t-h}, Y_{t-h}), \ldots, (X_t, Y_t)\},
\]
where $( X_i, Y_i )$ is the covariate-response pair. This sequence is passed into a fixed forecasting model \( \mu^{\tau}  \), which outputs a point prediction for the future response:
\begin{equation}
\label{ACI_task}
 \hat{Y}_{t+\tau} = \mu^{\tau}(X_{t-h:t}, Y_{t-h:t}).   
\end{equation}

\section{Online Split Conformal Prediction}
\label{sec:oscp}
In the standard split conformal setting, the calibration set is fixed, so the empirical distribution of conformity scores and its associated quantile function remain unchanged throughout deployment. In an online setting, the calibration set evolves as new data becomes available. At each time step $t$, we observe $Y_t$ and can therefore compute a conformity score $S_t$ to assess the accuracy of a past prediction made $\tau$ steps earlier. We use the signed residual
\begin{equation}
\label{ACI_res}
S_{t} = Y_{t} - \mu^{\tau}(X_{t-h-\tau:t-\tau}, Y_{t-h-\tau:t-\tau}) = Y_{t} - \hat{Y}_{t},
\end{equation}
which yields a possibly asymmetric interval that can adapt to skewed or directionally biased residual distributions. The absolute residual $|Y_t - \hat{Y}_t|$ is a common alternative that instead yields a symmetric interval, and either choice is compatible with everything that follows. To assess model performance over time, we maintain a sliding window of the most recent $R$ conformity scores. We refer to $\mathcal{D}_t$ both as the set of these $R$ scores and as the empirical distribution formed by assigning equal mass to each score:

\[
\mathcal{D}_t = \{ S_{t - R + 1}, S_{t - R + 2}, \ldots, S_t \}
\quad \text{or equivalently} \quad
\mathcal{D}_t = \frac{1}{R} \sum_{i = t - R + 1}^{t} \delta_{S_i},
\]
where \( \delta_{S_i} \) denotes the Dirac delta function centered at \( S_i \). Then, for each time step \( t \), the time-varying quantile function \( \hat{Q}_t(p) \) is defined as:
\begin{equation}
\label{ACI_Q}
\hat{Q}_t(p, \mathcal{D}_t ) := \inf \left\{ s : \frac{1}{R} \sum_{i=t-R+1}^{t} \mathbbm{1}\{ S_i  \leq s \} \geq p \right\}.
\end{equation}

Using this quantile function, the prediction set for $Y_{t+\tau}$ constructed at time $t$ is
\begin{equation}
\label{ACI_C}
\hat{\mathcal{C}}_{t+\tau}^{\tau}(\alpha) = \left[ \hat{Y}_{t+\tau} + \hat{Q}_t(\alpha/2, \mathcal{D}_t),\ \hat{Y}_{t+\tau} + \hat{Q}_t(1 - \alpha/2, \mathcal{D}_t) \right].
\end{equation}

The prediction set \( \hat{\mathcal{C}}_{t+\tau}^{\tau}(\alpha) \) is constructed using conformity scores computed from past predictions at times \( t' \leq t - \tau \), once the corresponding outcomes \( Y_{t'+\tau} \) have been observed. In other words, the residuals used to calibrate the quantile function \( \hat{Q}_t(\cdot) \) reflect prediction errors made by the forecast model at least \( \tau \) time steps ago. These residuals are then used to construct the prediction interval around the current forecast \( \hat{Y}_{t+\tau} = \mu^{\tau}(X_{t-h:t}, Y_{t-h:t}) \). As a result, when the forecast horizon \( \tau \) is large, the quantile function may be slow to adapt to recent changes in the data distribution, potentially leading to miscalibrated or suboptimal prediction intervals (lag in adaptation). This introduces a tradeoff between forecast horizon and calibration responsiveness in rapidly evolving environments. \\

Algorithm~\ref{alg:prediction-set-CP} gives the full procedure for constructing prediction sets online. Assuming observations begin at \( t = 1 \), the first forecast is generated at \( t = h + 1\), and the corresponding conformity score becomes available at \( t = h + \tau + 1\). Fitting $\hat{Q}_t$ requires a window of $R$ residuals, collected by $t=h+\tau+R$. We therefore define $t_0 = h+\tau+R$ and perform conformal updates for $t = t_0+1,\ldots,T$. For notational simplicity, in the algorithms that follow, we re-index time so that \( t_0 = 0 \).\\

\begin{algorithm}
\caption{Prediction Set Estimator (PSE)}
\label{alg:prediction-set-CP}
\begin{algorithmic}[1]
\State \textbf{Input:} $(X_{t-h-\tau:t}, Y_{t-h-\tau:t})$, forecasting model $\mu^{\tau}$, miscoverage level $\alpha \in (0, 1)$, calibration set $\mathcal{D}_{t-1}$
\State Compute point prediction: $\hat{Y}_{t+\tau} \gets \mu^{\tau}(X_{t-h:t}, Y_{t-h:t})$
\State Compute past prediction: $\hat{Y}_{t} \gets \mu^{\tau}(X_{t-h-\tau:t-\tau}, Y_{t-h-\tau:t-\tau})$
\State Compute conformity score: $S_t \gets Y_t - \hat{Y}_{t}$
\State Update calibration set: $\mathcal{D}_t \gets \{ S_{t-R+1}, \ldots, S_t \}$
\State Fit empirical quantile function:
\[
\hat{Q}_t(p, \mathcal{D}_t) := \inf \left\{ s : \frac{1}{R} \sum_{i=t-R+1}^{t} \mathbbm{1}\{ S_i \leq s \} \geq p \right\}
\]
\State Compute prediction interval:
\[
\hat{\mathcal{C}}_{t+\tau}^{\tau}(\alpha) \gets \left[ \hat{Y}_{t+\tau} + \hat{Q}_t(\alpha/2, \mathcal{D}_t),\ \hat{Y}_{t+\tau} + \hat{Q}_t(1 - \alpha/2, \mathcal{D}_t) \right]
\]
\State \Return $\hat{\mathcal{C}}_{t+\tau}^{\tau}(\alpha)$, $\mathcal{D}_t$
\end{algorithmic}
\end{algorithm}

\section{$\tau$-Delayed Adaptive Conformal Inference}

Adaptive Conformal Inference (ACI) addresses the challenge of making valid predictions in non-exchangeable, online environments by treating the coverage level as a tunable parameter that can be learned and updated over time. We begin by defining the miscoverage rate, which quantifies the probability that the true target value \( Y_t \) lies outside the prediction set generated \( \tau \) time steps earlier. Specifically, the miscoverage rate at time \( t \) is given by:
\[
\mathcal{M}_t(\alpha) := \mathbb{P}\left( Y_t \notin \hat{C}_{t}^{\tau}(\alpha) \right),
\]
where \( \hat{C}_{t}^{\tau}(\alpha) \) is the prediction set formed at time \( t - \tau \) for the target \( Y_t \). When using the signed residual, the miscoverage rate becomes \( \mathcal{M}_t(\alpha) = \mathbb{P}(S_t(y) < \hat{Q}_{t-\tau}(\alpha/2) \ \text{or} \ S_t(y) > \hat{Q}_{t-\tau}(1 - \alpha/2)) \).\\

In our time-series application, the data are non-exchangeable, therefore, the miscoverage rate \( \mathcal{M}_t(\alpha) \neq \alpha \). However, as noted by \citet{gibbs2021adaptive}, if the conformity scores used to define the quantile function \( \hat{Q}_{t-\tau}(\cdot) \) capture the bulk of the distribution, then it is possible to find a time-varying level \( \alpha_t^* \) such that \( \mathcal{M}_t(\alpha_t^*) = \alpha \), or approximately so. We define the optimal miscoverage level \( \alpha_t^* \) as the largest level for which the prediction set generated at time \( t - \tau \) would have yielded valid coverage at time \( t \):
\begin{equation}
\label{ACI_alpha_star}
\alpha_t^* := \sup \left\{ \beta \in [0,1] : \mathcal{M}_t(\beta) \leq \alpha \right\}.
\end{equation}

This definition relies on the assumption that the quantile function \( \hat{Q}_{t - \tau}(\cdot) \) is continuous, non-decreasing, and satisfies the boundary conditions \( \hat{Q}_{t - \tau}(0) = -\infty \) and \( \hat{Q}_{t - \tau}(1) = \infty \). Our definition of the quantile in Equation~\ref{ACI_Q} satisfies these assumptions once it is smoothed to eliminate discontinuities, extrapolated to extend to \( -\infty \) and \( \infty \), and adjusted to handle tied scores appropriately. Since the true miscoverage rate is unknown in practice, we estimate \( \alpha_t^* \) adaptively using a delayed online update rule:
\begin{equation}
\label{ACI_update}
\alpha_{t+\tau} = \alpha_t + \gamma \left( \alpha - \texttt{err}_t \right)
\quad \text{where} \quad
\texttt{err}_t = \mathbbm{1}\left\{ Y_t \notin \hat{\mathcal{C}}_t^\tau(\alpha_t) \right\},
\end{equation}

where \( \texttt{err}_t \) is a binary indicator that records whether the prediction set constructed at time \( t - \tau \) failed to cover the true observation \( Y_t \), and \( \gamma > 0 \) is a fixed step-size. The updated level \( \alpha_{t+\tau} \) is then used to construct the prediction set \( \hat{\mathcal{C}}_{t+\tau}^{\tau} \) at time \( t\) for \( Y_{t+\tau} \), ensuring temporal alignment between the prediction and the adaptive correction based on prior forecast performance. If the prediction set undercovers (i.e., \( \texttt{err}_t = 1 \)), \( \alpha_t \) is reduced to increase the width of future intervals. If the prediction set overcovers (i.e., \( \texttt{err}_t = 0 \)), \( \alpha_t \) is increased to reduce the interval width. We refer to the update rule in Eq~\ref{ACI_update} as $\tau$-Delayed Adaptive Conformal Inference ($\tau$-DACI). When the forecast horizon is set to \( \tau = 1 \), we recover the original ACI formulation proposed by \citet{gibbs2021adaptive}. $\tau$-DACI is showcased in Algorithm~\ref{alg:aci-tau-delay}.\\

\begin{algorithm}
\caption{Adaptive Conformal Inference (ACI) with \(\tau\)-Delayed Response}
\label{alg:aci-tau-delay}
\begin{algorithmic}[1]
\Require $(X_{t-h-\tau:t}, Y_{t-h-\tau:t})$, forecasting model $\mu^{\tau}$
\State \textbf{Inputs:} Initial level \( \alpha \), step-size \( \gamma > 0 \), forecast horizon \( \tau \), initial calibration set \( \mathcal{D}_{t_0} \)
\State \textbf{Initialize:} $\alpha^* \gets \alpha$, $\mathcal{D}_{\text{cal}} \gets \mathcal{D}_{t_0}$, empty queue \( Q \gets \emptyset\)
\For{$t = 1, \ldots, T$}
    \If{$t > \tau$}
        \State Dequeue past prediction set and level: \( (\hat{\mathcal{C}}_t^\tau, \alpha_t) \gets Q.\texttt{pop}() \)
        \State Evaluate coverage error: \( \texttt{err}_t \gets \mathbbm{1}\{ Y_t \notin \hat{\mathcal{C}}_t^\tau \} \)
        \State Update miscoverage level: \( \alpha^* \gets \alpha_t + \gamma(\alpha - \texttt{err}_t) \)
    \EndIf
    \State Compute prediction set for future time \( t + \tau \):
    \[
    \hat{\mathcal{C}}_{t+\tau}^{\tau},\ \mathcal{D}_{\text{cal}} \gets \textsc{PSE}((X_{t-h-\tau:t}, Y_{t-h-\tau:t}),\ \mu^{\tau},\ \alpha^*,\ \mathcal{D}_{\text{cal}})
    \]
    \State Enqueue: \( Q.\texttt{append}(\hat{\mathcal{C}}_{t+\tau}^{\tau}, \alpha^*) \)
\EndFor
\end{algorithmic}
\end{algorithm}

\subsection{Alternative Form of \texorpdfstring{$\tau$}{tau}-DACI}
\label{sec:alternative-form}

Although the delayed update is naturally expressed in the global time index \(t\), this representation obscures an important structural feature of the recursion. Since $\alpha_{t+\tau}$ depends on $\alpha_t$, the full sequence can be viewed as $\tau$ interleaved subsequences, each evolving at time points separated by $\tau$ steps. Reindexing these subsequences by phase makes this structure explicit. Within each phase, the delayed recursion becomes an ordinary one-step ACI update, with no explicit delay in the update index. This reformulation is useful theoretically because it allows standard ACI arguments, including boundedness and long-run coverage analysis, to be applied separately to each phase and then combined to recover statements for the full delayed process. To see this concretely, consider the case where $\tau = 5$. In this setting, the update at time $t = 6$ depends only on $\alpha_1$ and is independent of $\alpha_2, \alpha_3, \alpha_4,$ and $\alpha_5$:  
\[
\alpha_6 = \alpha_1 + \gamma(\alpha - \texttt{err}_1).
\]
More generally, for a fixed phase $\theta$, connected updates occur at times spaced exactly $\tau$ steps apart: $\theta = 1$ gives $t = 1, 6, 11, 16, \dots$; $\theta = 2$ gives $t = 2, 7, 12, 17, \dots$; and so on through $\theta = 5$, giving $t = 5, 10, 15, 20, \dots$. Varying $\theta$ over $\{1, \dots, \tau\}$ spans all time indices, ensuring that every $t$ belongs to exactly one phase. This motivates reindexing the process using two indices:
\begin{enumerate}
    \item a \textbf{phase index} $\theta \in \{1, \dots, \tau\}$, identifying the position within the cycle, and  
    \item an \textbf{update index} $m \geq 1$, counting how many updates have occurred within that phase.  
\end{enumerate}

Table~\ref{tab:theta_m_table} illustrates how global time indices are uniquely mapped to pairs of phase \(\theta\) and update index \(m\). Every time step $t$ can then be uniquely represented by a pair $(\theta, m)$, which forms the basis of the phase-based representation. 

\begin{equation}
\label{time_index_eq}
   t_\theta(m) := t_0 + \theta + (m-1)\tau.
\end{equation}
This implies
\[
t_\theta(m+1) = t_0 + \theta + m\tau = t_\theta(m) + \tau,
\]

\begin{table}[h!]
\centering
\begin{tabular}{c|cccc}
\toprule
\(\theta \, \backslash \, m\) & \(m = 1\) & \(m = 2\) & \(m = 3\) & \(m = 4\) \\ 
\midrule
\(\theta = 1\) & 1  & 6  & 11 & 16 \\ 
\(\theta = 2\) & 2  & 7  & 12 & 17 \\ 
\(\theta = 3\) & 3  & 8  & 13 & 18 \\ 
\(\theta = 4\) & 4  & 9  & 14 & 19 \\ 
\(\theta = 5\) & 5  & 10 & 15 & 20 \\ 
\bottomrule
\end{tabular}
\caption{Mapping of global time index \(t_\theta(m) = t_0 + \theta + (m-1)\tau\) for \(\tau = 5\) and \(t_0 = 0\) . 
Each row corresponds to a fixed phase \(\theta\), and each column to an update index \(m\).}
\label{tab:theta_m_table}
\end{table}

so the update rule (Equation~\ref{ACI_update}) at time \(t_\theta(m)\) becomes
\[
\alpha_{t_\theta(m+1)}
= \alpha_{t_\theta(m)} + \gamma\bigl(\alpha - \texttt{err}_{t_\theta(m)}\bigr)
\quad \text{where} \quad
\texttt{err}_{t_\theta(m)} = \mathbbm{1}\!\left\{ Y_{t_\theta(m)} \notin \hat{\mathcal{C}}_{t_\theta(m)}^\tau\!\bigl(\alpha_{t_\theta(m)}\bigr) \right\}.
\]

Defining phase-wise sequences
\[
\alpha^{(\theta)}_m := \alpha_{t_\theta(m)}, 
\quad e^{(\theta)}_m := \texttt{err}_{t_\theta(m)},
\]
we obtain the compact \(\tau\)-DACI form:
\begin{equation}
\label{decomposed_ACI}
   \boxed{
\alpha^{(\theta)}_{m+1} 
= \alpha^{(\theta)}_{m} + \gamma\bigl(\alpha - e^{(\theta)}_{m}\bigr)
},
\quad \theta \in \{1,\dots,\tau\},\; m\ge 1.
\end{equation}

In $t$-space, each phase updates once every $\tau$ steps, whereas in $m$-space (after decoupling) each sequence updates at every iteration, exactly like standard ACI. This preserves the link to the actual time index while removing the explicit delay from the update rule, enabling the direct application of standard ACI analysis tools phase-wise.\\

For \(t = t_0+1,\dots,T\), we define the phase of a time step \(t\) by 
\[ (t - t_0) \bmod \tau = \theta, \quad \theta \in \{1,\dots,\tau\}, \] 
with the convention that \(\theta = \tau\) when the remainder is zero. The initial time corresponds to \(\theta = 1\) and \(m = 1\), which gives \(t_1(1) = t_0+1\). The number \(K_\theta\) counts how many updates occur before or at the terminal time \(T\) in phase \(\theta\). This means we are looking for all integers \(m\) such that
\[
t_0 + \theta + (m-1)\tau \le T
\quad \iff \quad
m \le \frac{T - (t_0+\theta)}{\tau} + 1.
\]
Since \(m\) must be a positive integer, the largest allowable value of \(m\) is obtained by taking the integer part (floor) of the right-hand side. Therefore,
\begin{equation}
\label{K-theta}
    K_\theta = \left\lfloor \frac{T - (t_0+\theta)}{\tau} \right\rfloor + 1.
\end{equation}

The last phase index can be identified from the remainder when \(T - t_0\) is divided by \(\tau\):
\[
\theta^\star = (T - t_0) \bmod \tau,
\]
with the convention \(\theta^\star = \tau\) when the remainder is zero. This captures the fact that updates proceed in cycles of length \(\tau\), and \(\theta^\star\) marks the phase of the final update time within the last (possibly incomplete) cycle.

\subsection{Boundedness of $\alpha_m^{(\theta)}$}
\label{sec:boundedness}
Establishing bounds on $\alpha_m^{(\theta)}$ is essential for ensuring the stability of the $\tau$-DACI update process. Boundedness guarantees that the miscoverage levels remain within a controlled range, preventing divergence that could lead to unstable or degenerate prediction sets. For each fixed $\theta$, the phase-wise sequence $\alpha_m^{(\theta)}$ obeys the same Robbins--Monro style update as in standard ACI, and Appendix~\ref{app:Boundedness} shows that for all $m \geq 1$,
\begin{equation}
\label{eq:boundedness}
\alpha_m^{(\theta)} \in [-\gamma,\, 1+\gamma].
\end{equation}

This bound follows the same basic argument as Lemma~4.1 of \citet{gibbs2021adaptive}. If $\alpha_m^{(\theta)}$ moves beyond either boundary, the two-sided prediction interval induced by the signed-residual construction becomes degenerate: it equals all of $\mathbb{R}$ when $\alpha_m^{(\theta)}<0$ and is empty when $\alpha_m^{(\theta)}>1$. Consequently, the miscoverage indicator $e_m^{(\theta)}$ is forced to take the value that drives the subsequent update back toward the interior of the interval. Relative to the standard ACI argument, two modifications are required in the delayed setting. First, the relevant quantiles are those fitted at time $t_\theta(m-1)$ rather than at $t-1$, reflecting the $\tau$-step delay. Second, the argument is formulated in terms of a two-sided prediction interval rather than a one-sided threshold, as required by the signed-residual construction.

\subsection{Long-term Coverage Guarantees}
\label{sec:long-term-coverage}
We now state the long-term coverage guarantee for $\tau$-DACI, obtained by combining the phase-wise bound of Equation~\ref{eq:boundedness} with the phase-recombination argument of Appendix~\ref{app:LongTermCoverage}, following the reasoning of Proposition~4.1 in \citet{gibbs2021adaptive}. Appendix~\ref{app:LongTermCoverage} shows that
\begin{equation}
\label{eq:global-bound}
\left|\frac{1}{T}\sum_{t=1}^T \texttt{err}_t - \alpha \right|
\;\le\; \frac{1}{\gamma\, T} \sum_{\theta=1}^{\tau} \max\{\alpha^{(\theta)}_1,\;1-\alpha^{(\theta)}_1\}
\;+\; \frac{\tau}{T}.
\end{equation}
In particular, $\frac{1}{T}\sum_{t=1}^T \texttt{err}_t \to \alpha$ almost surely as $T \to \infty$.\\

The bound in \eqref{eq:global-bound} is \emph{distribution-free}: it makes no assumptions on the data-generating process beyond the phase-wise boundedness of the iterates established in Equation~\ref{eq:boundedness}. When $\tau = 1$, there is a single phase with $K_1 = T$, and \eqref{eq:global-bound} reduces exactly to the bound in Proposition~4.1 of \citet{gibbs2021adaptive}. The deviation from the target coverage $\alpha$ decreases as $T$ grows but is generally larger for longer forecast horizons (larger $\tau$), and also depends on the quality of the initialization $\alpha^{(\theta)}_1$.

\subsection{Approximate Marginal Coverage}
\label{marginal_coverage_section}
\label{sec:approx-marginal}
The long-run bound in Equation~\eqref{eq:global-bound} ensures that, on average over time, the empirical miscoverage (\(\frac{1}{T}\sum_{t=1}^T \texttt{err}_t\)) approaches the target level $\alpha$. However, this guarantee does not describe the behavior of the marginal coverage at a single time step. To study instantaneous performance, we adopt the same probabilistic modeling framework as \citet{gibbs2021adaptive}, generalized here to the $\tau$-delayed setting.

\paragraph{Problem setup:}  
We model the data as arising from a hidden Markov model. Let $\{A_t\}_{t\in\mathbb{N}} \subseteq \mathcal{A}$ denote the latent environment process. For the adaptive sequence $\{\alpha_t\}_{t\in\mathbb{N}}$, with values evolving within the bounded range \(\alpha_t \in [-\gamma,\,1+\gamma]\), we define the true conditional marginal miscoverage at time $t$ in environment $A_t$ as \(
M(\alpha_t \mid A_t) := \mathbb{P}\!\left( Y_t \notin \hat{\mathcal{C}}^{\tau}_t(\alpha_t) \,\middle|\, A_t \right)\).\\

In this setting, we assume that the process $\{(\alpha_t, A_t, \texttt{err}_t)\}_{t \in \mathbb{N}}$ is stationary, with
marginal stationary distribution $\pi$ for $(\alpha_t,A_t)$ on $[-\gamma,\,1+\gamma]\times\mathcal{A}$. In particular, the initialization
$(\alpha_1,A_1)$ is drawn from $\pi$. This stationarity assumption greatly simplifies the characterization of the behavior of the error indicators $\texttt{err}_t$. Finally, we impose two technical conditions:
\begin{enumerate}
    \item \textbf{Lipschitz continuity:} There exists a constant $L_{Lip}>0$ such that for all $a\in\mathcal{A}$ and all $\alpha_1,\alpha_2\in[-\gamma,\,1+\gamma]$,
    \[
    \big|M(\alpha_2\mid a) - M(\alpha_1\mid a)\big| \leq L_{Lip}\,|\alpha_2 - \alpha_1|.
    \]
    \item \textbf{Existence of an optimal level:} For each $a\in\mathcal{A}$, there exists a unique $\alpha^*_a \in (0,1)$ such that
    \[
    M(\alpha^*_a \mid a) = \alpha.
    \]
    We refer to $\alpha^*_{A_t}$ as the \emph{optimal miscoverage level} at time $t$, i.e., the value that would have achieved exact target coverage under environment $A_t$.
\end{enumerate}

\paragraph{Main result:}  
With these assumptions in place, we can quantify the deviation of the instantaneous marginal miscoverage from the target level. Specifically, Appendix~\ref{app:MarginalCoverage} shows that under stationarity, the expected squared deviation satisfies
\begin{equation}
\label{eq:marginal_miscoverage}
 \mathbb{E}\!\left[\big(M(\alpha_t \mid A_t) - \alpha\big)^2\right]
\;\leq\; \frac{L_{Lip}(1+\gamma)}{\gamma}\,
\mathbb{E}\!\left[\big|\alpha^*_{A_{t+\tau}} - \alpha^*_{A_t}\big|\right]
+ \frac{L_{Lip}}{2}\,\gamma,   
\end{equation}

where the expectation is taken with respect to the stationary distribution of $(\alpha_t, A_t)$.\\ 

The bound decomposes into two terms: the first measures the \emph{distribution shift} of the latent environment across $\tau$ steps, captured by the expected difference between the optimal levels $\alpha^*_{A_t}$ and $\alpha^*_{A_{t+\tau}}$, while the second reflects the \emph{adaptation penalty}, which grows linearly with the step-size $\gamma$. When $\tau = 1$, the expression reduces to the approximate marginal coverage bound of \citet{gibbs2021adaptive}; the delayed setting introduces an explicit dependence on shift over $\tau$ steps. Choosing
\[
\gamma_{\mathrm{opt}}
    = \sqrt{2\,\mathbb{E}\!\left[\,|\alpha^*_{A_{t+\tau}}
                                 - \alpha^*_{A_t}|\,\right]}
\]
minimizes the right-hand side of the bound in \ref{eq:marginal_miscoverage} and therefore yields the optimal
tradeoff between adaptation and stability.

\section{Evaluation Metrics}
\label{sec:eval-metrics}

Coverage alone does not provide a complete assessment of a conformal forecasting method. In particular, the nominal coverage level can be attained by producing arbitrarily wide prediction intervals that cover all observations without being informative. Also, measures of aggregate empirical coverage may obscure periods of substantial under- or overcoverage. We therefore evaluate our methods along complementary dimensions that capture overall calibration, the temporal stability of coverage, and the cost of calibration in terms of predictive sharpness (i.e., interval width).

\paragraph{Empirical coverage:} We report the empirical coverage
\begin{equation}
1 - \frac{1}{T}\sum_{t=1}^{T}\mathrm{err}_t,
\end{equation}
where $\mathrm{err}_t$ denotes the miscoverage indicator defined in \eqref{ACI_update}. A well-calibrated method should achieve empirical coverage close to the nominal target $1-\alpha$.

\paragraph{Worst-case local coverage error:} Global empirical coverage can mask temporal clustering of miscoverage. For example, a period of severe undercoverage following a distribution shift may have little effect on the overall average if it is preceded or followed by a long period of stable coverage. To quantify such local deviations from the target, we report the worst-case local coverage error (LCE) over all contiguous windows of length $k$:
\begin{equation}
\mathrm{LCE}_k
:=
\max_{\substack{1 \leq j \leq T-k+1}}
\left|
\alpha
-
\frac{1}{k}
\sum_{t=j}^{j+k-1}\mathrm{err}_t
\right|.
\label{eq:lce}
\end{equation}
Here, the average miscoverage rate within each window is compared with the nominal miscoverage level $\alpha$. Smaller values of $\mathrm{LCE}_k$ indicate that calibration is maintained more consistently over time, whereas larger values reveal intervals in which the method exhibits substantial local under- or overcoverage. Worst-case local coverage error is evaluated over windows of length $k\in\{100,\allowbreak 250\}$.

\paragraph{Interval score:} Coverage can be increased by widening prediction intervals, but this comes at the expense of informativeness. To evaluate the trade-off between calibration and sharpness, we report the interval score (IS) at level $\alpha$:
\begin{equation}
\mathrm{IS}_{\alpha}
\left(
[\hat{l}_t,\hat{u}_t], Y_t
\right)
:=
\left(\hat{u}_t-\hat{l}_t\right)
+
\frac{2}{\alpha}
\left(\hat{l}_t-Y_t\right)_{+}
+
\frac{2}{\alpha}
\left(Y_t-\hat{u}_t\right)_{+},
\label{eq:is}
\end{equation}
where $(x)_{+}=\max(x,0)$. When $Y_t$ lies within the prediction interval, IS reduces to the interval width $\left(\hat{u}_t-\hat{l}_t\right)$. When the interval fails to cover $Y_t$, IS additionally penalizes the magnitude of the miss, with the penalty scaled by $2/\alpha$. Thus, lower interval scores are better, and correspond to prediction intervals that are simultaneously narrow and well calibrated. We report the time-averaged interval score,
\begin{equation}
\frac{1}{T}
\sum_{t=1}^{T}
\mathrm{IS}_{\alpha}
\left(
[\hat{l}_t,\hat{u}_t], Y_t
\right),
\end{equation}
as our primary measure of predictive efficiency.
\begin{comment}
    We additionally report the median interval width as a complementary measure of sharpness that is less sensitive to occasional extremely wide intervals.
\end{comment}

\section{Simulated Residual Processes and Diagnostic Framework}
\label{sec:simulated-examples}

We evaluate $\tau$-DACI using directly simulated residual sequences
$\varepsilon_1,\ldots,\varepsilon_T$. This approach avoids introducing variation and un-needed complexity from a separate covariate--response model and its associated fitting procedure. Simulating sequences of residuals also allows the temporal dependence, persistence, and distributional dynamics of the residual process to be controlled independently. The resulting experiments isolate how these features interact with the feedback delay $\tau$. We consider six residual families, each designed to represent a distinct mechanism through which the exchangeability assumption may fail. The Gaussian family serves as an exchangeable reference case, while the remaining families introduce temporal dependence, heteroskedasticity, or distributional shift.\\

Two complementary diagnostics are used to evaluate the residual families. For the three families in which non-exchangeability arises from a persistent, time-varying process --- AR(1), GARCH(1,1), and Markov-switching --- we study whether the effect of delay can be summarized by a \emph{delay-to-memory ratio} $r=\tau/L$, where $L$ is a process-specific memory length measuring the time scale over which historical residual information remains informative about future residuals. We assess this using a \emph{curve-collapse diagnostic}. For each candidate horizontal-axis variable, $\tau$ or $r$, we pool across persistence configurations the interval scores attained at the selected step size $\gamma^\star$. $\gamma^\star$ is chosen using a coverage-aware criterion that approximately minimizes interval score, described in Section~\ref{sec:experimental-settings}. We divide the axis into 12 logarithmically spaced bins and compute the average within-bin standard deviation of the interval scores. A smaller average indicates tighter alignment across persistence configurations and hence a stronger collapse. We report the percentage reduction in scatter obtained by replacing $\tau$ with $r$. 
\begin{equation}
100\left(1 - \frac{S_r}{S_\tau}\right),
\label{eq:collapse-percentage}
\end{equation}
where $S_\tau$ and $S_r$ denote the average within-bin standard deviation of interval scores when the horizontal axis is $\tau$ and $r$, respectively. For the two families driven instead by an abrupt, deterministic change in the residual distribution --- mean shift and variance shift --- persistence is not the relevant axis. We therefore evaluate these settings using \emph{adaptation duration}, defined as the time required for coverage to recover following the shift.\\

Table~\ref{tab:families} summarizes the six families, the mechanism of non-exchangeability each represents, and the primary diagnostic applied to it.
\begin{table}[t]
\centering
\caption{Summary of the six simulated residual families used to evaluate $\tau$-DACI.}
\label{tab:families}
\begin{tabularx}{\textwidth}{@{}l l X@{}}
\toprule
Family & Source of non-exchangeability & Primary diagnostic \\
\midrule
Gaussian (i.i.d.) & None (exchangeable control) & Baseline: cost of adaptivity when none is required \\
AR(1) & Mean dependence, single memory scale & Curve collapse under $r_\varepsilon=\tau/L$ \\
\addlinespace
GARCH(1,1) & Conditional variance dependence & Curve collapse under $r_{\varepsilon^2}=\tau/L_{\text{vol}}$; value of residual normalization as delay approaches $L_{\text{vol}}$ \\
Mean shift & Abrupt, deterministic level change & Adaptation duration (first coverage-dip recovery) vs.\ $\tau$ and calibration window $R$ \\
Variance shift & Abrupt, deterministic dispersion change & Adaptation duration (first coverage-dip recovery) vs.\ $\tau$ and calibration window $R$ \\
Markov-switching & Stochastic, recurring regime change & Curve collapse under $r_{\mathrm{regime}}=\tau/L_{\mathrm{regime}}$; regime-mismatch pathology when the target regime differs from the regime at construction \\
\bottomrule
\end{tabularx}
\end{table}

\subsection{Experimental settings}
\label{sec:experimental-settings}

Unless stated otherwise, each simulation uses $T=7000$ time steps, target miscoverage level $\alpha=0.1$, and 10 independent random seeds. We consider delays $\tau\in\{1,\allowbreak 5,\allowbreak 10,\allowbreak 15,\allowbreak 20,\allowbreak 30,\allowbreak 40\}$ and step sizes
$\gamma\in\{0.001,\allowbreak 0.002,\allowbreak 0.004,\allowbreak 0.008,\allowbreak 0.016,\allowbreak 0.032,\allowbreak 0.064,\allowbreak 0.128,\allowbreak 0.16\}$. For select analyses, this grid is extended or refined with additional $\tau$ or $\gamma$ values, or a longer simulation horizon to better resolve a specific pattern. Such extensions are visible in the corresponding figures. Any family-specific extensions to these grids are described in the corresponding subsection.\\

For each residual family and delay, we select the step size $\gamma^\star$ using a coverage-aware rule based on the interval score, averaged over the $10$ random seeds. When comparisons span multiple persistence values, shift magnitudes, calibration-window lengths, or normalization constructions, $\gamma^\star$ is selected separately for each configuration. We first retain all step sizes whose average interval score is within $1\%$ of the minimum over the candidate grid. Among these near-optimal values, we select the step size with the lowest interval score and empirical coverage of at least $0.89$, whenever such a candidate exists. If no near-optimal candidate satisfies this coverage threshold, we select the unrestricted interval-score minimizer. This rule favors solutions with approximately adequate coverage while avoiding a hard coverage constraint.\\

The calibration-window length $R$ is chosen based on the residual distribution. For the Gaussian, AR(1), and GARCH(1,1) families, the unconditional distribution is stationary, so residuals from the distant past remain informative. We use $R=500$ to reduce variability in the estimated conformal quantile. For the mean-shift and variance-shift families, pre-shift residuals may no longer represent the current distribution. For Markov-switching, a long window mixes residuals across regimes. In both cases, we use the shorter window $R=250$ to limit contamination and provide an additional source of adaptation alongside $\gamma$.\\

All reported results are averaged over the 10 seeds. Unless noted otherwise, the same simulation settings and evaluation procedure are used across residual families.

%For numerical stability, $\alpha$ was clipped to $$([tol,1-tol])$$ after each update. Consequently, the implementation differs slightly from the theoretical recursion, which permits temporary excursions outside ([0,1]). This safeguard may attenuate extreme adaptive responses near the boundaries, but it is applied uniformly across all forecast delays and residual regimes.

\subsection{Gaussian --- exchangeable baseline}
\label{sec:gaussian_methods}

We first consider independent Gaussian residuals,
\begin{equation}
\varepsilon_t \overset{\mathrm{i.i.d.}}{\sim}
\mathcal{N}(\mu,\sigma^2),
\end{equation}
with $\mu=0$ and $\sigma=1$. This setting satisfies the exchangeability assumption and therefore
serves as the baseline against which the adaptive methods are evaluated. It represents a stable, well-specified forecasting setting in which the residual variation consists solely of independent, irreducible noise.

\subsection{AR(1) --- temporal dependence}
\label{sec:ar_methods}

To study the effect of serial dependence, we generate residuals from the stationary AR(1) process
\begin{equation}
\varepsilon_{t+1}
=
\varphi \varepsilon_t + \eta_{t+1},
\qquad
\eta_t \overset{\mathrm{i.i.d.}}{\sim}
\mathcal{N}(0,\sigma_\eta^2).
\end{equation}

Such dependence may arise when the forecasting model fails to capture a persistent component of the data-generating process, such as a slowly varying trend or a seasonal pattern. The omitted structure may then be inherited by the residual sequence. For a stationary AR(1) process, the lag-$k$ autocorrelation is 
\[
\rho(k)=\operatorname{Corr}(\varepsilon_t,\varepsilon_{t+k})=\varphi^k. 
\]

Writing $\varphi^k=e^{-k/L}$ defines the memory length $L:=-1/\log\varphi$, at which the autocorrelation has decayed to $e^{-1}$ of its lag-zero value. We characterize the feedback delay relative to this intrinsic time scale using the delay-to-memory ratio $r$: 
\begin{equation}
\label{eq:ratio_ar}
    r_\varepsilon := \frac{\tau}{L}, \qquad L := -\frac{1}{\log \varphi}.
\end{equation}

The autocorrelation remaining across the delay is then $\rho(\tau)=e^{-r_\varepsilon}$. Thus, when $r_\varepsilon\ll1$, substantial dependence remains between the residuals available for calibration and the residual to be covered. When $r_\varepsilon\gg1$, this dependence has largely dissipated before feedback becomes available.\\

We vary $\varphi\in\{0.1,\allowbreak 0.6,\allowbreak 0.85,\allowbreak 0.95,\allowbreak 0.99\}$, corresponding to memory lengths $L\approx\{0.43,\allowbreak 1.96,\allowbreak 6.15,\allowbreak 19.50,\allowbreak 99.50\}$ steps. The resulting configurations span delay-to-memory ratios $r$ ranging from approximately $0.01$, for $\varphi=0.99$ and $\tau=1$, to approximately $92$, for $\varphi=0.1$ and $\tau=40$. The grid therefore includes regimes in which the delay is much shorter than, comparable to, and much longer than the dependence time scale.\\

To isolate the effect of persistence from changes in unconditional variance, we set the innovation standard deviation to $\sigma_\eta=\sqrt{1-\varphi^2}$. Equivalently, the one-step conditional variance is $\operatorname{Var}(\varepsilon_{t+1}\mid \varepsilon_t)=\sigma_\eta^2=1-\varphi^2$. Under stationarity, this choice fixes the unconditional variance at $\operatorname{Var}(\varepsilon_t)=\sigma_\eta^2/(1-\varphi^2)=1$ for every value of $\varphi$.

\subsection{GARCH(1,1) --- volatility clustering}
\label{sec:garch_methods}

To study time variation in residual scale, we generate residuals from a GARCH(1,1) process,
\begin{equation}
\label{eq:garch}
\varepsilon_t=\sigma_t \eta_t,
\qquad
\sigma_t^2
=
\omega+\alpha\varepsilon_{t-1}^2+\beta\sigma_{t-1}^2,
\qquad
\eta_t\overset{\mathrm{i.i.d.}}{\sim}\mathcal{N}(0,1).
\end{equation}
This process produces volatility clustering: large residuals tend to be followed by residuals of large magnitude, although their signs remain unpredictable. Such behavior may arise when the difficulty of the forecasting problem varies over time and persists across consecutive periods.\\

The distinction from the AR(1) setting concerns the feature of the residual distribution that is predictable. In the AR(1) process, dependence appears in the conditional mean of the residual. In the GARCH process, the conditional mean remains zero, while the conditional variance $\sigma_t^2$ evolves over time. Consequently, temporal information is relevant primarily for adapting the width of the prediction interval rather than its location.\\

The squared residuals of a GARCH(1,1) process admit an ARMA(1,1) representation. Let $\nu_t:=\varepsilon_t^2-\sigma_t^2$ denote the volatility innovation, which has
conditional mean zero. Substituting $\sigma_t^2=\varepsilon_t^2-\nu_t$ into \eqref{eq:garch} gives
\begin{equation}
\varepsilon_t^2
=
\omega+(\alpha+\beta)\varepsilon_{t-1}^2
+\nu_t-\beta\nu_{t-1}.
\end{equation}
Thus, the dependence in the squared residuals decays geometrically at a rate governed by $\alpha+\beta$. We therefore define the volatility memory length and the corresponding delay-to-memory ratio as
\begin{equation}
\label{eq:ratio_garch}
    r_{\varepsilon^2}:=\frac{\tau}{L_{\mathrm{vol}} } \qquad
    L_{\mathrm{vol}}:= \frac{-1}{\log(\alpha+\beta)}.
\end{equation}

Unlike $L$ in the AR(1) setting, $L_{\mathrm{vol}}$ measures persistence in residual magnitude rather than in the residual values themselves. We vary $\alpha+\beta\in\{0.5,\allowbreak 0.9,\allowbreak 0.95,\allowbreak 0.99\}$, corresponding to volatility memory lengths $L_{\mathrm{vol}}\approx\{1.44,\allowbreak 9.49,\allowbreak 19.50,\allowbreak 99.50\}$ steps. We fix $\alpha=0.1$ and set $\beta=(\alpha+\beta)-0.1$, so that persistence varies through a single parameter. To fix the unconditional variance across configurations, we set $\omega=1-\alpha-\beta$. Under stationarity, $\operatorname{Var}(\varepsilon_t)=\omega/(1-\alpha-\beta)$, so this choice yields $\operatorname{Var}(\varepsilon_t)=1$ for every persistence level.\\

A common refinement is to normalize residuals by an estimate of their conditional scale before forming \(\mathcal{D}_t\), and then rescale the resulting quantiles when constructing the prediction interval. This can reduce heteroskedasticity in the residuals and make their distribution more stable over time \citep{gibbs2021adaptive}. The GARCH family provides a natural setting for evaluating this construction because, unlike the Gaussian and AR(1) families, its conditional variance evolves over time. In this setting, normalization can reduce the effect of volatility clustering by expressing each residual relative to its conditional standard deviation. However, with delayed forecasting, an additional issue arises: the conditional scale at the time the interval is constructed may differ from the scale at the target time \(t+\tau\). To separate these two effects, we compare the unnormalized construction with two oracle normalization methods, summarized in Table~\ref{tab:garch-normalization}. The \emph{current-scale oracle} uses the true conditional standard deviation \(\sigma_t\) both to normalize the residual and to rescale the target interval. The \emph{target-scale oracle} also normalizes using \(\sigma_t\), but rescales the interval using the realized target-time scale \(\sigma_{t+\tau}\). The target-scale oracle is included specifically as a diagnostic to isolate whether deterioration in the current-scale oracle is caused by the delay-induced mismatch between the scale at interval construction and the scale realized at the target time, rather than by normalization itself. Because both methods rely on knowledge of the true conditional scale, they are non-deployable and are used only for diagnostic purposes.\\

To examine the normalization behavior once the feedback delay approaches and exceeds the volatility memory length, we additionally evaluate the most persistent configuration, \(\alpha+\beta=0.99\), at \(\tau\in\{100,300\}\), corresponding to \(r_{\varepsilon^2}\approx1.0\) and \(r_{\varepsilon^2}\approx3.0\), respectively. These extended-delay runs are used only for the normalization diagnostic and use a longer simulation horizon of \(T=12{,}000\).\\

\begin{table}[htbp]
\centering
\caption{Scale information used by the GARCH normalization experiments.}
\label{tab:garch-normalization}
\begin{tabular}{llll}
\toprule
Method
& Calibration divisor
& Target-interval multiplier
& Deployable \\
\midrule
Unnormalized
& $1$
& $1$
& Yes \\

Current-scale oracle
& $\sigma_t$
& $\sigma_t$
& No \\

Target-scale oracle
& $\sigma_t$
& $\sigma_{t+\tau}$
& No \\
\bottomrule
\end{tabular}
\end{table}

\subsection{Mean and Variance Shift --- Abrupt Changes in Residual Location and Scale}
\label{sec:mean_variance_shift_methods}

To study adaptation to abrupt changes in the residual distribution, we generate piecewise Gaussian residuals over three $800$-step segments: a baseline period, a shifted period, and a return to baseline. This setup captures both adaptation after the onset of a shift and recovery once the residual distribution returns to its initial state. Each shift magnitude or ratio is simulated separately, so the response to one shift does not affect the evaluation of another. The segment length is sufficient for the calibration window to discard pre-transition residuals and for post-transition behavior to be observed beyond the longest delay considered.\\

We use a primary calibration window of $R=250$. Unlike the stationary Gaussian, AR(1), and GARCH(1,1) settings, observations from before a change point may no longer be representative of the current residual distribution. A shorter window therefore provides an additional source of adaptation by removing all pre-transition residuals within $R$ steps. To examine whether faster turnover improves adaptation under repeated changes, we also repeat each combined-shift trajectory below with $R=100$, holding all other settings fixed.\\

We measure \emph{adaptation speed} by the duration of the first coverage dip following each shift, defined as the number of steps between the first time rolling empirical coverage, computed over a $100$-observation window, falls below $0.89$ and the first subsequent time it returns to at least $0.89$. Only the first dip is considered. If coverage does not recover before the segment ends, the duration is recorded at the segment boundary and treated as right-censored. We compute this measure for each combination of shift magnitude and $\tau$ using the corresponding $\gamma^\star$, selected separately for $R=250$ and $R=100$.\\

\paragraph{Mean shift:} We hold the standard deviation fixed at $\sigma=1$ and shift the mean,
\begin{equation}
\mu(t)=
\begin{cases}
0,      & 0 \leq t < 800,\\
\Delta, & 800 \leq t < 1600,\\
0,      & 1600 \leq t < 2400,
\end{cases}
\label{eq:mean-shift-segments}
\end{equation}
with $\Delta\in\{0.5,\allowbreak 1,\allowbreak 2,\allowbreak 4\}$, ranging from shifts small relative to the nominal Gaussian interval half-width, $z_{0.95}\sigma\approx1.645$, to shifts that substantially exceed it. To examine tracking under repeated shifts, we construct a combined-shift trajectory alternating baseline and each magnitude, $[\,0,\Delta_1,0,\Delta_2,0,\Delta_3,0,\Delta_4,0\,]$, with $800$-step segments and total horizon $T=7200$.\\

\paragraph{Variance shift:} We hold the mean fixed at $\mu=0$ and shift the standard deviation by a post-to-pre ratio $r=\sigma_{\mathrm{post}}/\sigma_{\mathrm{pre}}\in\{1.5,\allowbreak 2,\allowbreak 3,\allowbreak 5,\allowbreak 8\}$. Unlike an extreme mean shift, a large increase in scale does not force a miss at every time point, since the residual distribution remains centered at zero. The combined-shift trajectory alternates baseline and each ratio, $[\,1,r_1,1,r_2,1,r_3,1,r_4,1,r_5,1\,]$, with $800$-step segments and total horizon $T=8800$.\\ 

\subsection{Markov-switching --- stochastic and recurring regime changes}
\label{sec:markov_methods}

To study recurring changes in the residual distribution, we generate residuals from a
two-state hidden Markov model \citep{hamilton1994timeseries,zucchini2009hmm}. Let
$A_t\in\{0,1\}$ denote the latent regime at time $t$, with transition matrix
\[
P=
\begin{pmatrix}
p_{00} & 1-p_{00}\\
1-p_{11} & p_{11}
\end{pmatrix},
\qquad
P_{ij}=\Pr(A_t=j\mid A_{t-1}=i).
\]
Conditional on the active regime, the residual is generated according to
\begin{equation}
\varepsilon_t\mid A_t=j
\sim
\mathcal N(\mu_j,\sigma_j^2).
\label{eq:markov-emission}
\end{equation}
Conditional on the state sequence, the residual draws are independent over time.\\

The chain has stationary distribution $\pi=(\pi_0,\pi_1)$, where $\pi_j$ denotes the
long-run marginal probability of occupying state $j$. The stationary distribution satisfies $\pi P=\pi$ and $\pi_0+\pi_1=1$. Solving these
conditions gives
\[
\pi_0=\frac{1-p_{11}}{2-p_{00}-p_{11}},
\qquad
\pi_1=\frac{1-p_{00}}{2-p_{00}-p_{11}}.
\]
In the simulations, the latent chain is initialized in state $0$. We generate and
discard 200 burn-in observations before retaining the $T$ observations used in the
experiment. This burn-in reduces the influence of the fixed initial state and allows
the state distribution to approach $\pi$ before evaluation begins.\\

Unlike the deterministic mean- and variance-shift families, regime changes in the Markov-switching setting occur randomly and repeatedly. The construction represents an environment that alternates between operating conditions that tend to remain active for multiple consecutive time steps, such as baseline and shifted periods or calm and volatile periods. This distinction is particularly relevant under delayed feedback. An interval issued at time $t$ for target time $t+\tau$ is constructed using information available at or before time $t$, but the target residual is generated under state $A_{t+\tau}$, which may differ from the state at construction.\\

\noindent\textbf{Delay-to-memory ratio:}
For the two-state transition matrix above, the unique non-unit eigenvalue is
$\lambda_2=p_{00}+p_{11}-1$. Appendix~\ref{app:regime-autocorr} derives the corresponding
lag-$k$ state autocorrelation,
\begin{equation}
\operatorname{Corr}(A_t,A_{t+k})
=
\lambda_2^k,
\qquad
k\geq0.
\label{eq:regime-autocorr}
\end{equation}
Here, persistence refers specifically to the rate at which information about the
current latent state is lost across time. A larger positive value of $\lambda_2$
corresponds to slower decay and hence greater persistence. All transition matrices used
in the experiments satisfy $0<\lambda_2<1$, so the state autocorrelation decays
monotonically with the lag.\\

Following the same convention used for the AR(1) and GARCH(1,1) families, we define the
regime memory length and delay-to-memory ratio as
\begin{equation}
\label{eq:ratio_markov}
   r_{\mathrm{regime}}:=\frac{\tau}{L_{\mathrm{regime}}} \qquad L_{\mathrm{regime}}:=\frac{-1}{\log|\lambda_2|}
\end{equation}

The latent-state dependence remaining
across the delay is therefore
$|\lambda_2|^\tau=e^{-r_{\mathrm{regime}}}$. Thus,
$r_{\mathrm{regime}}\ll1$ indicates that substantial dependence remains between the
states at interval construction and at the target time, whereas
$r_{\mathrm{regime}}\gg1$ indicates that the chain has largely forgotten its earlier
state.\\

\noindent\textbf{Markov-switching regimes:}
We consider three state-dependent residual constructions:
\begin{equation}
\begin{array}{lll}
\text{Mean switching:}
& \mu_0=0,\quad \mu_1=\Delta,
& \sigma_0=\sigma_1=1,\\[2mm]
\text{Variance switching:}
& \mu_0=\mu_1=0,
& \sigma_0=1,\quad \sigma_1=r,\\[2mm]
\text{Joint switching:}
& \mu_0=0,\quad \mu_1=\Delta,
& \sigma_0=1,\quad \sigma_1=r.
\end{array}
\label{eq:markov-families}
\end{equation}
We fix $\Delta=4$ and $r=5$.\\

Appendix~\ref{app:residual-feature-autocov} derives how latent-state persistence appears in the observed residuals and their transformations. Under mean and joint switching, the residual level inherits the geometric decay factor $\lambda_2^k$, although its autocorrelation magnitude is attenuated relative to the latent-state autocorrelation by within-state randomness. Under variance switching, the signed residual has zero autocorrelation because its conditional mean is zero in both states; the regime dependence instead appears in $|\varepsilon_t|$ and $\varepsilon_t^2$. Thus, across the three switching families, the relevant memory length remains $L_{\mathrm{regime}}$, determined by the latent state process rather than by the attenuation in any particular observed feature. The corresponding positive-lag autocorrelation expressions are summarized in Table~\ref{tab:markov-feature-correlations} in Appendix~\ref{app:residual-feature-autocov}.\\

\noindent\textbf{Symmetric transitions:}
For the primary experiment, we use symmetric transition matrices,
\begin{equation}
P(p)=
\begin{pmatrix}
p & 1-p\\
1-p & p
\end{pmatrix},
\qquad
p\in\{0.80,0.90,0.95,0.97,0.99\}.
\label{eq:markov-symmetric}
\end{equation}
Symmetry fixes $\pi_0=\pi_1=0.5$, so the memory of the state process can be varied without changing the long-run frequency of either regime. Since $\mu_0,\mu_1,\sigma_0,\sigma_1$ are held fixed across the grid, this also fixes the marginal residual distribution across persistence levels, isolating the effect of regime memory within each switching family. The corresponding non-unit eigenvalues are $\lambda_2=2p-1\in\{0.60,0.80,0.90,0.94,0.98\}$, giving memory lengths
$L_{\mathrm{regime}}\approx\{1.96,4.48,9.49,16.16,49.50\}$, ranging from frequent switching to long regime episodes. The memory length $L_{\mathrm{regime}}$, which governs autocorrelation decay, is distinct from the mean dwell time in a regime. The dwell time, $t_{\mathrm{dwell}}=1/(1-p)$, describes the expected duration of one uninterrupted regime episode. The corresponding dwell times are $t_{\mathrm{dwell}}\in\{5,10,20,33.3,100\}$ across the same grid.\\

\noindent\textbf{Normalization:}
We apply the same two normalization constructions used in the GARCH experiments (Table~\ref{tab:garch-normalization}), replacing the conditional scales $\sigma_t$ and $\sigma_{t+\tau}$ with the state-dependent scales $\sigma_{A_t}$ and $\sigma_{A_{t+\tau}}$. The current-state oracle uses $\sigma_{A_t}$ for both residual normalization and target-interval scaling. The target-state oracle also normalizes by $\sigma_{A_t}$, but rescales the interval using $\sigma_{A_{t+\tau}}$. Comparing the two isolates the effect of the delay-induced mismatch between the scale at interval construction and the scale at the target time, assuming perfect knowledge of the current state. We restrict this experiment to the variance- and joint-switching regimes.\\

\noindent\textbf{Regime-transition diagnostic:}
To examine whether interval performance deteriorates when the target regime differs from the regime at interval construction, we compare interval scores according to whether the latent state changes across the forecast horizon, defining the mismatch indicator $M_{t,\tau}:=\mathbf{1}\{A_t\neq A_{t+\tau}\}$. Because the interval depends on a rolling calibration window rather than solely on $A_t$, mismatch status is an explanatory diagnostic rather than a complete characterization of the information used to construct
the interval. Under stationarity,
\begin{equation}
\Pr(M_{t,\tau}=1)
=
2\pi_0\pi_1\left(1-\lambda_2^\tau\right),
\label{eq:markov-mismatch-probability}
\end{equation}
which for a symmetric chain reduces to $\Pr(M_{t,\tau}=1)=\tfrac12(1-e^{-r_{\mathrm{regime}}})$,
connecting the delay-to-memory ratio directly to the mismatch frequency.\\

\section{The Effect of Delay on ACI Across Residual Regimes}
\label{sec:delay-effects}
We organize the results by residual family. Throughout, we attend to how the step-size interacts with delay and persistence jointly to shape coverage, efficiency, and local calibration. 

\subsection{Gaussian --- exchangeable baseline}
\label{sec:gaussian_results}

\begin{figure}[htbp]
    \centering
    \includegraphics[width=0.72\textwidth]{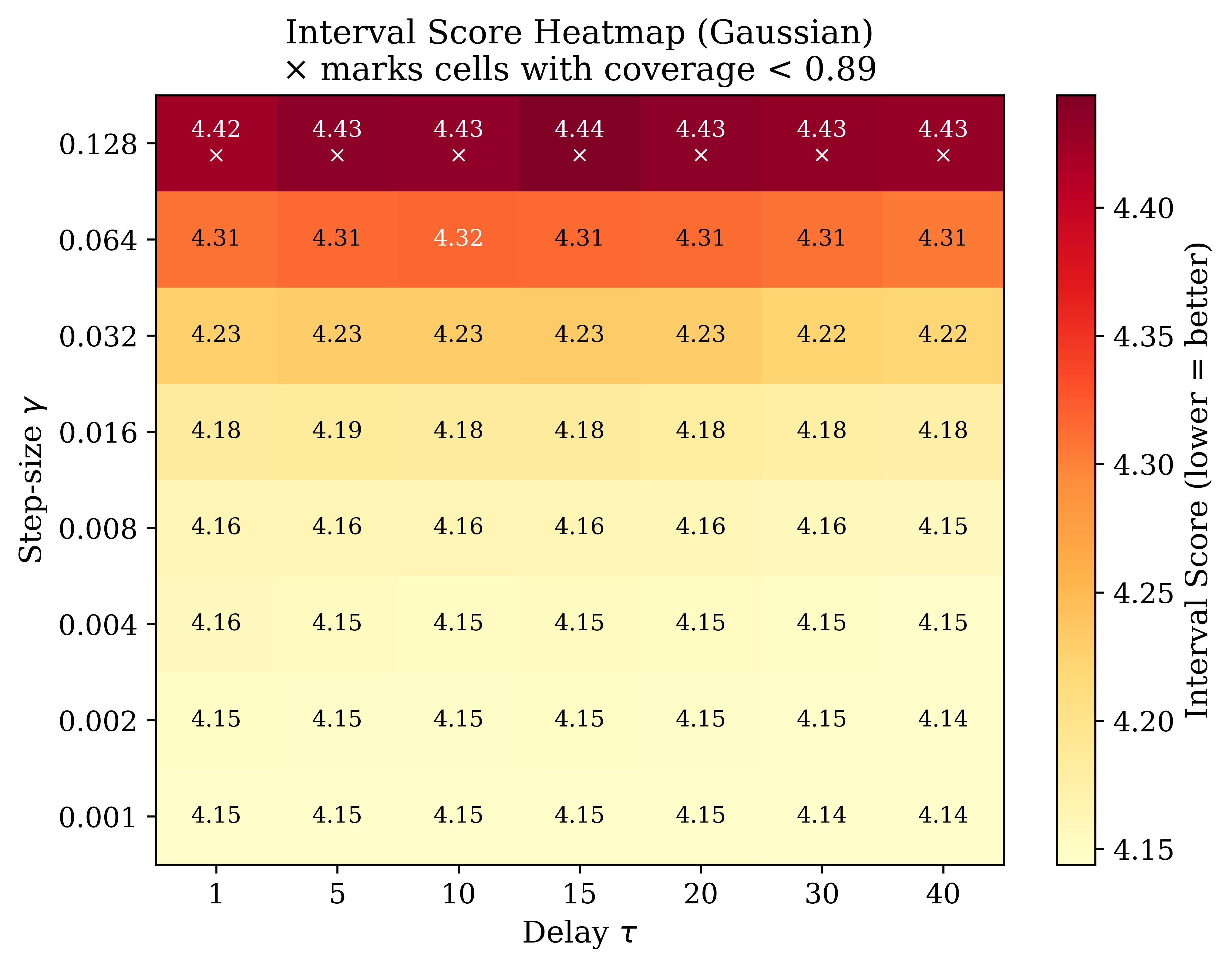}
    \caption{Interval score heatmap for Gaussian (i.i.d.) residuals across the $\gamma \times \tau$ grid. Markers (\texttt{×}) indicate coverage below 0.89.}
    \label{fig:gaussian_heatmap}
\end{figure}

Figure~\ref{fig:gaussian_heatmap} confirms the known one-step-ahead result of \citet{zaffran2022adaptive} in the $\tau$-delayed setting. Under exchangeable i.i.d.\ residuals, increasing $\gamma$ monotonically worsens the interval score, from approximately $4.15$ at $\gamma = 0.001$ to roughly $4.43$ at $\gamma = 0.128$. This amounts to a $7\%$ efficiency loss with no coverage benefit. The effect is invariant to the delay: every row of the heatmap is flat across all seven values of $\tau$. When residuals are exchangeable and contain no temporal dependence, past residuals provide no information about future residuals, so delaying their feedback neither improves nor worsens performance. Coverage failure, marked $\times$ in the figure, occurs only for $\gamma = 0.128$ and uniformly across all $\tau$, indicating that it is caused by an overly large step size rather than by delay. This panel therefore provides a baseline in which $\tau$-DACI recovers classical ACI behavior when there is no temporal dependence for the delay to interact with.\\

\subsection{AR(1) --- temporal dependence}
\label{sec:ar_results}

\begin{figure}[htbp]
    \centering
    \includegraphics[width=\textwidth]{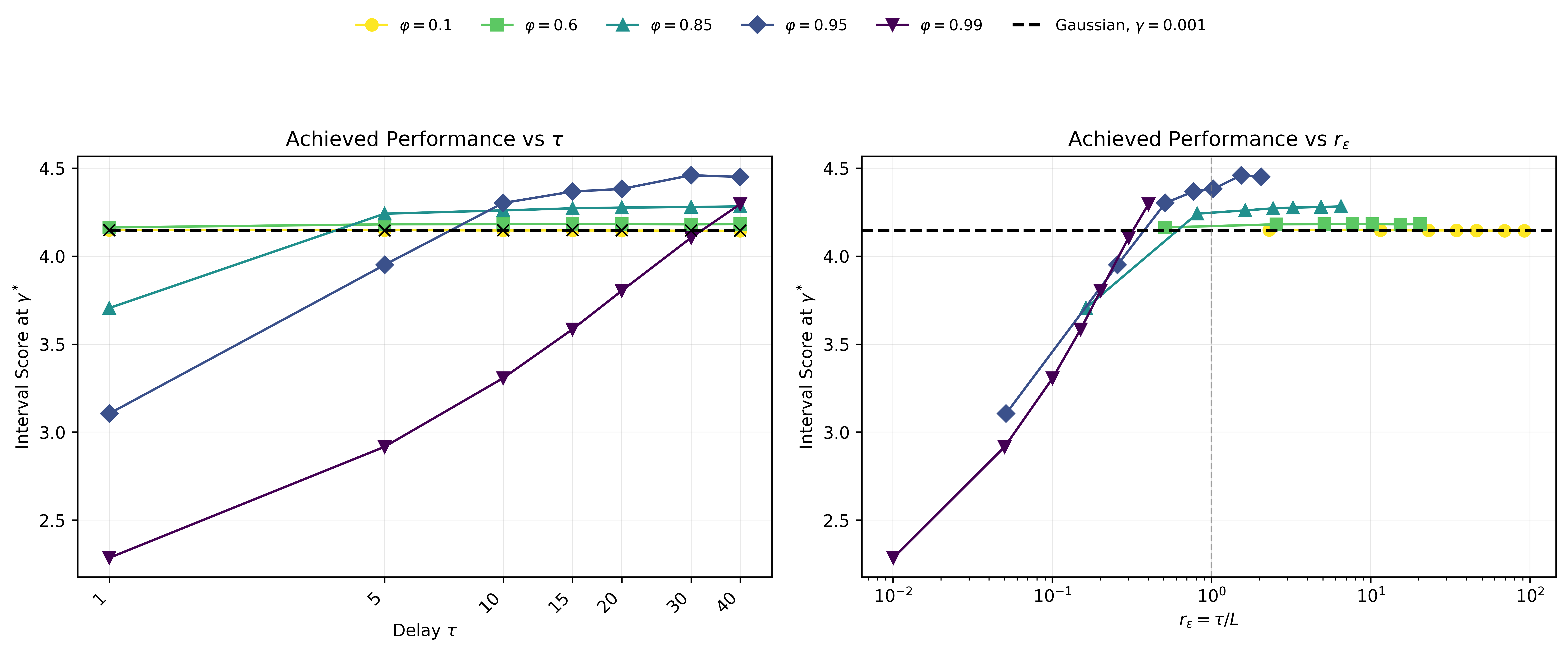}
    \caption{Achieved interval score for AR(1) residuals at the selected step size $\gamma^\star$, shown by persistence level as a function of the raw delay $\tau$ (left) and the delay-to-memory ratio $r_\varepsilon$ (right). The dashed black line shows the i.i.d.\ Gaussian reference at $\gamma=0.001$.}
    \label{fig:ar_performance_collapse}
\end{figure}

Figure~\ref{fig:ar_performance_collapse} reports the achieved interval score at the approximately score-minimizing step size $\gamma^\star$, selected using the coverage-aware criterion described in Section~\ref{sec:experimental-settings}. In the left panel, plotted against the delay $\tau$, each persistence level traces a distinct curve. Higher persistence is associated with lower scores at $\tau=1$ and a steeper increase as the delay grows, indicating that performance depends jointly on $\varphi$ and $\tau$. In the right panel, the score is plotted against the ratio $r_\varepsilon=\tau/L$. This reparameterization produces a substantial collapse across persistence levels. Plotting against $r_\varepsilon$ reduces within-bin scatter by approximately $79\%$ relative to plotting against $\tau$. This suggests that $r_\varepsilon$ is a useful scaling variable for performance under delay, although some persistence-dependent variation remains.\\

The collapsed relationship shows a transition near $r_\varepsilon\approx1$. For $r_\varepsilon\lesssim1$, the score is generally below the Gaussian baseline. The method attains a lower score than under exchangeable residuals with the same marginal variance, which is consistent with the approximately score-minimizing step size exploiting dependence between the residuals available for calibration and the residual $\tau$ steps ahead that the interval must cover. The advantage is largest at small $r_\varepsilon$; for example, at $\varphi=0.99$ and $\tau=1$, the score is approximately $2.3$, compared with a baseline near $4.15$. This advantage holds without a coverage cost at $\varphi=0.95$, where coverage remains close to the nominal target. At $\varphi=0.99$, coverage in the same region falls to $0.83$--$0.86$, indicating that at the highest persistence level, part of the efficiency gain comes at the cost of coverage. For $r_\varepsilon\gtrsim1$, the score approaches or exceeds the baseline. This is consistent with the relevant dependence having largely decayed once the delay reaches the memory scale. Under weak persistence, the score levels off near the Gaussian baseline. Under strong persistence, particularly for $\varphi=0.95$, it remains above the baseline. One possible explanation is that serial dependence reduces the effective number of independent observations in the calibration window.\\

\begin{figure}[htbp]
    \centering
    \includegraphics[width=\textwidth]{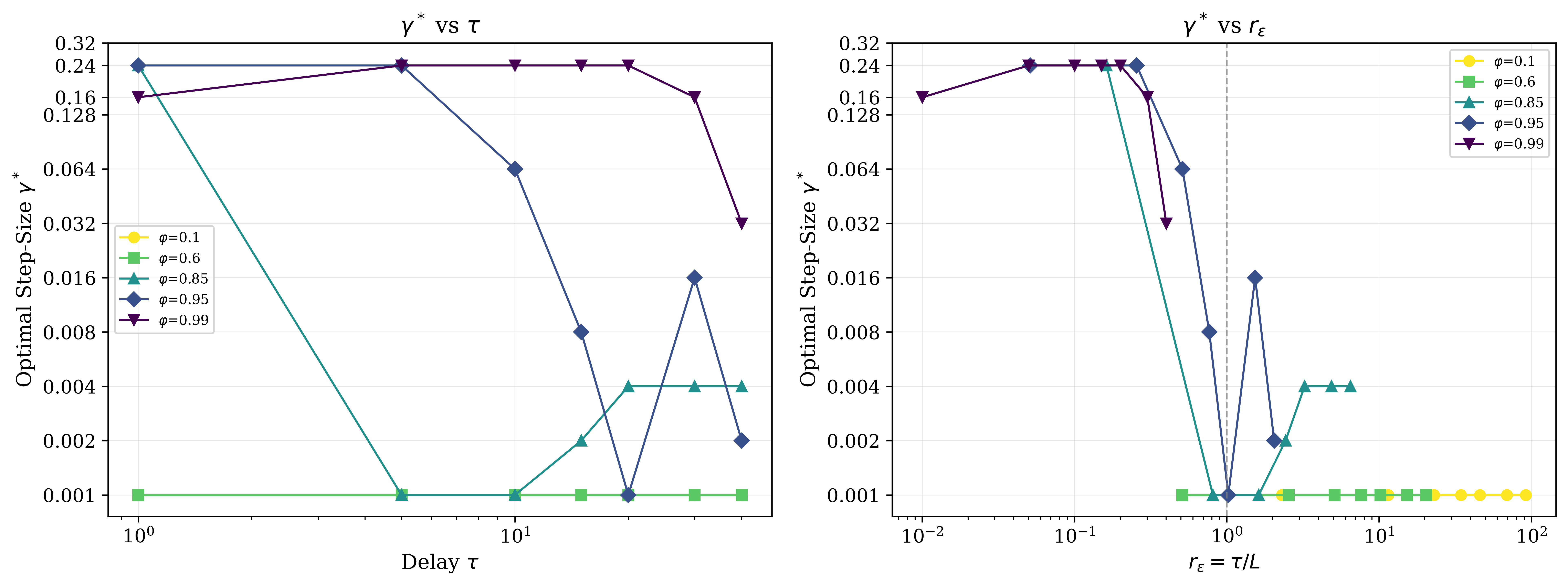}
    \caption{Approximately score-minimizing step-size $\gamma^\star$ for AR(1) residuals, against the delay $\tau$ (left) and the delay-to-memory ratio $r_\varepsilon$ (right).}
    \label{fig:ar_gamma_star}
\end{figure}

Figure~\ref{fig:ar_gamma_star} reports the optimal step size $\gamma^\star$. As with the achieved score, $\gamma^\star$ is organized more clearly by the ratio $r_\varepsilon$ than by $\tau$ alone. In the left panel, its dependence on $\tau$ differs markedly across persistence levels, and the curves do not share a common transition. Against $r_\varepsilon$, the behavior is more systematic. For $r_\varepsilon \ll 1$, large step sizes are optimal. As $r_\varepsilon$ approaches one, $\gamma^\star$ falls sharply, often reaching the lower boundary of the grid. This decline occurs near $r_\varepsilon \approx 1$, where the achieved interval score also begins to level off, indicating a change in the optimal adaptation regime as the delay approaches the process memory length. When correlation survives the delay, a large step size allows the procedure to adapt to exploitable structure. Once that dependence has decayed, the residuals available for calibration carry little information about the target, and performance converges toward the exchangeable-residual case (Section~\ref{sec:gaussian_results}), where increasing $\gamma$ monotonically worsens the interval score with no coverage benefit. This is consistent with the smallest step size minimizing the interval score once $r_\varepsilon$ exceeds roughly one. Beyond the transition, however, $\gamma^\star$ does not remain at the grid floor for the more persistent processes but returns to a slightly larger value. This reflects the coverage-aware selection rule: where the score-minimizing step size undercovers, the rule selects a larger step that restores coverage in most cases.\\

\begin{figure}[htbp]
    \centering
    \includegraphics[width=\textwidth]{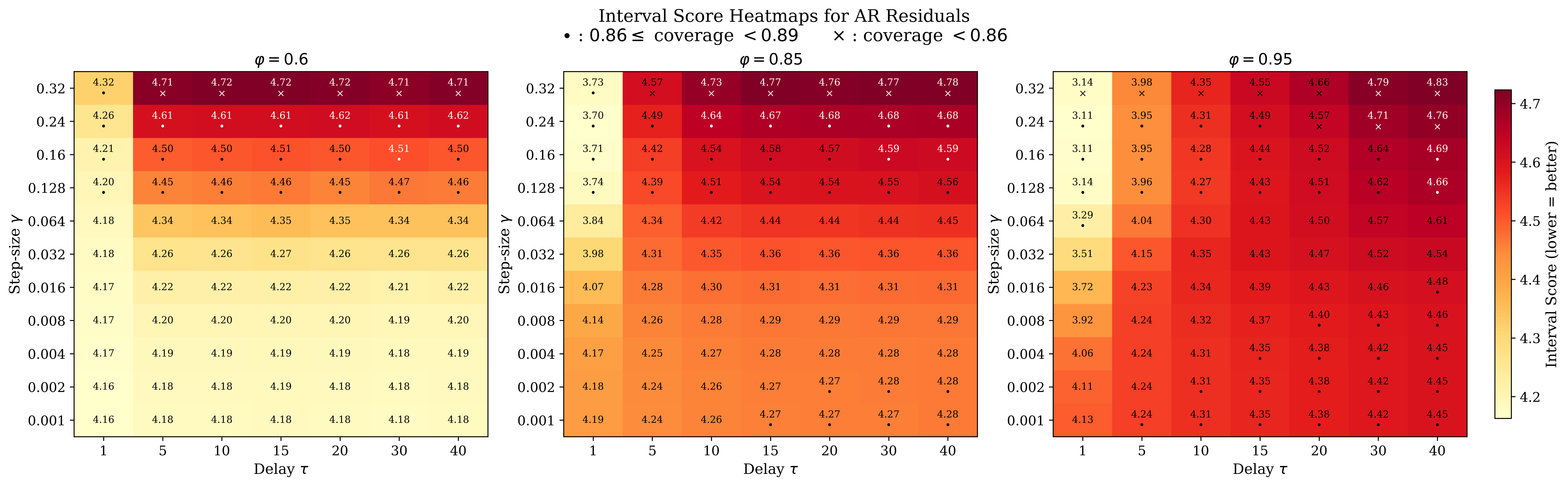}
    \caption{Interval score across the $\gamma \times \tau$ grid for AR(1) residuals at three persistence levels ($\varphi = 0.6$, $0.85$, $0.95$). A dot marks mild undercoverage ($0.86 \leq \text{coverage} < 0.89$) and a cross marks ($\text{coverage} < 0.86$); unmarked cells meet coverage.}
    \label{fig:ar_heatmaps}
\end{figure}

Figure~\ref{fig:ar_heatmaps} reports the interval score over the $(\gamma,\tau)$ grid for three persistence levels. Dots indicate mild undercoverage, $0.86 \leq \mathrm{coverage} < 0.89$, while crosses indicate coverage below $0.86$. The coverage-feasible region, consisting of unmarked cells, shrinks as $\varphi$ increases. For $\varphi=0.6$, undercoverage is concentrated at large step sizes and becomes substantial only at $\gamma=0.32$. For $\varphi=0.85$ and $\varphi=0.95$, mild undercoverage advances from two directions: from large $\gamma$ throughout the grid and from small $\gamma$ at large $\tau$. These heatmap figures show why step-size selection should not rely on interval score alone. At higher persistence, the score-minimizing cells themselves undercover, though only mildly (marked by dots rather than crosses). Because undercoverage at the absolute minimum is mild, a coverage-compliant near-optimal step size is typically available, allowing coverage to be favored at little cost in efficiency. Substantial undercoverage, marked by crosses, occurs away from the score-minimizing region, where more aggressive updating worsens both calibration and interval score.\\

Local coverage error exhibits a different pattern from interval score. Although it generally increases as the delay grows relative to the memory length, the curves do not collapse across persistence levels, indicating that \(r_\varepsilon\) alone does not explain differences in the magnitude of local calibration error. The full results are reported in Appendix~\ref{app:ar-results} (Figure~\ref{fig:ar_lce}).\\

\subsection{GARCH(1,1) --- volatility clustering}
\label{sec:garch_results}

\begin{figure}[htbp]
    \centering
    \includegraphics[width=\textwidth]{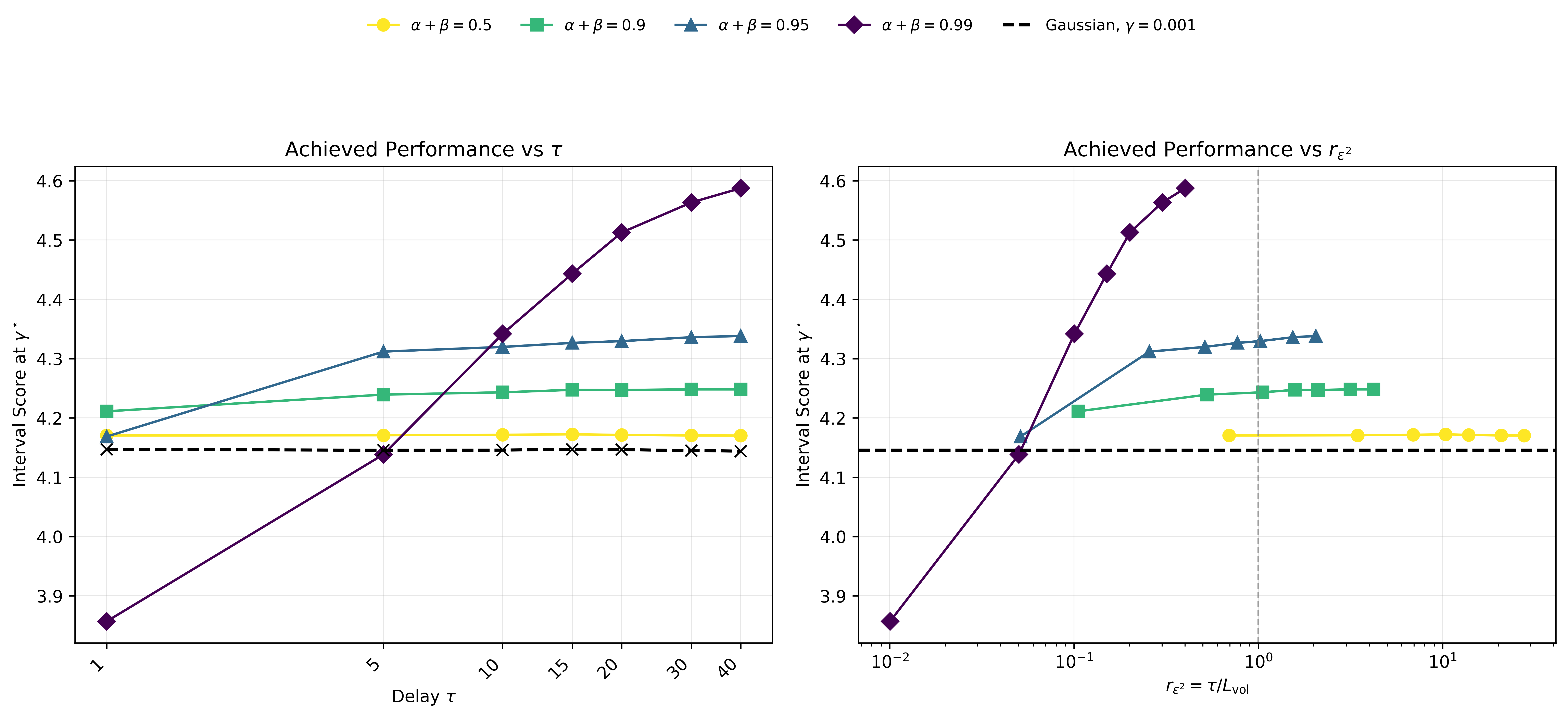}
    \caption{Achieved interval score at the selected step-size $\gamma^\star$ for GARCH(1,1) residuals (unnormalized
    construction), against the delay $\tau$ (left) and the delay-to-memory
    ratio $r_{\varepsilon^2}$ (right), across
    persistence levels.}
    \label{fig:garch-performance-collapse}
\end{figure}

Figure~\ref{fig:garch-performance-collapse} reports the achieved interval
score for the unnormalized-residual construction at the optimal step size
$\gamma^\star$, plotted against the delay $\tau$ in the left panel and the
delay-to-memory ratio $r_{\varepsilon^2}$ in the
right panel. In terms of the raw delay, the persistence regimes exhibit
distinct behavior. For $\alpha+\beta=0.99$, the interval score increases
from approximately $3.86$ at $\tau=1$, below the i.i.d.\ Gaussian
baseline, to approximately $4.59$ at $\tau=40$, crossing the baseline
between $\tau=5$ and $\tau=10$. The $\alpha+\beta=0.95$ curve shows a
similar but weaker deterioration, whereas the scores for
$\alpha+\beta\in\{0.5,\allowbreak 0.9\}$ vary only mildly with $\tau$ and
remain close to the Gaussian benchmark.\\

Reparameterizing the horizontal axis by $r_{\varepsilon^2}$ improves the
alignment across persistence regimes, reducing the collapse metric by
approximately $46\%$ relative to the raw-delay representation. The
improvement is nevertheless less pronounced than in the AR(1) setting,
where the corresponding reduction is approximately $79\%$. Moreover,
unlike the AR(1) curves, the GARCH(1,1) curves provide limited evidence of
a common change in behavior near $r_{\varepsilon^2}=1$. A substantial
improvement over the i.i.d.\ Gaussian baseline is observed only for the
most persistent process, $\alpha+\beta=0.99$, at the shortest delay. At
the remaining delays and persistence levels, achieved performance is
generally close to or worse than the Gaussian baseline.  The larger
gains observed under AR(1) may indicate that dependence in the signed
residuals is more directly exploitable by the adaptive procedure than
dependence confined to the conditional variance, although the present
experiments do not isolate this mechanism. Thus, the delay-to-memory ratio organizes
the variation across GARCH persistence levels more effectively than the
raw delay, but it does not identify a sharp threshold separating regimes
in which volatility dependence is and is not beneficial.\\

\begin{figure}[htbp]
    \centering
    \includegraphics[width=0.85\textwidth]{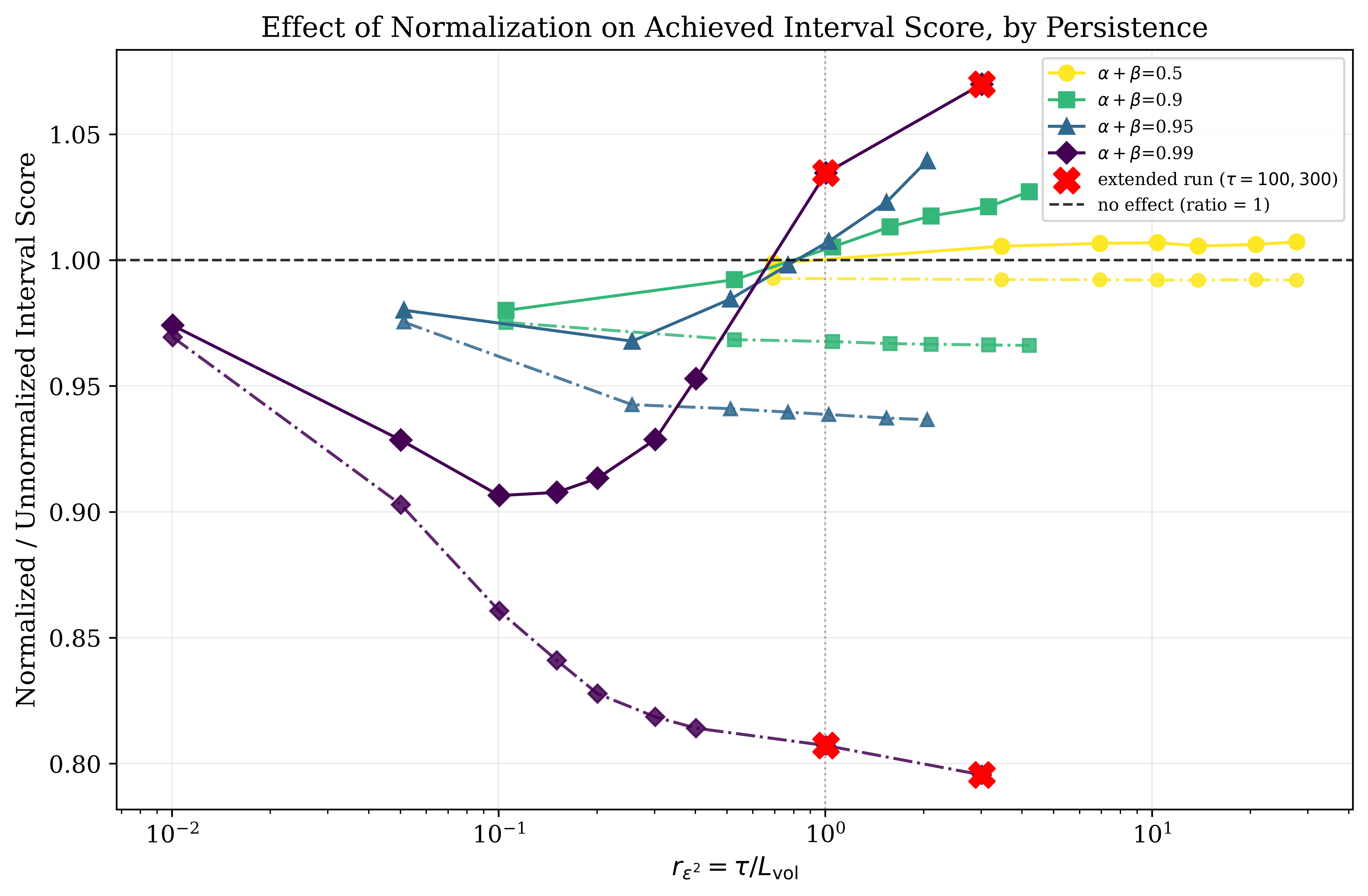}
    \caption{Ratio of the interval score obtained with normalized residuals to that obtained with unnormalized residuals for the GARCH(1,1) process, shown by persistence level as a function of $r_{\varepsilon^2}$. Solid lines correspond to the current-scale oracle, while dash-dotted lines correspond to the target-scale oracle.}
    \label{fig:garch-normalization-ratio-oracle}
\end{figure}

Figure~\ref{fig:garch-normalization-ratio-oracle} reports the ratio of the interval score obtained with normalized residuals to that obtained with unnormalized residuals as a function of $r_{\varepsilon^2}$. Ratios below one indicate that normalization improves interval performance, whereas ratios above one indicate that it worsens performance. For the current-scale oracle, replacing $\tau$ with $r_{\varepsilon^2}$ in the curve-collapse diagnostic reduces the within-bin scatter by approximately $50\%$, indicating that the benefit of normalization is strongly related to the delay relative to the volatility memory length.\\

For the current-scale oracle, normalization is generally beneficial when $r_{\varepsilon^2}<1$, particularly for the more persistent processes, whereas its effect is negligible for $\alpha+\beta=0.5$. As the delay approaches and exceeds the volatility memory length, the score ratios rise toward or above one. The extended-delay results for $\alpha+\beta=0.99$ reinforce this pattern: normalization becomes disadvantageous near $r_{\varepsilon^2}=1$ and deteriorates further by $r_{\varepsilon^2}\approx3$. To assess whether this deterioration is driven by scale mismatch rather than by normalization itself, we use the target-scale oracle as a diagnostic. Its score ratios remain below one throughout the plotted range. The improvement is modest for the lower-persistence processes but becomes substantial for $\alpha+\beta=0.99$, where the ratio continues to decline over the extended-delay runs and reaches approximately $0.80$ by $r_{\varepsilon^2}\approx3$. The widening separation between the current-scale and target-scale oracle curves indicates that the deterioration at longer delays is largely attributable to using the prediction-time scale $\sigma_t$ to rescale an interval intended for time $t+\tau$. As prediction-time scale information becomes less informative at longer delays, access to the realized target-time scale becomes increasingly valuable. Although this oracle information is unavailable in practice, the result suggests that more accurate forecasts of $\sigma_{t+\tau}$ could recover part of this benefit.\\

Normalization also generally improves local coverage stability. Across most GARCH configurations, the current-scale oracle yields lower worst-case local coverage error than the unnormalized construction, including at delays where its interval-score advantage has disappeared. This suggests that the delay-induced scale mismatch affects interval efficiency more strongly than local coverage stability. Full results are reported in Appendix~\ref{app:garch-results} (Figure~\ref{fig:garch-lce-unnorm-vs-norm}).

\subsection{Mean and variance shifts}
\label{sec:mean_variance_shift_results}

\begin{figure}[htbp]
    \centering
    \includegraphics[width=\textwidth]{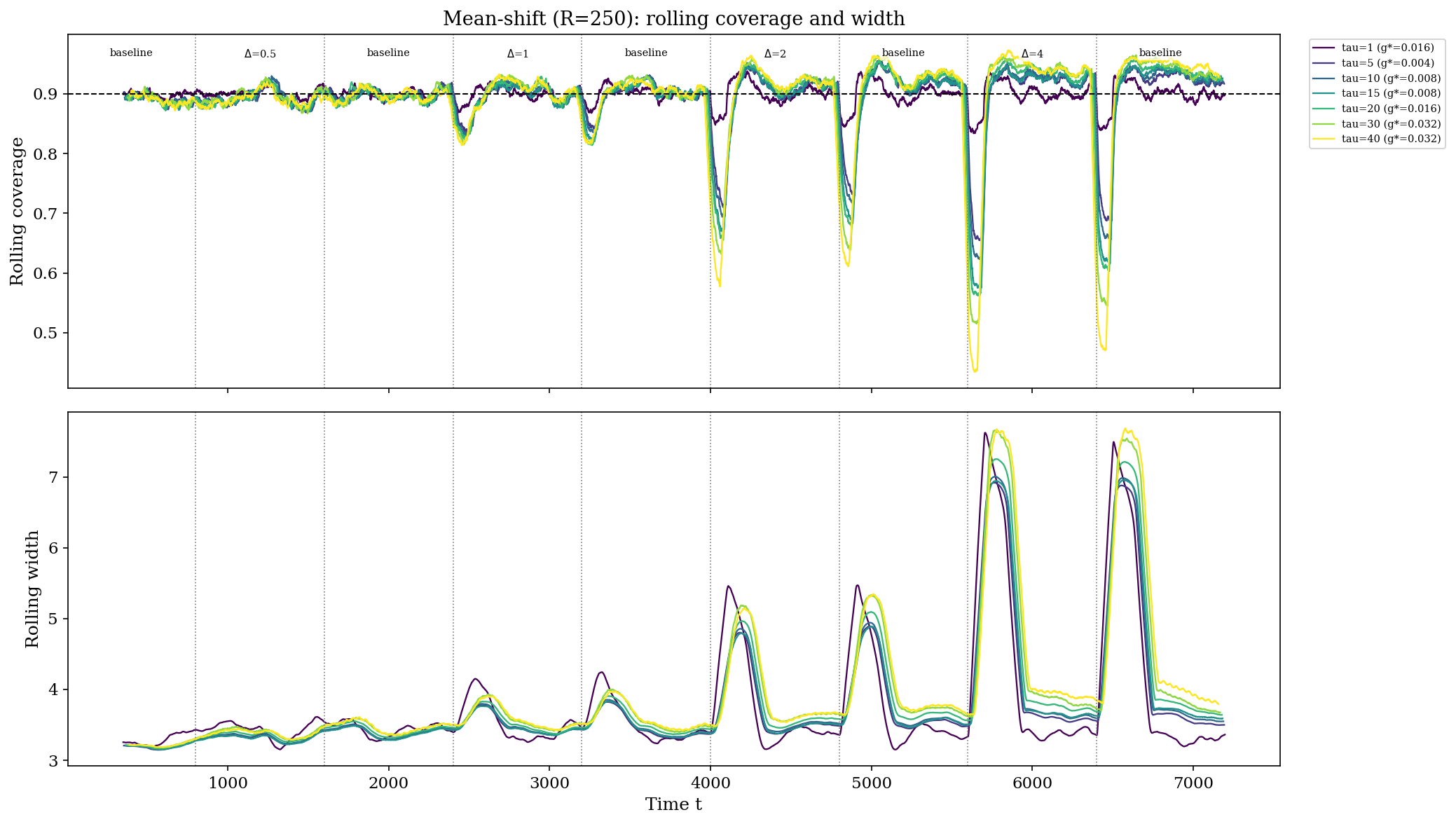}
    \caption{Rolling coverage (top) and interval width (bottom) along the combined mean-shift trajectory using calibration-window length $R=250$, computed over a window of $100$ observations.}
    \label{fig:mean-shift-coverage-width}
\end{figure}

\begin{figure}[htbp]
    \centering
    \includegraphics[width=\textwidth]{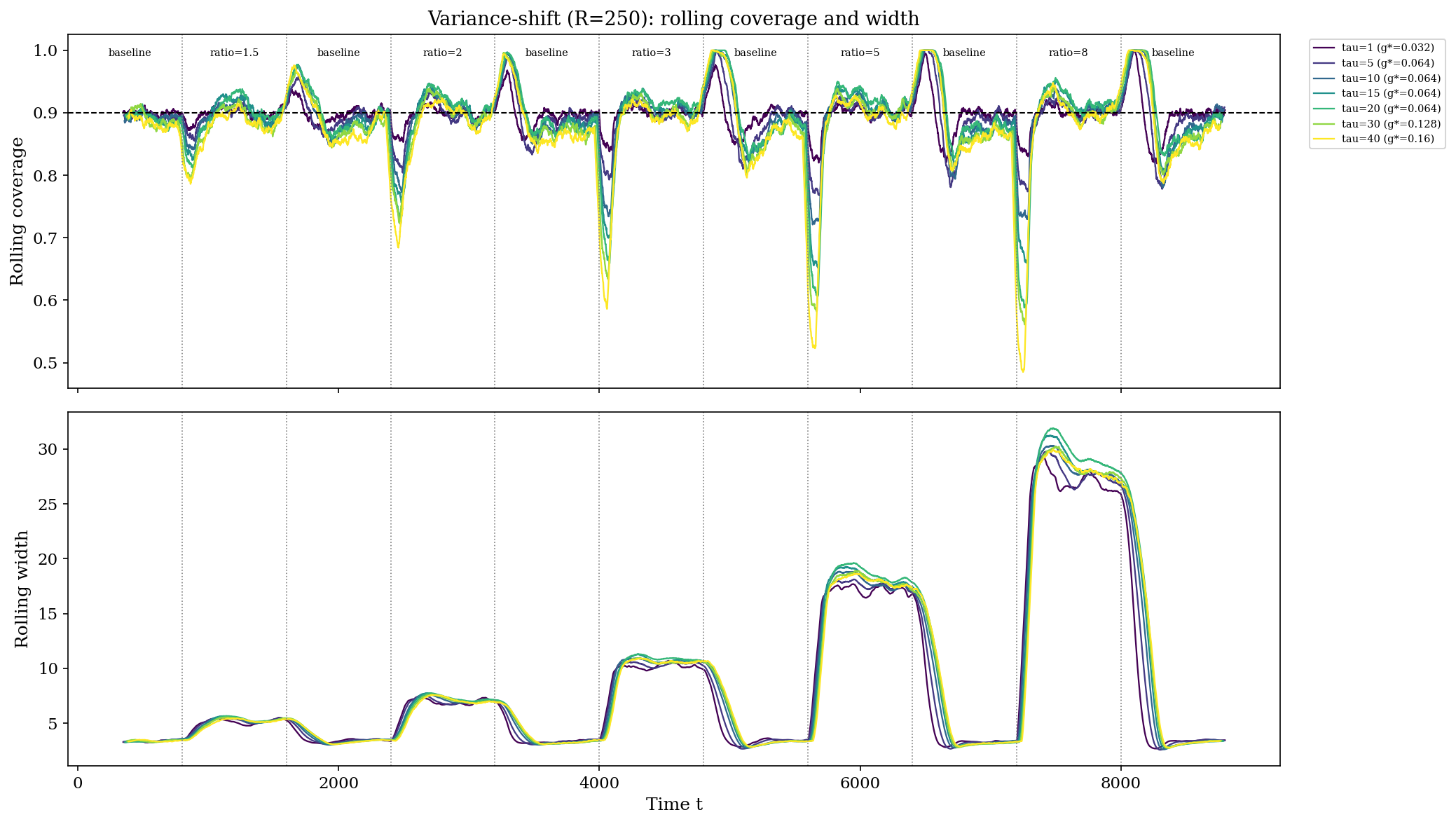}
    \caption{Rolling coverage (top) and interval width (bottom) along the combined variance-shift trajectory using calibration-window length $R=250$, computed over a window of $100$ observations.}
    \label{fig:variance-shift-coverage-width}
\end{figure}

Figures~\ref{fig:mean-shift-coverage-width} and \ref{fig:variance-shift-coverage-width} show rolling coverage and width along the combined-shift trajectories described in Section~\ref{sec:mean_variance_shift_methods}, in which successive regime changes increase in magnitude before returning to the baseline regime. Results are shown using the primary calibration-window length $R=250$. In both families, regime transitions produce localized coverage failures whose depth generally increases with shift magnitude and feedback delay. Longer delays produce more severe transient miscoverage, although their effect on the duration of these failures is less systematic. The corresponding width trajectories reveal an important structural difference between changes in residual location (mean shifts) and changes in residual scale (variance shifts). These qualitative coverage and width patterns remain visible when the rolling window is shortened from $100$ to $20$ observations (Figures~\ref{fig:mean-shift-coverage-width-window20} and \ref{fig:variance-shift-coverage-width-window20} in the appendix \ref{app:mean-variance-shift-results}). Mean shifts produce sharp width expansions followed by gradual contraction as the interval adapts to the new residual location. Variance shifts instead produce sustained, regime-specific width plateaus that increase with residual scale. During variance shifts, after the process returns to baseline, the temporarily oversized intervals generate overcoverage while width contracts. Thus, mean shifts produce transient width spikes at regime transitions, whereas variance shifts produce sustained, regime-specific width plateaus.\\

Although coverage troughs generally deepen with delay, adaptation duration is not monotonic in $\tau$ for either shift family. For example, under $R=250$,
adaptation duration decreases from $131$ steps at $\tau=5$ to $114$ steps at $\tau=40$ for $\Delta=4$, and from $165$ to $143$ steps for $r=8$ over the
same delays; see Tables~\ref{tab:mean-shift-adaptation-speed} and \ref{tab:variance-shift-adaptation-speed}. This is not contradictory because
adaptation duration measures the time required for rolling coverage to return above $0.89$, not the depth of the coverage failure. The delay-specific
selection of $\gamma^\star$ may partially offset delayed feedback: the larger step sizes selected at the longest delays permit stronger corrections once feedback becomes available. This may limit adaptation duration even when the initial coverage loss is deeper. The selected step sizes are shown in Figure~\ref{fig:shift-optimal-step-size} in
Appendix~\ref{app:mean-variance-shift-results}.\\

Shortening the calibration window has different effects across the two families. For mean shifts, $R=100$ produces the shorter adaptation duration in $24$ of the $28$ configurations in Table~\ref{tab:mean-shift-adaptation-speed}, so the improvement extends across most shift magnitudes and delays. The comparison is particularly informative for $\tau\in\{10,15,20\}$, where $R=100$ and $R=250$ select the same $\gamma^\star$ (Figure~\ref{fig:shift-optimal-step-size}, Appendix~\ref{app:mean-variance-shift-results}). The shorter adaptation durations at these delays can therefore be attributed primarily to faster removal of residuals from the preceding location regime. At $\tau\in\{5,30,40\}$, $R=100$ also improves adaptation duration in most configurations despite selecting a smaller $\gamma^\star$ than $R=250$. The results for $\Delta=0.5$ are less stable because coverage remains close to the threshold, making adaptation duration sensitive to brief crossings. For variance shifts, the pattern reverses. Table~\ref{tab:variance-shift-adaptation-speed} shows that $R=250$ produces the
shorter duration in $27$ of the $35$ configurations. Unlike the mean-shift case, $R=100$ selects a substantially smaller $\gamma^\star$ at every delay (Figure~\ref{fig:shift-optimal-step-size}, Appendix \ref{app:mean-variance-shift-results}). The resulting weaker
feedback update likely contributes to the longer coverage dips.\\

\begin{table}[htbp]
\centering
\caption{Adaptation duration for the mean-shift family, by shift magnitude $\Delta$ and delay $\tau$, comparing calibration window sizes $R=250$ and $R=100$. Bold indicates the lower (faster) value.}
\label{tab:mean-shift-adaptation-speed}
\resizebox{\textwidth}{!}{%
\begin{tabular}{lcccccccccccccc}
\toprule
$\Delta$ & \multicolumn{2}{c}{$\tau=1$} & \multicolumn{2}{c}{$\tau=5$} & \multicolumn{2}{c}{$\tau=10$} & \multicolumn{2}{c}{$\tau=15$} & \multicolumn{2}{c}{$\tau=20$} & \multicolumn{2}{c}{$\tau=30$} & \multicolumn{2}{c}{$\tau=40$} \\
 & $R=250$ & $R=100$ & $R=250$ & $R=100$ & $R=250$ & $R=100$ & $R=250$ & $R=100$ & $R=250$ & $R=100$ & $R=250$ & $R=100$ & $R=250$ & $R=100$ \\
\midrule
0.5 & 10 & \textbf{6} & 35 & \textbf{1} & 100 & \textbf{70} & 1 & 1 & 129 & \textbf{3} & \textbf{46} & 129 & 139 & \textbf{115} \\
1 & 99 & \textbf{79} & 151 & \textbf{128} & 155 & \textbf{111} & 188 & \textbf{120} & 192 & \textbf{113} & 184 & \textbf{124} & 178 & \textbf{128} \\
2 & \textbf{101} & 108 & 158 & \textbf{119} & 154 & \textbf{111} & 151 & \textbf{113} & 150 & \textbf{114} & 140 & \textbf{120} & 143 & \textbf{122} \\
4 & \textbf{102} & 112 & 131 & \textbf{117} & 128 & \textbf{111} & 129 & \textbf{107} & 119 & \textbf{103} & 112 & \textbf{109} & 114 & \textbf{111} \\
\bottomrule
\end{tabular}
}
\end{table}

\begin{table}[htbp]
\centering
\caption{Adaptation duration for the variance-shift family, by shift ratio $r$ and delay $\tau$, comparing $R=250$ and $R=100$. Bold indicates the lower (faster) value.}
\label{tab:variance-shift-adaptation-speed}
\resizebox{\textwidth}{!}{%
\begin{tabular}{lcccccccccccccc}
\toprule
$r$ & \multicolumn{2}{c}{$\tau=1$} & \multicolumn{2}{c}{$\tau=5$} & \multicolumn{2}{c}{$\tau=10$} & \multicolumn{2}{c}{$\tau=15$} & \multicolumn{2}{c}{$\tau=20$} & \multicolumn{2}{c}{$\tau=30$} & \multicolumn{2}{c}{$\tau=40$} \\
 & $R=250$ & $R=100$ & $R=250$ & $R=100$ & $R=250$ & $R=100$ & $R=250$ & $R=100$ & $R=250$ & $R=100$ & $R=250$ & $R=100$ & $R=250$ & $R=100$ \\
\midrule
1.5 & \textbf{1} & 83 & \textbf{158} & 165 & \textbf{153} & 230 & \textbf{205} & 229 & \textbf{216} & 218 & \textbf{242} & 269 & \textbf{243} & 245 \\
2 & 136 & \textbf{103} & \textbf{162} & 207 & \textbf{165} & 203 & \textbf{197} & 220 & \textbf{195} & 214 & \textbf{209} & 225 & 223 & \textbf{220} \\
3 & 114 & \textbf{99} & \textbf{138} & 146 & \textbf{137} & 151 & \textbf{148} & 149 & 158 & \textbf{145} & 162 & \textbf{156} & \textbf{158} & 159 \\
5 & 121 & \textbf{102} & \textbf{146} & 149 & \textbf{127} & 152 & \textbf{142} & 147 & 156 & \textbf{142} & \textbf{165} & 201 & \textbf{152} & 201 \\
8 & 132 & \textbf{103} & \textbf{165} & 166 & \textbf{133} & 171 & \textbf{154} & 172 & \textbf{155} & 161 & \textbf{152} & 184 & \textbf{143} & 175 \\
\bottomrule
\end{tabular}
}
\end{table}

\begin{figure}[htbp]
    \centering
    \includegraphics[width=\textwidth]{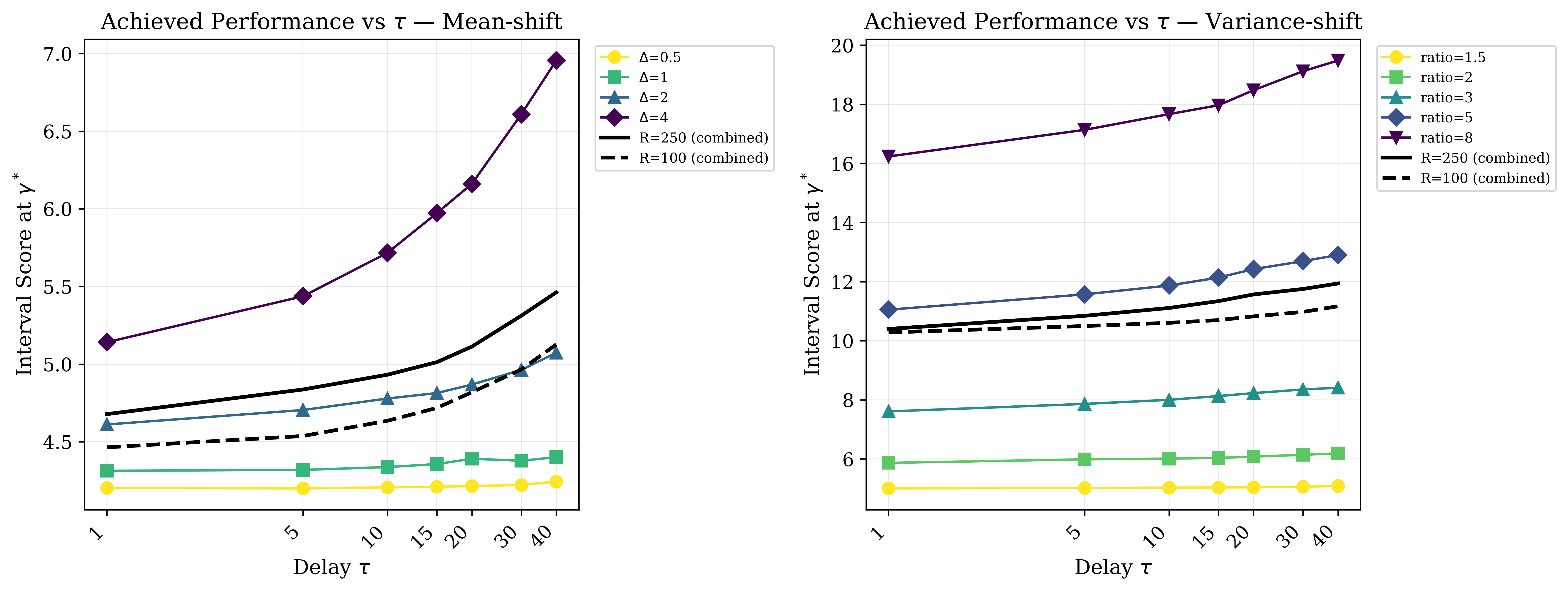}
    \caption{Achieved interval score at the selected step size
    $\gamma^\star(\tau)$ as a function of feedback delay for the mean-shift
    family (left) and variance-shift family (right). Colored curves show the
    isolated-shift experiments using $R=250$. The solid and dashed black curves show the combined-shift experiments using $R=250$ and $R=100$, respectively. $\gamma^\star(\tau)$ is selected separately for each delay,
    shift family, calibration-window length, and experiment using the coverage-aware interval-score criterion.}
    \label{fig:shift-achieved-performance}
\end{figure}

Figure~\ref{fig:shift-achieved-performance} shows that larger shifts produce
higher interval scores and greater sensitivity to delay. They also generally
produce higher local coverage errors and favor larger optimal step sizes,
reflecting the need for more aggressive feedback following severe coverage
failures. This pattern is clearest for $\Delta=4$ and for the variance ratios
$r=5$ and $r=8$. Nevertheless, even with larger selected values of
$\gamma^\star$, severe shifts remain more difficult to accommodate than mild
shifts. The corresponding step-size and local-coverage results are reported in
Figures~\ref{fig:shift-optimal-step-size} and
\ref{fig:shift-local-coverage-error} in
Appendix~\ref{app:mean-variance-shift-results}.\\

Shortening the calibration window has a much clearer interval-score benefit for
mean shifts than for variance shifts. For mean shifts, $R=100$ consistently
produces a noticeably lower score than $R=250$, in agreement with its generally
shorter adaptation durations. For variance shifts, the reduction in interval
score is comparatively modest, particularly at shorter delays, and occurs
despite generally slower restoration of coverage under $R=100$. The two metrics
emphasize different aspects of performance: interval score balances width and
miscoverage over the complete trajectory, whereas adaptation duration measures
local recovery following a transition. The LCE results in
Figure~\ref{fig:shift-local-coverage-error} reinforce this distinction. Under
variance shifts, LCE is generally higher for $R=100$ than for $R=250$,
consistent with the longer adaptation durations.\\

\subsection{Markov-switching --- stochastic and recurring regime changes}
\label{sec:markov_results}

 Using the curve-collapse diagnostic defined in Section~\ref{sec:simulated-examples}, replacing $\tau$ with $r_{\mathrm{regime}}$ reduces the within-bin interval-score scatter by approximately $9\%$, $27\%$, and $10\%$ under mean, variance, and joint switching, respectively. Although the reduction is modest for mean and joint switching, it is positive for all three families and is strongest under variance switching. The normalization results exhibit substantially stronger alignment: the scatter in the normalized-to-unnormalized interval-score ratio decreases by approximately $83\%$ and $84\%$ for the current-state oracle under variance and joint switching, respectively. We therefore present the main interval-score and normalization results against $r_{\mathrm{regime}}$. The corresponding results against the raw delay $\tau$ are shown in Figures~\ref{fig:markov-symmetric-performance-mismatch-tau} and~\ref{fig:markov-symmetric-normalization-tau} in Appendix~\ref{app:markov-results}.\\

Figure~\ref{fig:markov-symmetric-performance-mismatch} reports the interval score and the mismatch-minus-match score difference for the symmetric chains. Unlike the Gaussian, AR(1), and GARCH(1,1) families, the Markov-switching families were not standardized to unit unconditional variance. Under the symmetric design, \(\pi_0=\pi_1=0.5\), and with \(\Delta=4\) and \(r=5\), the unconditional residual variances are \(5\), \(13\), and \(17\) for the mean-, variance-, and joint-switching families, respectively. Their larger absolute interval scores therefore partly reflect greater marginal residual dispersion and are not directly comparable with those of the other residual families. Because $\pi_0=\pi_1=0.5$ for every value of $p$, the marginal residual distribution is fixed within each Markov-switching family, allowing persistence levels to be compared directly. The top row shows that $p=0.99$ produces the lowest interval score at the shortest delay but deteriorates rapidly as $r_{\mathrm{regime}}$ increases. Under mean switching, the scores for $p\leq0.95$ remain nearly constant, whereas the score for $p=0.97$ is slightly higher and the score for $p=0.99$ continues to increase. Under variance and joint switching, the configurations with $p\leq0.95$ approach similar plateaus before $r_{\mathrm{regime}}=1$, while $p=0.97$ and $p=0.99$ remain higher.\\

The mismatch analysis tests whether the deterioration in interval performance is specifically associated with the target falling in a different regime from the one occupied at interval construction. A positive mismatch-minus-match difference therefore indicates an additional interval-score penalty when \(A_t\neq A_{t+\tau}\). This difference is close to zero for the less persistent chains, remains positive for \(p=0.97\), and is largest for \(p=0.99\). When switching is frequent (low persistence), the rolling calibration window contains many observations from both regimes and therefore more closely represents their stationary mixture. The resulting intervals are less specific to the regime occupied at the forecast origin, so changing states between the forecast origin and target has relatively little additional effect. Under greater persistence, the calibration window contains fewer regime changes and can become more strongly influenced by the current regime during interval construction. A target generated under the opposite regime is then poorly represented by the recent residual history. This explains both the larger mismatch penalty and the higher interval scores for \(p=0.97\) and \(p=0.99\) at larger delays.\\

The mismatch penalty is particularly large at the shortest delay for the more persistent chains. At \(\tau=1\), a mismatch corresponds to an immediate regime transition, which occurs with probability \(1-p\); for \(p\in\{0.95,0.97,0.99\}\), these probabilities are only \(0.05\), \(0.03\), and \(0.01\), respectively. Because \(\gamma^\star\) is selected using the average interval score, these rare mismatches have relatively little influence on the selected step size, so intervals can perform well within the current regime while performing poorly immediately after a switch. As \(\tau\) increases, mismatches become more common according to Equation~\eqref{eq:markov-mismatch-probability}, and the mismatch-minus-match gap generally narrows even as the average interval score can increase.\\

\begin{figure}[htbp]
    \centering
    \includegraphics[width=\textwidth]{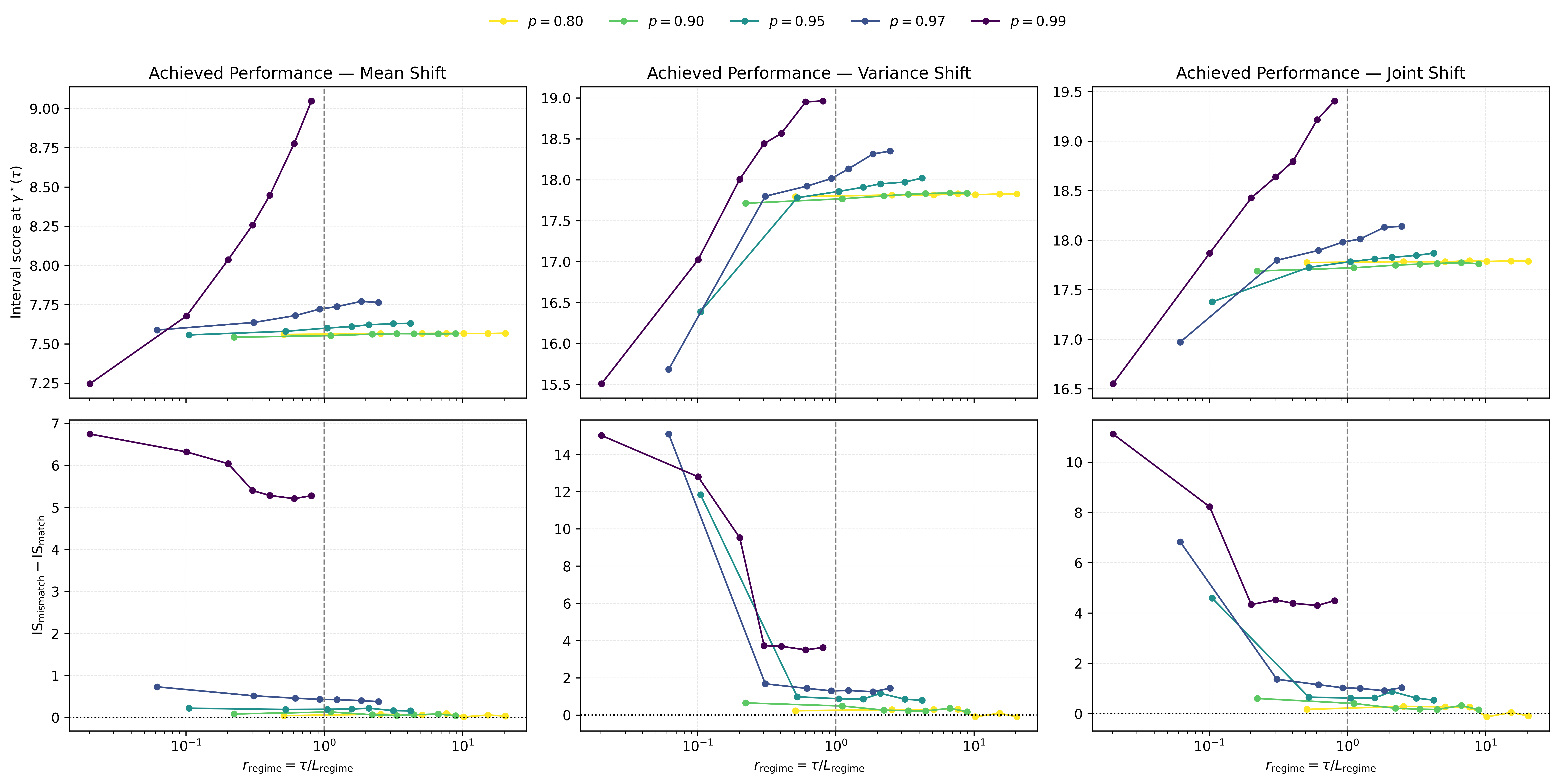}
    \caption{Achieved interval score (top) and mismatch-minus-match conditional interval
    score (bottom) under symmetric Markov switching with $R=250$.}
    \label{fig:markov-symmetric-performance-mismatch}
\end{figure}

\begin{figure}[htbp]
    \centering
    \includegraphics[width=\textwidth]{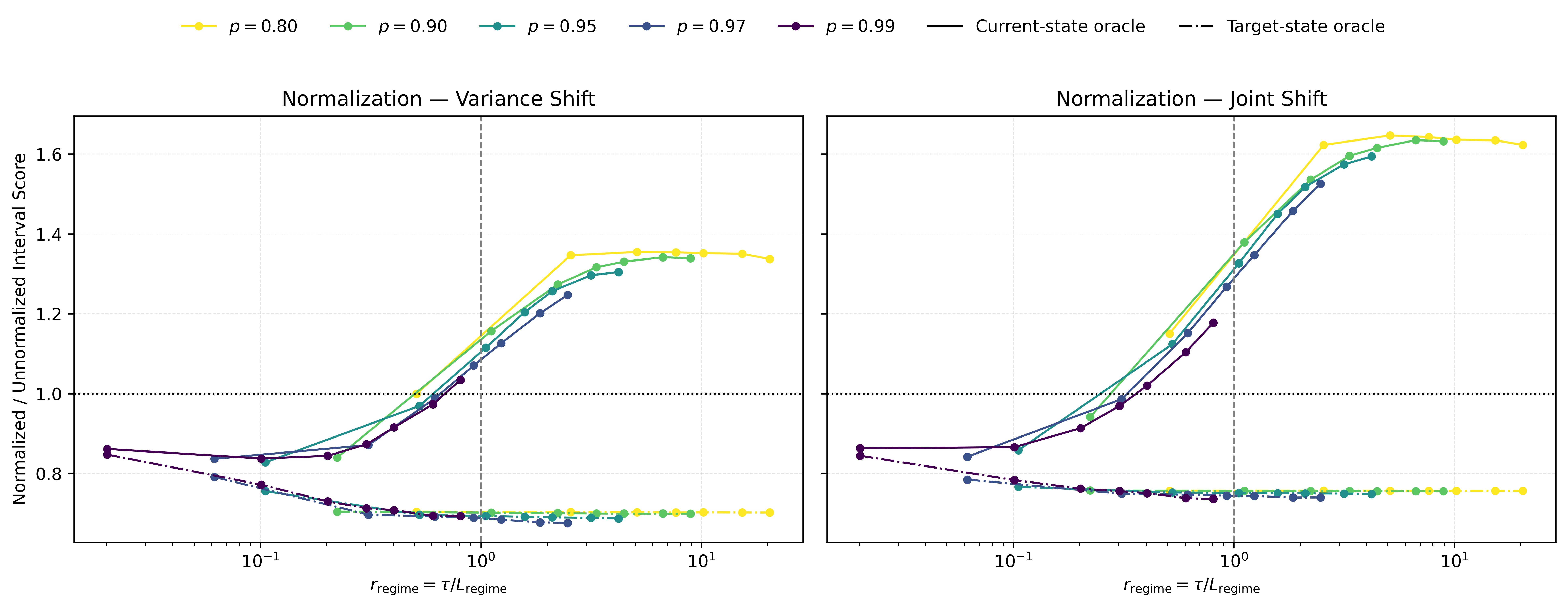}
    \caption{Normalized-to-unnormalized interval-score ratio for the two oracle
    normalization constructions under symmetric Markov switching with $R=250$.}
    \label{fig:markov-symmetric-normalization}
\end{figure}

Figure~\ref{fig:markov-symmetric-normalization} shows a pattern similar to the GARCH results: current-state normalization becomes less effective as \(r_{\mathrm{regime}}\) increases. The target-state oracle does not exhibit the same deterioration, indicating that the loss in current-state normalization is driven by mismatch between the scale at interval construction and the scale at the target time. Because the target-state oracle uses future state information, it is not deployable; rather, it provides a benchmark for the potential gains from better forecasts of the target-time regime or scale. The normalization gain is smaller under joint switching because normalization addresses the state-dependent variance but not the state-dependent mean.\\

\subsection{Cross-family comparison of delay-to-memory scaling}
\label{sec:collapse-summary}

We conclude the simulation results by comparing how well the delay-to-memory
reparameterization organizes performance across the persistent residual
families. Table~\ref{tab:collapse-summary-raw} reports the reduction in within-bin interval-score scatter obtained by replacing the raw delay $\tau$ with the
corresponding delay-to-memory ratio.\\

\begin{table}[htbp]
\centering
\caption{Reduction in within-bin interval-score scatter obtained by replacing
the raw delay $\tau$ with the corresponding delay-to-memory ratio $r$.}
\label{tab:collapse-summary-raw}
\begin{tabular}{llc}
\toprule
Family & Ratio & Scatter reduction \\
\midrule
AR(1) & $r_\varepsilon = \tau/L$ & $79\%$ \\
GARCH(1,1) & $r_{\varepsilon^2} = \tau/L_{\mathrm{vol}}$ & $46\%$ \\
Markov-switching (mean) & $r_{\mathrm{regime}} = \tau/L_{\mathrm{regime}}$ & $9\%$ \\
Markov-switching (variance) & $r_{\mathrm{regime}} = \tau/L_{\mathrm{regime}}$ & $27\%$ \\
Markov-switching (joint) & $r_{\mathrm{regime}} = \tau/L_{\mathrm{regime}}$ & $10\%$ \\
\bottomrule
\end{tabular}
\end{table}

The strongest collapse occurs for AR(1), where dependence is carried directly
by the residual level. The reduction remains substantial but smaller under
GARCH(1,1), where persistence appears in the conditional variance rather than
the signed residual. For Markov switching, the raw interval score is only
weakly to moderately organized by $r_{\mathrm{regime}}$. Thus, although the
delay-to-memory ratio improves on the raw delay across all three persistent
process classes, the extent of the improvement depends on the form of
non-exchangeability.\\

The delay-to-memory ratio can be more informative when the outcome of interest is the relative benefit of normalization. Table~\ref{tab:collapse-summary-normratio} reports the same diagnostic applied to the ratio of normalized to unnormalized interval score. The delay-to-memory ratio reduces scatter by $50\%$ for the GARCH current-scale oracle and by more than $80\%$ for the current-state oracle under both variance and joint Markov switching. The contrast between Tables~\ref{tab:collapse-summary-raw} and
\ref{tab:collapse-summary-normratio} shows that the usefulness of the
delay-to-memory ratio depends on the quantity being examined. Under Markov
switching, $r_{\mathrm{regime}}$ provides only a limited collapse of overall
interval performance, but strongly organizes the effect of using
state-dependent scale information. The ratio therefore captures the relevance of state-dependent scale information across the forecast delay more clearly than it characterizes overall performance.

\begin{table}[htbp]
\centering
\caption{Reduction in within-bin scatter of the normalized-to-unnormalized
interval-score ratio obtained by replacing the raw delay $\tau$ with the corresponding delay-to-memory ratio $r$.}
\label{tab:collapse-summary-normratio}
\begin{tabular}{llc}
\toprule
Family & Construction & Scatter reduction \\
\midrule
GARCH(1,1) & Current-scale oracle & $50\%$ \\
Markov-switching (variance) & Current-state oracle & $83\%$ \\
Markov-switching (joint) & Current-state oracle & $84\%$ \\
\bottomrule
\end{tabular}
\end{table}

\section{Conclusion}

This paper studies adaptive conformal inference when coverage feedback is delayed by a fixed forecast horizon $\tau$. The phase-wise representation of the delayed recursion yields a long-run coverage bound whose finite-sample deviation depends explicitly on the delay, while the marginal coverage bound provides a complementary characterization by relating coverage deviation to changes in the underlying environment across the forecast horizon and the adaptation rate $\gamma$. This motivates asking how the persistence of the residual signal relative to $\tau$ governs finite-sample performance. The simulations investigate this question by introducing a delay-to-memory ratio and testing whether it organizes performance across several mechanistically distinct forms of residual dependence and distributional change.\\

A central finding is that the effect of feedback delay is more informative when interpreted relative to the memory of the residual process than in raw time steps alone. At the same time, the delay-to-memory ratio is not a universal characterization of performance. Its usefulness depends on which feature of the residual distribution carries temporal dependence and on the quantity being evaluated. The relevant memory scale may therefore concern the residual level, its magnitude or variance, or a latent regime. More generally, the ratio is best viewed as a diagnostic of whether information available at prediction time is likely to remain relevant at the forecast target.\\

The persistence and shift experiments also expose a limitation of using a single fixed step size throughout an online forecasting sequence. The appropriate $\gamma$ depends on both the residual dynamics and the relation between delay and persistence. Under exchangeable residuals, aggressive adaptation reduces efficiency without improving coverage, whereas under settings with informative temporal structure or abrupt distributional change, faster adaptation can be useful. A step size suited to one setting may therefore be too conservative when the distribution changes suddenly and too aggressive once the residual signal becomes weak or uninformative.\\

This motivates coupling delayed ACI with procedures that do not require committing to a single learning rate in advance. Online Expert Aggregation on ACI (AgACI) \citep{zaffran2022adaptive} and Fully Adaptive Conformal Inference (FACI) \citep{gibbs2024conformal} use aggregation across candidate learning rates to adapt the step size automatically, providing natural templates for this direction. A complementary direction is to exploit temporal structure more directly. Estimates of residual persistence could help determine how strongly delayed feedback should influence current updates, while dependence across forecast horizons could provide additional information about future errors. Related work has begun to exploit temporal dependence across forecast horizons in online conformal prediction \citep{wang2024online, yang2024bellman, li2026optimization}. A natural question is whether estimates of residual persistence could also be used to determine how strongly delayed feedback should influence the ACI update, while retaining its coverage guarantees. In this broader setting, the delay-to-memory ratio would continue to serve as a diagnostic for when delayed feedback is likely to remain informative.\\
\appendix
\section{Proofs for Coverage Guarantees Under Delayed Feedback}

This appendix collects the full derivations supporting the boundedness, long-run coverage, and approximate marginal coverage results for the $\tau$-delayed ACI recursion stated in the main text (Sections~\ref{sec:boundedness}--\ref{sec:approx-marginal}).

\subsection{Boundedness of $\alpha_m^{(\theta)}$}
\label{app:Boundedness}

For each fixed $\theta$, the phase-wise sequence $\alpha_m^{(\theta)}$ obeys the same Robbins--Monro style update as in standard ACI. 
Since $\alpha \in [0,1]$ and $e_m^{(\theta)} \in \{0,1\}$, the change between consecutive updates satisfies
\begin{equation}
    \label{step_change}
    \bigl|\alpha_{m+1}^{(\theta)} - \alpha_m^{(\theta)}\bigr| 
    = \gamma\bigl|\alpha - e_m^{(\theta)}\bigr| \le \gamma.
\end{equation}

From \eqref{step_change}, if $\alpha_m^{(\theta)} \in [0,1]$ then it follows that $-\gamma \le \alpha_{m+1}^{(\theta)} \le 1 + \gamma$. Moreover, the two boundary cases ensure that $\alpha_m^{(\theta)}$ cannot drift further outward:
\begin{itemize}
    \item If $\alpha_m^{(\theta)} < 0$, then $\alpha_m^{(\theta)}/2 < 0$ and $1-\alpha_m^{(\theta)}/2 > 1$, so both thresholds diverge: $\hat{Q}_{t_\theta(m-1)}(\alpha_m^{(\theta)}/2) = -\infty$ and $\hat{Q}_{t_\theta(m-1)}(1-\alpha_m^{(\theta)}/2) = +\infty$. The coverage interval is then all of $\mathbb{R}$, which forces $e_m^{(\theta)}=0$. Hence
    \[
        \alpha_{m+1}^{(\theta)} - \alpha_m^{(\theta)} = \gamma\alpha \ge 0,
        \quad \Rightarrow \quad
        \alpha_{m+1}^{(\theta)} \ge \alpha_m^{(\theta)}.
    \]
    \item If $\alpha_m^{(\theta)} > 1$, then $\alpha_m^{(\theta)}/2 > 1-\alpha_m^{(\theta)}/2$, and monotonicity of $\hat{Q}_{t_\theta(m-1)}(\cdot)$ gives $\hat{Q}_{t_\theta(m-1)}(\alpha_m^{(\theta)}/2) \ge \hat{Q}_{t_\theta(m-1)}(1-\alpha_m^{(\theta)}/2)$, so the coverage interval is empty (or a single point). This forces $e_m^{(\theta)}=1$. Hence
    \[
        \alpha_{m+1}^{(\theta)} - \alpha_m^{(\theta)} = \gamma(\alpha - 1) \le 0,
        \quad \Rightarrow \quad
        \alpha_{m+1}^{(\theta)} \le \alpha_m^{(\theta)}.
    \]
\end{itemize}

Therefore, for all $m \ge 1$ we have
\[
\alpha_m^{(\theta)} \in [-\gamma,\, 1+\gamma],
\]
exactly as in Lemma~4.1 of \citet{gibbs2021adaptive}, with the difference being that here $e_m^{(\theta)}$ is determined by the quantiles $\hat{Q}_{t_\theta(m-1)}(\alpha_{m}^{(\theta)}/2)$ and $\hat{Q}_{t_\theta(m-1)}(1 - \alpha_{m}^{(\theta)}/2)$ fit at time $t_\theta(m-1)$ rather than at $t-1$, reflecting the $\tau$-step delay, and it is determined by a two-sided coverage interval rather than a one-sided threshold, reflecting the signed-residual construction.

\subsection{Long-Term Coverage Guarantees}
\label{app:LongTermCoverage}
We now derive the long-term coverage guarantees for $\tau$-delayed ACI, following the reasoning of Proposition~4.1 in \citet{gibbs2021adaptive}. For simplicity, we set $t_0 = 0$ throughout.

\paragraph{Step 1: Phase-wise bound:}
Expanding the recursion in Equation~\ref{decomposed_ACI} for a fixed phase $\theta$ gives
\[
\sum_{m=1}^K \bigl(\alpha^{(\theta)}_{m+1} - \alpha^{(\theta)}_m\bigr)
= \gamma \sum_{m=1}^K \bigl(\alpha - e^{(\theta)}_m\bigr)
\quad\Longrightarrow\quad
\frac{1}{K}\sum_{m=1}^K \bigl(e^{(\theta)}_m - \alpha\bigr)
= \frac{\alpha^{(\theta)}_1 - \alpha^{(\theta)}_{K+1}}{\gamma\,K}.
\]
Since $\alpha^{(\theta)}_1 \in [0,1]$ and $\alpha^{(\theta)}_{K+1} \in [-\gamma,\,1+\gamma]$,
\[
\bigl|\alpha^{(\theta)}_1 - \alpha^{(\theta)}_{K+1}\bigr|
\;\le\; \max\{\alpha^{(\theta)}_1 - (-\gamma),\;1+ \gamma -\alpha^{(\theta)}_1\},
\]
which yields the phase-wise bound
\begin{equation}
\label{eq:phasewise-bound}
\left|\frac{1}{K}\sum_{m=1}^K e^{(\theta)}_m - \alpha \right|
\;\le\; \frac{\max\{\alpha^{(\theta)}_1,\;1-\alpha^{(\theta)}_1\}}{\gamma\,K} \;+\; \frac{1}{K}.
\end{equation}

\paragraph{Step 2: Phase sets and notation:}
For $\theta \in \{1,\dots,\tau\}$, let
\[
P_\theta \;:=\; \{\, t \in \{1,\dots,T\} : (t - \theta) \bmod \tau = 0 \,\}
\]
be the set of time indices where the $\theta$-sequence is updated, then $K_\theta := |P_\theta|$ is the number of updates in phase $\theta$. Denote $\bar{e}^{(\theta)} := \frac{1}{K_\theta} \sum_{m=1}^{K_\theta} e^{(\theta)}_m$ as the average empirical miscoverage in phase $\theta$. Therefore, \eqref{eq:phasewise-bound} yields
\[
\bigl|\bar{e}^{(\theta)} - \alpha \bigr|
\;\le\; \frac{\max\{\alpha^{(\theta)}_1,\;1-\alpha^{(\theta)}_1\}}{\gamma\,K_\theta} \;+\; \frac{1}{K_\theta}.
\]

\paragraph{Step 3: From phase-wise to global coverage:}
Let \(\texttt{err}_t := \mathbbm{1}\{Y_t \notin \hat{\mathcal C}^\tau_t(\alpha_t)\}\) denote the indicator of miscoverage at time t. Partitioning the index set \(\{1,\dots,T\}\) by the phase sets \(P_\theta\) and using the phase timeline \(t_\theta(m)\), we obtain
\[
\frac{1}{T}\sum_{t=1}^T \texttt{err}_t
= \frac{1}{T}\;
\Biggl( \sum_{\theta=1}^{\tau} \sum_{t \in P_{\theta}} \texttt{err}_{t_{\theta}(m)} \Biggr)
= \sum_{\theta=1}^{\tau} \frac{K_\theta}{T}\;
\Biggl( \frac{1}{K_\theta}\sum_{m=1}^{K_\theta} e^{(\theta)}_m \Biggr) = \sum_{\theta=1}^{\tau} \frac{K_\theta}{T} \bar{e}^{(\theta)}.
\]

The long-term coverage is calculated
\[
\left|\frac{1}{T}\sum_{t=1}^T \texttt{err}_t - \alpha\right|
\le \sum_{\theta=1}^{\tau} \frac{K_\theta}{T}\,\left|\bar{e}^{(\theta)} - \alpha\right| \\
\le \sum_{\theta=1}^{\tau} \frac{K_\theta}{T}\left(
\frac{\max\{\alpha^{(\theta)}_1,\;1-\alpha^{(\theta)}_1\}}{\gamma\,K_\theta} + \frac{1}{K_\theta}
\right)\\
\]

Therefore, the global coverage bound is
\begin{equation}
\label{eq:global-bound-app}
\left|\frac{1}{T}\sum_{t=1}^T \texttt{err}_t - \alpha \right|
\;\le\; \frac{1}{\gamma\, T} \sum_{\theta=1}^{\tau} \max\{\alpha^{(\theta)}_1,\;1-\alpha^{(\theta)}_1\}
\;+\; \frac{\tau}{T}.
\end{equation}
In particular, $\frac{1}{T}\sum_{t=1}^T \texttt{err}_t \to \alpha$ almost surely as $T \to \infty$.\\

\subsection{Approximate Marginal Coverage}
\label{app:MarginalCoverage}
We provide the proof of the $\tau$-delayed analogue of Theorem~4.2 in \citet{gibbs2021adaptive}, which bounds the mean-squared deviation of the marginal miscoverage $M(\alpha_t \mid A_t)$ from the target $\alpha$.\\

\noindent\textbf{Approximate marginal coverage for $\tau$-Delayed ACI:}
\label{thm:tau-delayed-M}
Under the setup of Section~\ref{marginal_coverage_section}, assume $(\alpha_t,A_t,\texttt{err}_t)$ is stationary, $M(\cdot\mid a)$ is $L_{Lip}$-Lipschitz for all $a\in\mathcal A$, and each $a$ admits a unique $\alpha^*_a$ such that $M(\alpha^*_a \mid a)=\alpha$. Then
\[
\mathbb{E}\!\left[ \bigl(M(\alpha_t\mid A_t) - \alpha\bigr)^2 \right]
\;\le\; \frac{L_{Lip}(1+\gamma)}{\gamma}\,
\mathbb{E}\!\left[ \bigl| \alpha^*_{A_{t+\tau}} - \alpha^*_{A_t} \bigr| \right]
+ \frac{L_{Lip} \gamma}{2},
\]
where the expectation is taken with respect to the stationary distribution of the process.\\

We begin from the $\tau$-delayed recursion:
\[
\alpha_{t+\tau} = \alpha_t + \gamma(\alpha - \texttt{err}_t).
\]
Subtracting $\alpha^*_{A_t}$ and squaring both sides gives
\[
(\alpha_{t+\tau} - \alpha^*_{A_t})^2
= (\alpha_t - \alpha^*_{A_t})^2
+ 2\gamma(\alpha - \texttt{err}_t)(\alpha_t - \alpha^*_{A_t})
+ \gamma^2(\alpha - \texttt{err}_t)^2.
\]
\paragraph{Bounding the cross term $\mathbb{E}\!\left[(\alpha-\texttt{err}_t)(\alpha_t-\alpha^*_{A_t})\right]$:}
By the law of total expectation and the definition of $M(\cdot \mid A_t)$, noting that
$\mathbb{E}[\texttt{err}_t \mid A_t, \alpha_t] = M(\alpha_t \mid A_t)$, we have
\begin{align}
\mathbb{E}\!\left[(\alpha-\texttt{err}_t)(\alpha_t-\alpha^*_{A_t})\right]
&= \mathbb{E}\!\left[\,
\mathbb{E}\!\left[(\alpha-\texttt{err}_t)(\alpha_t-\alpha^*_{A_t}) \,\middle|\, A_t,\,\alpha_t\right]
\right] \nonumber\\
&= \mathbb{E}\!\left[(\alpha-\mathbb{E}[\texttt{err}_t\mid A_t,\alpha_t])(\alpha_t-\alpha^*_{A_t})\right] \nonumber\\
&= \mathbb{E}\!\left[(\alpha-M(\alpha_t\mid A_t))(\alpha_t-\alpha^*_{A_t})\right]. \label{eq:cross-start}
\end{align}

\noindent\textbf{Lipschitz continuity and sign alignment:}
Assume there exists $L_{Lip}>0$ such that for all $a\in\mathcal A$ and $\alpha_1,\alpha_2\in[-\gamma,1+\gamma]$,
\[
\big|M(\alpha_2\mid a) - M(\alpha_1\mid a)\big| \leq L_{Lip}\,|\alpha_2 - \alpha_1|.
\]
Taking $\alpha_1=\alpha_t$ and $\alpha_2=\alpha^*_{A_t}$ and using $M(\alpha^*_{A_t}\mid A_t)=\alpha$ gives
\begin{align}
|M(\alpha_t \mid A_t) - \alpha|
&\;\le\; L_{Lip}\,|\alpha_t - \alpha^*_{A_t}| \notag\\[6pt]
\Longrightarrow\;&\; |\alpha_t - \alpha^*_{A_t}|
   \;\ge\; \tfrac{1}{L_{Lip}}\,|M(\alpha_t \mid A_t) - \alpha| \notag\\[6pt]
\Longrightarrow\;&\; |\alpha_t - \alpha^*_{A_t}|\,|M(\alpha_t \mid A_t) - \alpha|
   \;\ge\; \tfrac{1}{L_{Lip}}\,\big(M(\alpha_t \mid A_t) - \alpha\big)^2
\label{eq:lipschitz-core}
\end{align}

Moreover, $M(\cdot\mid A_t)$ is nondecreasing with $\alpha_t$ and $M(\alpha^*_{A_t}\mid A_t)=\alpha$, hence
\[
\mathrm{sign}\!\big(M(\alpha_t\mid A_t)-\alpha\big)=\mathrm{sign}\!\big(\alpha_t-\alpha^*_{A_t}\big),
\]
which implies
\begin{equation}
\label{eq:abs-to-prod-clean}
|M(\alpha_t\mid A_t)-\alpha|\,|\alpha_t-\alpha^*_{A_t}|
=(M(\alpha_t\mid A_t)-\alpha)(\alpha_t-\alpha^*_{A_t}).
\end{equation}

\noindent\textbf{Cross-term bound:}
Combining \eqref{eq:lipschitz-core} and \eqref{eq:abs-to-prod-clean},
\[
(\alpha - M(\alpha_t\mid A_t))(\alpha_t-\alpha^*_{A_t})
= -\,|M(\alpha_t\mid A_t)-\alpha|\,|\alpha_t-\alpha^*_{A_t}|
\;\le\; -\,\frac{1}{L_{Lip}}\,\big(M(\alpha_t\mid A_t)-\alpha\big)^2.
\]
Taking expectations and using \eqref{eq:cross-start} yields
\begin{equation}
\label{eq:cross-final}
\mathbb{E}\!\left[(\alpha-\texttt{err}_t)(\alpha_t-\alpha^*_{A_t})\right]
\;\le\; -\,\frac{1}{L_{Lip}}\,\mathbb{E}\!\left[\big(M(\alpha_t\mid A_t)-\alpha\big)^2\right].
\end{equation}

\noindent\textbf{From the expansion to an inequality:}
Using \eqref{eq:cross-final} in the squared expansion, we get
\begin{align}
\mathbb{E}\!\left[(\alpha_{t+\tau}-\alpha^*_{A_t})^2\right]
&\le \mathbb{E}\!\left[(\alpha_t-\alpha^*_{A_t})^2\right]
- \frac{2\gamma}{L_{Lip}}\,\mathbb{E}\!\left[(M(\alpha_t\mid A_t)-\alpha)^2\right]
+ \gamma^2\,\mathbb{E}\!\left[(\alpha-\texttt{err}_t)^2\right].
\label{eq:onestep}
\end{align}
Rearranging,
\begin{align}
\frac{2\gamma}{L_{Lip}}\,\mathbb{E}\!\left[(M(\alpha_t\mid A_t)-\alpha)^2\right]
\;\le\;
\mathbb{E}\!\left[(\alpha_t-\alpha^*_{A_t})^2\right]
- \mathbb{E}\!\left[(\alpha_{t+\tau}-\alpha^*_{A_t})^2\right]
+ \gamma^2\,\mathbb{E}\!\left[(\alpha-\texttt{err}_t)^2\right].
\label{eq:onestep-rearranged}
\end{align}

\noindent\textbf{Summation and averaging:}
Summing \eqref{eq:onestep-rearranged} over $t=1,\dots,T$ and dividing by $T$,
\begin{align}
\frac{2\gamma}{L_{Lip}}\cdot \frac{1}{T}\sum_{t=1}^{T}\mathbb{E}\!\left[(M(\alpha_t\mid A_t)-\alpha)^2\right]
\;\le\;
\frac{1}{T}\sum_{t=1}^{T}\Big\{
\mathbb{E}\!\left[(\alpha_t-\alpha^*_{A_t})^2\right]
- \mathbb{E}\!\left[(\alpha_{t+\tau}-\alpha^*_{A_t})^2\right]
\Big\}
+ \gamma^2\,\frac{1}{T}\sum_{t=1}^{T}\mathbb{E}\!\left[(\alpha-\texttt{err}_t)^2\right].
\label{eq:sum-pre}
\end{align}

\noindent
Since $\texttt{err}_t\in\{0,1\}$ and $\alpha\in[0,1]$, we have $(\alpha-\texttt{err}_t)^2\le 1$, hence
\begin{equation}
\label{eq:errsq-bound}
\frac{1}{T}\sum_{t=1}^{T}\mathbb{E}\!\left[(\alpha-\texttt{err}_t)^2\right]\le 1.
\end{equation}

By adding and subtracting $\frac{1}{T}\sum_{t=1}^{T}\mathbb{E}\!\left[(\alpha_{t+\tau}-\alpha^*_{A_{t+\tau}})^2\right]$, we write
\begin{align}
&\frac{1}{T}\sum_{t=1}^{T}\Big\{
\mathbb{E}\!\left[(\alpha_t-\alpha^*_{A_t})^2\right]
- \mathbb{E}\!\left[(\alpha_{t+\tau}-\alpha^*_{A_t})^2\right]
\Big\} \nonumber\\
&\quad=\;
\underbrace{\frac{1}{T}\sum_{t=1}^{T}\Big\{
\mathbb{E}\!\left[(\alpha_t-\alpha^*_{A_t})^2\right]
- \mathbb{E}\!\left[(\alpha_{t+\tau}-\alpha^*_{A_{t+\tau}})^2\right]
\Big\}}_{\text{(I)}}
\;+\;
\underbrace{\frac{1}{T}\sum_{t=1}^{T}\mathbb{E}\!\left[
(\alpha_{t+\tau}-\alpha^*_{A_{t+\tau}})^2 - (\alpha_{t+\tau}-\alpha^*_{A_t})^2
\right]}_{\text{(II)}}.
\label{eq:split}
\end{align}

\noindent\textbf{(I) term:}
Reindexing the second sum in (I) by $s=t+\tau$,
\[
\frac{1}{T}\sum_{t=1}^{T}\mathbb{E}\!\left[(\alpha_{t+\tau}-\alpha^*_{A_{t+\tau}})^2\right]
= \frac{1}{T}\sum_{s=\tau+1}^{T+\tau}\mathbb{E}\!\left[(\alpha_{s}-\alpha^*_{A_{s}})^2\right].
\]
Hence,
\begin{align}
\text{(I)}
&= \frac{1}{T}\Bigg(
\sum_{t=1}^{T}\mathbb{E}\!\left[(\alpha_t-\alpha^*_{A_t})^2\right]
- \sum_{s=\tau+1}^{T+\tau}\mathbb{E}\!\left[(\alpha_{s}-\alpha^*_{A_{s}})^2\right]
\Bigg) \nonumber\\
&= \frac{1}{T}\Bigg(
\sum_{t=1}^{\tau}\mathbb{E}\!\left[(\alpha_t-\alpha^*_{A_t})^2\right]
- \sum_{s=T+1}^{T+\tau}\mathbb{E}\!\left[(\alpha_{s}-\alpha^*_{A_{s}})^2\right]
\Bigg).
\label{eq:telescoping}
\end{align}
Because $\alpha_t\in[-\gamma,1+\gamma]$ and $\alpha^*_{A_t}\in(0,1)$, the squared terms $(\alpha_t-\alpha^*_{A_t})^2$ are bounded
by $(1+\gamma)^2$, so $\text{(I)}=O(\tau/T)$. Under stationarity, letting $T\to\infty$ gives
\begin{equation}
\label{eq:I-vanish}
\text{(I)} \;\xrightarrow[T\to\infty]{}\; 0.
\end{equation}

\noindent\textbf{(II) term.}
For each $t$,
\begin{equation}
(\alpha_{t+\tau}-\alpha^*_{A_{t+\tau}})^2 - (\alpha_{t+\tau}-\alpha^*_{A_t})^2
= 2\alpha_{t+\tau}\Big(\alpha^*_{A_t}-\alpha^*_{A_{t+\tau}}\Big) - \Big((\alpha^*_{A_t})^2-(\alpha^*_{A_{t+\tau}})^2\Big)
\end{equation}
Hence,
\begin{align}
\text{(II)} = \frac{1}{T}\Bigg(
\sum_{t=1}^{T}\mathbb{E}\!\Big[2\alpha_{t+\tau}\Big(\alpha^*_{A_t}-\alpha^*_{A_{t+\tau}}\Big) \Big]
- \sum_{t=1}^{T}\mathbb{E}\!\Big[(\alpha^*_{A_t})^2-(\alpha^*_{A_{t+\tau}})^2\Big]
\Bigg)
\label{eq:term2}
\end{align}

\noindent\emph{The second term in \text{(II)}}:

\begin{align}
  &\frac{1}{T}\Bigg(\sum_{t=1}^{T}\mathbb{E}\!\left[(\alpha^*_{A_t})^2-(\alpha^*_{A_{t+\tau}})^2\right]\Bigg) \nonumber\\
  &= \frac{1}{T}\sum_{t=1}^{T}\mathbb{E}\!\left[(\alpha^*_{A_t})^2\right] -\frac{1}{T}\sum_{s=1+\tau}^{T+\tau}\mathbb{E}\!\left[(\alpha^*_{A_{s}})^2\right] \nonumber\\
  &= \frac{1}{T}\sum_{t=1}^{\tau}\mathbb{E}\!\left[(\alpha^*_{A_t})^2\right] -\frac{1}{T}\sum_{s=T+1}^{T+\tau}\mathbb{E}\!\left[(\alpha^*_{A_{s}})^2\right]
  \label{eq:2-term2}
\end{align}

Because $\alpha^*_{A_t}\in(0,1)$, the squared terms $(\alpha^*_{A_t})^2$ are bounded
by $1$, so \eqref{eq:2-term2} $=O(\tau/T)$. Under stationarity, letting $T\to\infty$ gives \eqref{eq:2-term2} $\;\xrightarrow[T\to\infty]{}\; 0.$\\

\noindent\emph{The first term in \text{(II)}}: Since $\alpha_{t+\tau} \in [-\gamma, 1+\gamma]$
\begin{align}
&\frac{1}{T}
\sum_{t=1}^{T}\mathbb{E}\!\Big[2\alpha_{t+\tau}\Big(\alpha^*_{A_t}-\alpha^*_{A_{t+\tau}}\Big) \Big]
\le \frac{1}{T}\sum_{t=1}^{T}2(1+\gamma)\mathbb{E}\!\Big[\Big|\alpha^*_{A_{t+\tau}}-\alpha^*_{A_t}\Big|\Big]
\label{eq:2-term1}
\end{align}

\medskip

\noindent\textbf{Let $T\to\infty$ and conclude:}
Combine \eqref{eq:sum-pre}, \eqref{eq:errsq-bound}, the vanishing of (I), the vanishing of the second term in (II), and \eqref{eq:2-term1}, then let $T\to\infty$:
\begin{align}
\frac{2\gamma}{L_{Lip}}\,\mathbb{E}\!\left[(M(\alpha_t\mid A_t)-\alpha)^2\right]
&\le 2(1+\gamma)\,\mathbb{E}\!\left[\big|\alpha^*_{A_{t+\tau}}-\alpha^*_{A_t}\big|\right]
+ \gamma^2.
\end{align}
Hence,
\begin{equation}
\label{eq:final-you}
\boxed{\;
\mathbb{E}\!\left[(M(\alpha_t\mid A_t)-\alpha)^2\right]
\;\le\; \frac{L_{Lip}(1+\gamma)}{\gamma}\,\mathbb{E}\!\left[\big|\alpha^*_{A_{t+\tau}}-\alpha^*_{A_t}\big|\right]
\;+\; \frac{L_{Lip}}{2}\,\gamma.
\;}
\end{equation}

This is the stated bound in Theorem~\ref{thm:tau-delayed-M}.  Intuitively, the first term measures the \emph{distribution shift} across $\tau$ steps by the change in the optimal level $\alpha^*_{A_t}$, and the second term is the \emph{adaptation penalty} controlled by the step size $\gamma$.

\section{Autocorrelation Structure of the Markov-Switching Process}
\label{app:markov-autocorr}

\subsection{Autocorrelation of the two-state regime indicator}
\label{app:regime-autocorr}

This appendix derives $\operatorname{Corr}(A_t,A_{t+k})=\lambda_2^{\,k}$ for the two-state
regime indicator $A_t\in\{0,1\}$ used in Section~\ref{sec:simulated-examples}, starting
from the transition matrix.

\paragraph{Step 1: Eigenvalues}
The transition matrix is
\[
P = \begin{pmatrix} p_{00} & 1-p_{00} \\ 1-p_{11} & p_{11} \end{pmatrix}.
\]
Since $P$ is row-stochastic, $P\mathbf{1}=\mathbf{1}$, so $\lambda_1=1$ is an eigenvalue
with eigenvector $\mathbf{1}=(1,1)^\top$. The characteristic polynomial of a
$2\times2$ matrix is $\lambda^2-\operatorname{tr}(P)\lambda+\det(P)=0$, so the two
eigenvalues satisfy $\lambda_1+\lambda_2=\operatorname{tr}(P)=p_{00}+p_{11}$. With
$\lambda_1=1$,
\[
\lambda_2 = p_{00}+p_{11}-1.
\]

\paragraph{Step 2: Eigenvectors}
The eigenvector for $\lambda_1=1$ is $v_1=(1,1)^\top$, as above. For $\lambda_2$,
solve $(P-\lambda_2 I)v=0$. Using $p_{00}-\lambda_2=1-p_{11}$ and
$p_{11}-\lambda_2=1-p_{00}$,
\[
P-\lambda_2 I = \begin{pmatrix} 1-p_{11} & 1-p_{00} \\ 1-p_{11} & 1-p_{00} \end{pmatrix},
\]
which has identical rows (confirming $\det(P-\lambda_2I)=0$), giving
\[
v_2 = \begin{pmatrix} 1-p_{00} \\ -(1-p_{11}) \end{pmatrix}.
\]
Writing $V=(v_1\mid v_2)$, we have 
\[
V = \big(v_1 \mid v_2\big) = \begin{pmatrix} 1 & 1-p_{00} \\ 1 & -(1-p_{11}) \end{pmatrix}.
\]
$V$ is invertible, with
\[
V^{-1} = \frac{1}{D}\begin{pmatrix} 1-p_{11} & 1-p_{00} \\ 1 & -1 \end{pmatrix}
\]
where $D:=(2-p_{00} -p_{11}) > 0$.\\

\paragraph{Step 3: Closed form of \(P^k\)}
Let \(\Lambda=\operatorname{diag}(1,\lambda_2)\). The eigendecomposition
\(P=V\Lambda V^{-1}\) gives
\[
P^k=V\Lambda^kV^{-1}.
\]
Carrying out the multiplication and substituting the stationary distribution
\[
\pi_0=\frac{1-p_{11}}{D},
\qquad
\pi_1=\frac{1-p_{00}}{D}
\]
gives
\[
P^k=
\begin{pmatrix}
\pi_0+\pi_1\lambda_2^k
&
\pi_1(1-\lambda_2^k)\\
\pi_0(1-\lambda_2^k)
&
\pi_1+\pi_0\lambda_2^k
\end{pmatrix}.
\]
Every row sums to one, as required.

\paragraph{Step 4: Moments of \(A_t\)}
Under stationarity, \(A_t\) is Bernoulli with success probability \(\pi_1\).
Therefore,
\[
\mathbb E[A_t]=\pi_1,
\qquad
\operatorname{Var}(A_t)
=
\pi_1(1-\pi_1)
=
\pi_0\pi_1.
\]

\paragraph{Step 5: Joint expectation}
Because \(A_tA_{t+k}\) equals one only when both indicators equal one,
\begin{align*}
\mathbb E[A_tA_{t+k}]
&=
\Pr(A_t=1,A_{t+k}=1)\\
&=
\Pr(A_t=1)
\Pr(A_{t+k}=1\mid A_t=1).
\end{align*}
The second probability is the lower-right entry of \(P^k\). Hence,
\[
\mathbb E[A_tA_{t+k}]
=
\pi_1\left(\pi_1+\pi_0\lambda_2^k\right)
=
\pi_1^2+\pi_0\pi_1\lambda_2^k.
\]

\paragraph{Step 6: Covariance}
By stationarity,
\(\mathbb E[A_t]=\mathbb E[A_{t+k}]=\pi_1\). It follows that
\begin{align*}
\operatorname{Cov}(A_t,A_{t+k})
&=
\mathbb E[A_tA_{t+k}]
-
\mathbb E[A_t]\mathbb E[A_{t+k}]\\
&=
\left(
\pi_1^2+\pi_0\pi_1\lambda_2^k
\right)
-\pi_1^2\\
&=
\pi_0\pi_1\lambda_2^k.
\end{align*}

\paragraph{Step 7: Correlation}
By stationarity,
\[
\operatorname{Var}(A_t)
=
\operatorname{Var}(A_{t+k})
=
\pi_0\pi_1.
\]
Therefore,
\begin{align*}
\operatorname{Corr}(A_t,A_{t+k})
&=
\frac{\operatorname{Cov}(A_t,A_{t+k})}
{\sqrt{\operatorname{Var}(A_t)
       \operatorname{Var}(A_{t+k})}}\\
&=
\frac{\pi_0\pi_1\lambda_2^k}
{\pi_0\pi_1}\\
&=
\lambda_2^k,
\qquad k\geq0.
\end{align*}
The factor \(\pi_0\pi_1\) cancels between the covariance and the
normalization. Thus, the latent state indicator has no attenuation factor:
its autocorrelation is exactly \(\lambda_2^k\). If \(\lambda_2>0\), as in
the transition matrices used in our experiments, the autocorrelation decays
monotonically. More generally, a negative \(\lambda_2\) produces an
alternating sign, while its magnitude decays as \(|\lambda_2|^k\).

\subsection{Autocorrelation of state-dependent residual features}
\label{app:residual-feature-autocov}

This section derives how the dependence in the latent regime process induces
dependence in the residuals and their transformations. The derivation applies to the
mean-switching, variance-switching, and joint-switching residual families considered in
Section~\ref{sec:simulated-examples}.\\

\paragraph{Step 1: State-dependent conditional means}
Let
\[
Y_t=g(\varepsilon_t)
\]
denote a transformation of the residual with a finite second moment. Define its
state-dependent conditional mean by
\[
m_j
:=
\mathbb E[Y_t\mid A_t=j],
\qquad
j\in\{0,1\}.
\]
Because \(A_t\) is binary, the conditional mean of \(Y_t\) can be written as
\[
\mathbb E[Y_t\mid A_t]
=
m_0+(m_1-m_0)A_t.
\]
This expression equals \(m_0\) when \(A_t=0\) and \(m_1\) when \(A_t=1\).

\paragraph{Step 2: Law of total covariance}
For \(k>0\), apply the law of total covariance with $Y_t$, $Y_{t+k}$, and the conditioning variable \(Z =(A_t,A_{t+k})\), a random vector containing the states at both time points. Because \(A_t,A_{t+k}\in\{0,1\}\), \(Z\) takes one of the four possible values \( (0,0), (0,1),(1,0), (1,1)\). The law of total covariance gives
\begin{align*}
\operatorname{Cov}(Y_t,Y_{t+k})
&=
\mathbb E\!\left[
\operatorname{Cov}
\left(
Y_t,Y_{t+k}\mid A_t,A_{t+k}
\right)
\right]\\
&\quad+
\operatorname{Cov}\!\left(
\mathbb E[Y_t\mid A_t,A_{t+k}],
\mathbb E[Y_{t+k}\mid A_t,A_{t+k}]
\right).
\end{align*}

Conditional on \(A_t\) and \(A_{t+k}\), the residual features at the two time points can be written as
\[
Y_t
=
g\!\left(\mu_{A_t}+\sigma_{A_t}\eta_t\right),
\qquad
Y_{t+k}
=
g\!\left(
\mu_{A_{t+k}}+\sigma_{A_{t+k}}\eta_{t+k}
\right).
\]
The noise variables \(\eta_t\) and \(\eta_{t+k}\) are independent. Therefore,
\(Y_t\) and \(Y_{t+k}\) are conditionally independent once \(A_t\) and
\(A_{t+k}\) are fixed, and
\[
\operatorname{Cov}
\left(
Y_t,Y_{t+k}\mid A_t,A_{t+k}
\right)
=
0.
\]
The first term in the total-covariance decomposition is consequently zero.\\

Once \(A_t\) is known, knowledge of \(A_{t+k}\) provides no additional information about \(Y_t\), because the conditional distribution of \(Y_t\) depends only on the state at time \(t\). Similarly, once \(A_{t+k}\) is known, knowledge of \(A_t\) provides no additional information about \(Y_{t+k}\). Hence, 
\[
\mathbb E[Y_t\mid A_t,A_{t+k}]
=
\mathbb E[Y_t\mid A_t]
=
m_{A_t}
\qquad
\mathbb E[Y_{t+k}\mid A_t,A_{t+k}]
=
\mathbb E[Y_{t+k}\mid A_{t+k}]
=
m_{A_{t+k}}.
\]
The law of total covariance therefore reduces to

\begin{align*}
\operatorname{Cov}(Y_t,Y_{t+k})
&=
\operatorname{Cov}(m_{A_t},m_{A_{t+k}})\\
&=
\operatorname{Cov}\!\left(
m_0+(m_1-m_0)A_t,
m_0+(m_1-m_0)A_{t+k}
\right)\\
&=
(m_1-m_0)^2
\operatorname{Cov}(A_t,A_{t+k}).
\end{align*}
From Section~\ref{app:regime-autocorr},
\[
\operatorname{Cov}(A_t,A_{t+k})
=
\pi_0\pi_1\lambda_2^k.
\]
Therefore,
\begin{equation}
\operatorname{Cov}(Y_t,Y_{t+k})
=
(m_1-m_0)^2
\pi_0\pi_1\lambda_2^k,
\qquad
k>0.
\label{eq:feature-autocovariance}
\end{equation}
This is the two-state specialization of the general hidden Markov model moment formula
in \citet[Exercise~3(e)--(f), p.~42]{zucchini2009hmm}. Any residual feature whose
conditional mean differs between the two states therefore inherits the geometric factor
\(\lambda_2^k\).

\paragraph{Step 3: Autocorrelation and attenuation}
Let
\[
V_Y:=\operatorname{Var}(Y_t).
\]
By stationarity,
\(\operatorname{Var}(Y_{t+k})=V_Y\), so dividing
Equation~\eqref{eq:feature-autocovariance} by \(V_Y\) gives
\begin{equation}
\operatorname{Corr}(Y_t,Y_{t+k})
=
c_Y\lambda_2^k,
\qquad
c_Y
:=
\frac{(m_1-m_0)^2\pi_0\pi_1}{V_Y},
\qquad
k>0.
\label{eq:feature-autocorrelation}
\end{equation}
The numerator of \(c_Y\) is
\[
\operatorname{Var}\!\left(
\mathbb E[Y_t\mid A_t]
\right)
=
(m_1-m_0)^2\pi_0\pi_1.
\]
Thus, \(c_Y\) is the fraction of the variance of \(Y_t\) attributable to differences
in its conditional mean between the two states. It changes the magnitude of the
autocorrelation but not its geometric decay factor.

\paragraph{Step 4: Mean-switching residuals}
For the mean-switching model,
\[
\varepsilon_t
=
\mu_0+\Delta A_t+\eta_t,
\qquad
\Delta:=\mu_1-\mu_0,
\qquad
\eta_t\overset{\mathrm{iid}}{\sim}\mathcal N(0,1),
\]
where \(\{\eta_t\}\) is independent of the state process. Taking
\(Y_t=\varepsilon_t\), the state-dependent conditional means are
\[
m_0=\mu_0,
\qquad
m_1=\mu_1,
\]
so \(m_1-m_0=\Delta\). Equation~\eqref{eq:feature-autocovariance} gives
\begin{equation}
\operatorname{Cov}(\varepsilon_t,\varepsilon_{t+k})
=
\Delta^2\pi_0\pi_1\lambda_2^k,
\qquad
k>0.
\label{eq:mean-switching-autocovariance}
\end{equation}
Because
\[
\operatorname{Var}(\varepsilon_t)
=
1+\Delta^2\pi_0\pi_1,
\]
the positive-lag autocorrelation is
\begin{equation}
\operatorname{Corr}(\varepsilon_t,\varepsilon_{t+k})
=
\frac{\Delta^2\pi_0\pi_1}
{1+\Delta^2\pi_0\pi_1}
\lambda_2^k,
\qquad
k>0.
\label{eq:mean-switching-autocorrelation}
\end{equation}
The independent Gaussian noise reduces the magnitude of the residual autocorrelation
but does not change its geometric decay factor.

\paragraph{Step 5: Variance-switching residuals}
For the variance-switching model,
\[
\varepsilon_t
=
\sigma_{A_t}\eta_t,
\qquad
\eta_t\overset{\mathrm{iid}}{\sim}\mathcal N(0,1),
\]
with \(\mu_0=\mu_1=0\). Taking \(Y_t=\varepsilon_t\) gives
\[
m_0=m_1=0.
\]
Equation~\eqref{eq:feature-autocovariance} therefore gives
\begin{equation}
\operatorname{Cov}(\varepsilon_t,\varepsilon_{t+k})
=
0,
\qquad
k>0.
\label{eq:variance-switching-signed-autocovariance}
\end{equation}
Thus, the signed residuals are serially uncorrelated even though they are not
independent. Their linear autocorrelation does not detect the regime dependence because
the conditional residual mean is zero in both states.\\

Taking \(Y_t=\varepsilon_t^2\), we have
\[
m_j
=
\mathbb E[\varepsilon_t^2\mid A_t=j]
=
\sigma_j^2\mathbb E[\eta_t^2]
=
\sigma_j^2,
\]
because \(\mathbb E[\eta_t^2]=\operatorname{Var}(\eta_t)+\mathbb E[\eta_t]^2=1\).
Thus,
\[
m_1-m_0=\sigma_1^2-\sigma_0^2,
\]
and Equation~\eqref{eq:feature-autocovariance} gives
\begin{equation}
\operatorname{Cov}(\varepsilon_t^2,\varepsilon_{t+k}^2)
=
(\sigma_1^2-\sigma_0^2)^2
\pi_0\pi_1\lambda_2^k,
\qquad k>0.
\label{eq:variance-switching-squared-autocovariance}
\end{equation}\\

Similarly, taking \(Y_t=|\varepsilon_t|\), we have
\[
m_j
=
\mathbb E[|\varepsilon_t|\mid A_t=j]
=
\sigma_j\mathbb E[|\eta_t|]
=
\sigma_j\sqrt{\frac{2}{\pi}},
\]
where \(\mathbb E[|\eta_t|]=\sqrt{2/\pi}\) for
\(\eta_t\sim\mathcal N(0,1)\). Thus,
\[
m_1-m_0
=
(\sigma_1-\sigma_0)\sqrt{\frac{2}{\pi}},
\]
and
\begin{equation}
\operatorname{Cov}(|\varepsilon_t|,|\varepsilon_{t+k}|)
=
\frac{2}{\pi}
(\sigma_1-\sigma_0)^2
\pi_0\pi_1\lambda_2^k,
\qquad k>0.
\label{eq:variance-switching-absolute-autocovariance}
\end{equation}\\
Both the absolute and squared residuals reveal the regime dependence that is absent
from the autocorrelation of the signed residuals.

\paragraph{Step 6: Joint-switching residuals}
For the joint-switching model,
\[
\varepsilon_t
=
\mu_{A_t}+\sigma_{A_t}\eta_t,
\]
so both the conditional mean and conditional variance depend on the state. Taking
\(Y_t=\varepsilon_t\) in Equation~\eqref{eq:feature-autocovariance} gives
\begin{equation}
\operatorname{Cov}(\varepsilon_t,\varepsilon_{t+k})
=
(\mu_1-\mu_0)^2
\pi_0\pi_1\lambda_2^k,
\qquad
k>0.
\label{eq:joint-switching-autocovariance}
\end{equation}
By the law of total variance,
\[
\operatorname{Var}(\varepsilon_t)
=
\mathbb E\!\left[
\operatorname{Var}(\varepsilon_t\mid A_t)
\right]
+
\operatorname{Var}\!\left(
\mathbb E[\varepsilon_t\mid A_t]
\right).
\]
The first term is
\begin{align*}
\mathbb E\!\left[
\operatorname{Var}(\varepsilon_t\mid A_t)
\right]
&=
\sum_{j=0}^1
\Pr(A_t=j)
\operatorname{Var}(\varepsilon_t\mid A_t=j)\\
&=
\pi_0\sigma_0^2+\pi_1\sigma_1^2.
\end{align*}
The second term is
\begin{align*}
\operatorname{Var}\!\left(
\mathbb E[\varepsilon_t\mid A_t]
\right)
&=
\operatorname{Var}(\mu_{A_t})\\
&=
\pi_0\pi_1(\mu_1-\mu_0)^2.
\end{align*}
Therefore,
\begin{equation}
\operatorname{Var}(\varepsilon_t)
=
\pi_0\sigma_0^2
+
\pi_1\sigma_1^2
+
\pi_0\pi_1(\mu_1-\mu_0)^2.
\label{eq:joint-switching-variance}
\end{equation}
The first two terms represent the average variance within the two states, while the
final term represents the variance of the state-dependent conditional mean.\\

Therefore,
\begin{equation}
\operatorname{Corr}(\varepsilon_t,\varepsilon_{t+k})
=
\frac{
(\mu_1-\mu_0)^2\pi_0\pi_1
}{
\pi_0\sigma_0^2
+
\pi_1\sigma_1^2
+
\pi_0\pi_1(\mu_1-\mu_0)^2
}
\lambda_2^k,
\qquad
k>0.
\label{eq:joint-switching-autocorrelation}
\end{equation}
The state-dependent variance changes the attenuation coefficient but not the geometric
factor \(\lambda_2^k\).\\

Table~\ref{tab:markov-feature-correlations} summarizes the resulting positive-lag autocorrelation structure. The attenuation coefficient \(c_Y\leq 1\) affects only the magnitude of the observed-feature autocorrelation, while the geometric decay rate remains \(\lambda_2^k\). In particular, under variance switching, the signed residual is uncorrelated across time, while \(|\varepsilon_t|\) and \(\varepsilon_t^2\) retain the latent regime dependence.

\begin{table}[H]
\centering
\small
\caption{Positive-lag autocorrelation of the residual features that reveal each
Markov-switching regime.}
\label{tab:markov-feature-correlations}
\begin{tabular}{llll}
\toprule
Residual family
& Feature $Y_t$
& Attenuation coefficient
& $\operatorname{Corr}(Y_t,Y_{t+k})$, $k>0$ \\
\midrule

Mean switching
& $\varepsilon_t$
& $\displaystyle
c_{\varepsilon}^{(\mathrm{mean})}
=
\frac{\Delta^2\pi_0\pi_1}
{1+\Delta^2\pi_0\pi_1}$
& $c_{\varepsilon}^{(\mathrm{mean})}\lambda_2^k$ \\[3mm]

Variance switching
& $\varepsilon_t$
& $0$
& $0$ \\[2mm]

Variance switching
& $|\varepsilon_t|$
& $\displaystyle
c_{|\varepsilon|}
=
\frac{\frac{2}{\pi}(\sigma_1-\sigma_0)^2\pi_0\pi_1}
{V_{|\varepsilon|}}$
& $c_{|\varepsilon|}\lambda_2^k$ \\[3mm]

Variance switching
& $\varepsilon_t^2$
& $\displaystyle
c_{\varepsilon^2}
=
\frac{(\sigma_1^2-\sigma_0^2)^2\pi_0\pi_1}
{V_{\varepsilon^2}}$
& $c_{\varepsilon^2}\lambda_2^k$ \\[3mm]

Joint switching
& $\varepsilon_t$
& $\displaystyle
c_{\varepsilon}^{(\mathrm{joint})}
=
\frac{\Delta^2\pi_0\pi_1}
{\pi_0\sigma_0^2+\pi_1\sigma_1^2
+\Delta^2\pi_0\pi_1}$
& $c_{\varepsilon}^{(\mathrm{joint})}\lambda_2^k$ \\[3mm]

\bottomrule
\end{tabular}
\end{table}
\section{Additional Results}

\subsection{AR(1)}
\label{app:ar-results}

\begin{figure}[H]
    \centering
    \includegraphics[width=\textwidth]{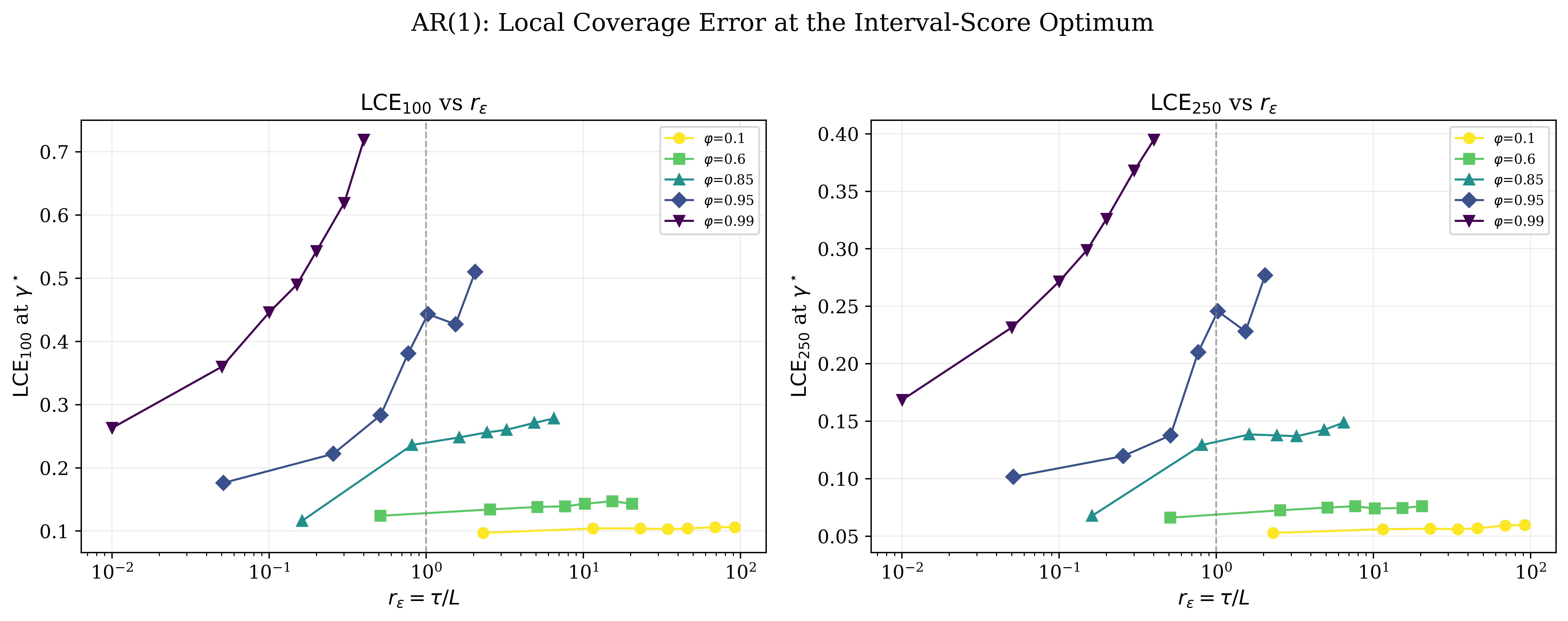}
    \caption{Worst-case local coverage error at the approximately score-minimizing step size
    $\gamma^\star$ for AR(1) residuals, against the delay-to-memory ratio
    $r_\varepsilon$, for window lengths $k = 100$ (left) and $k = 250$ (right).}
    \label{fig:ar_lce}
\end{figure}

Figure~\ref{fig:ar_lce} reports the worst-case local coverage error at the interval-score-optimal step size $\gamma^\star$ as a function of delay-to-memory ratio $r_\varepsilon$ for both window lengths. Within each persistence level, the error generally increases with $r_\varepsilon$, indicating that local calibration deteriorates as the delay grows relative to the memory length. Unlike the achieved interval score, however, the curves do not collapse across $\varphi$. Instead, they remain vertically separated, with stronger persistence associated with larger local coverage error over the overlapping range of $r_\varepsilon$. Thus, $r_\varepsilon$ organizes the variation in local coverage error within a persistence level, but does not account for its magnitude across persistence levels, which remains strongly dependent on $\varphi$.

\subsection{GARCH(1,1)}
\label{app:garch-results}

\begin{figure}[H]
    \centering
    \includegraphics[width=\textwidth]{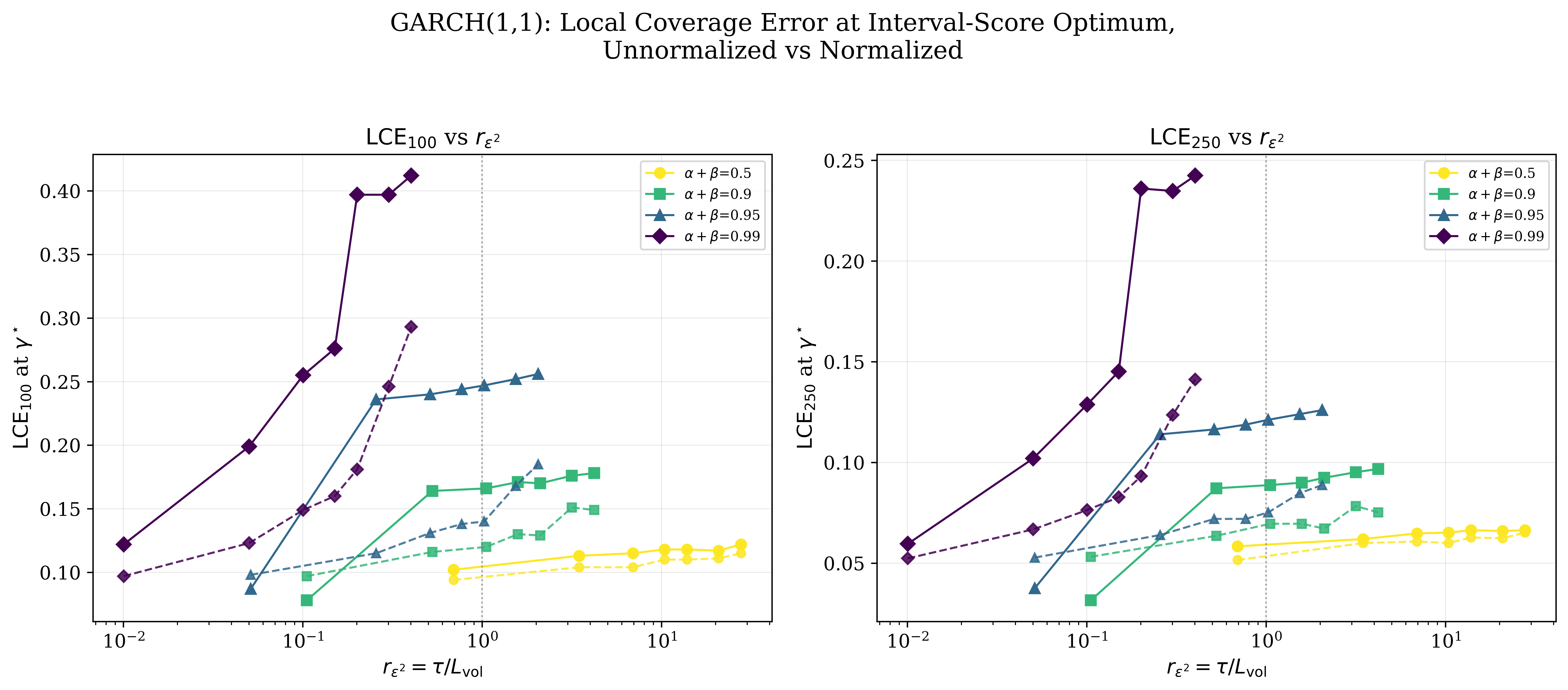}
    \caption{Worst-case local coverage error at the approximately score-minimizing
    step-size $\gamma^\star$ for GARCH(1,1) residuals, against the
    delay-to-memory ratio $r_{\varepsilon^2}$, for
    window lengths $k=100$ (left) and $k=250$ (right). Solid lines use the unnormalized-residual construction and dashed lines use the current-scale oracle normalization.}
    \label{fig:garch-lce-unnorm-vs-norm}
\end{figure}

Figure~\ref{fig:garch-lce-unnorm-vs-norm} compares the worst-case local coverage error of the unnormalized and current-scale oracle normalized constructions for window lengths $k=100$ and $k=250$. Normalization reduces local coverage error across nearly all configurations, with the only exceptions occurring at $\tau=1$ for $\alpha+\beta=0.9$ and $\alpha+\beta=0.95$. The gains are most pronounced for the more persistent processes: for $\alpha+\beta=0.95$ and $\alpha+\beta=0.99$, normalization substantially attenuates the increase in local coverage error as $r_{\varepsilon^2}$ grows. Local coverage errors are lower for $k=250$ than for $k=100$, but the qualitative benefit of normalization is consistent across both window lengths. Notably, the current-scale oracle retains a local-coverage advantage even when its interval-score advantage disappears for $r_{\varepsilon^2}\gtrsim1$. This suggests that the delay-induced mismatch in conditional scale affects interval efficiency more strongly than local coverage stability.

\subsection{Mean and Variance Shift}
\label{app:mean-variance-shift-results}

\begin{figure}[H]
    \centering
    \includegraphics[width=\textwidth]{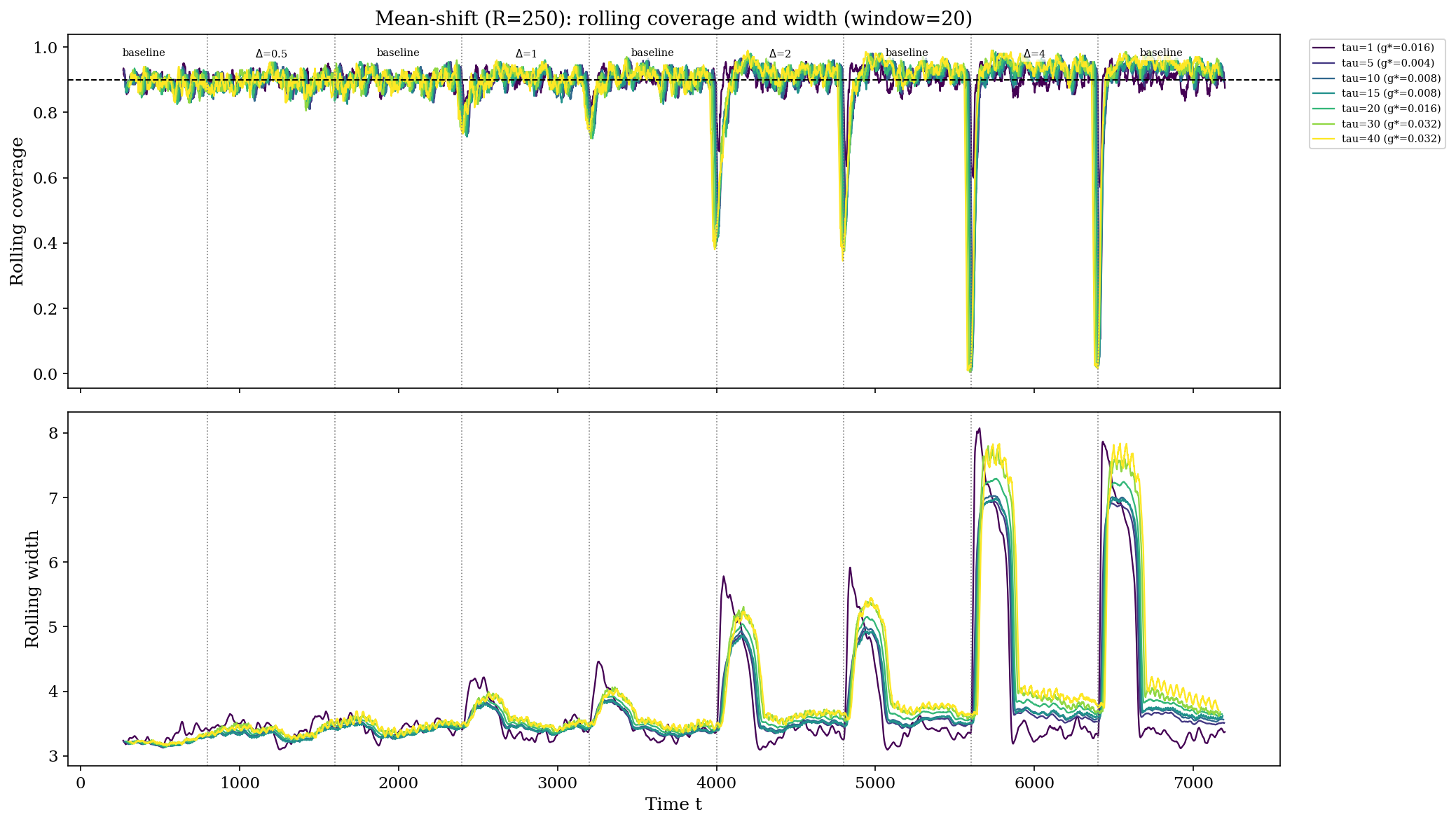}
    \caption{Rolling coverage (top) and interval width (bottom) along the combined mean-shift trajectory using calibration-window length $R=250$. Rolling quantities are computed over a window of $20$ observations.}
    \label{fig:mean-shift-coverage-width-window20}
\end{figure}

\begin{figure}[H]
    \centering
    \includegraphics[width=\textwidth]{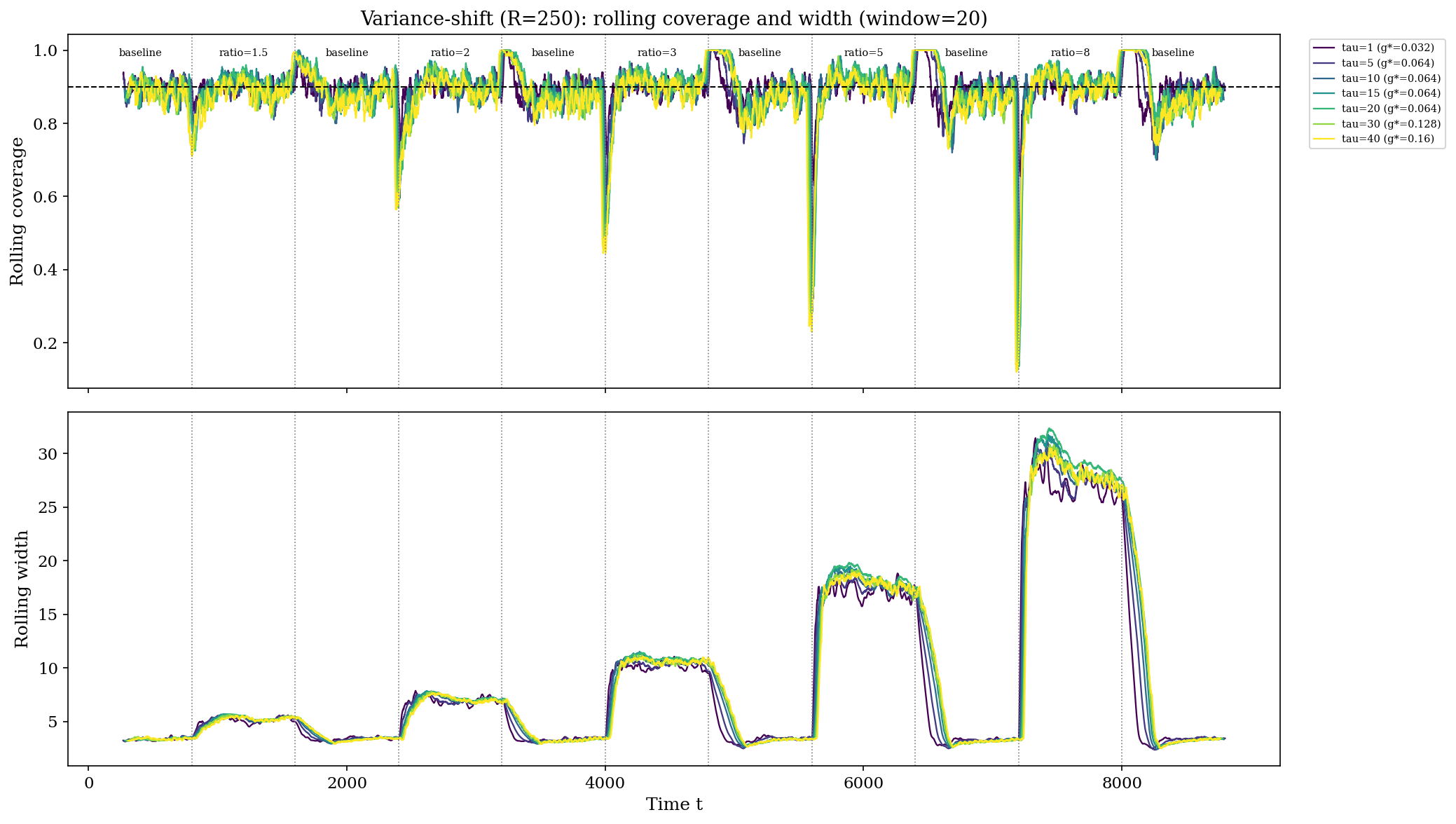}
    \caption{Rolling coverage (top) and interval width (bottom) along the combined variance-shift trajectory using calibration-window length $R=250$. Rolling quantities are computed over a window of $20$ observations.}
    \label{fig:variance-shift-coverage-width-window20}
\end{figure}

\begin{figure}[H]
    \centering
    \includegraphics[width=\textwidth]{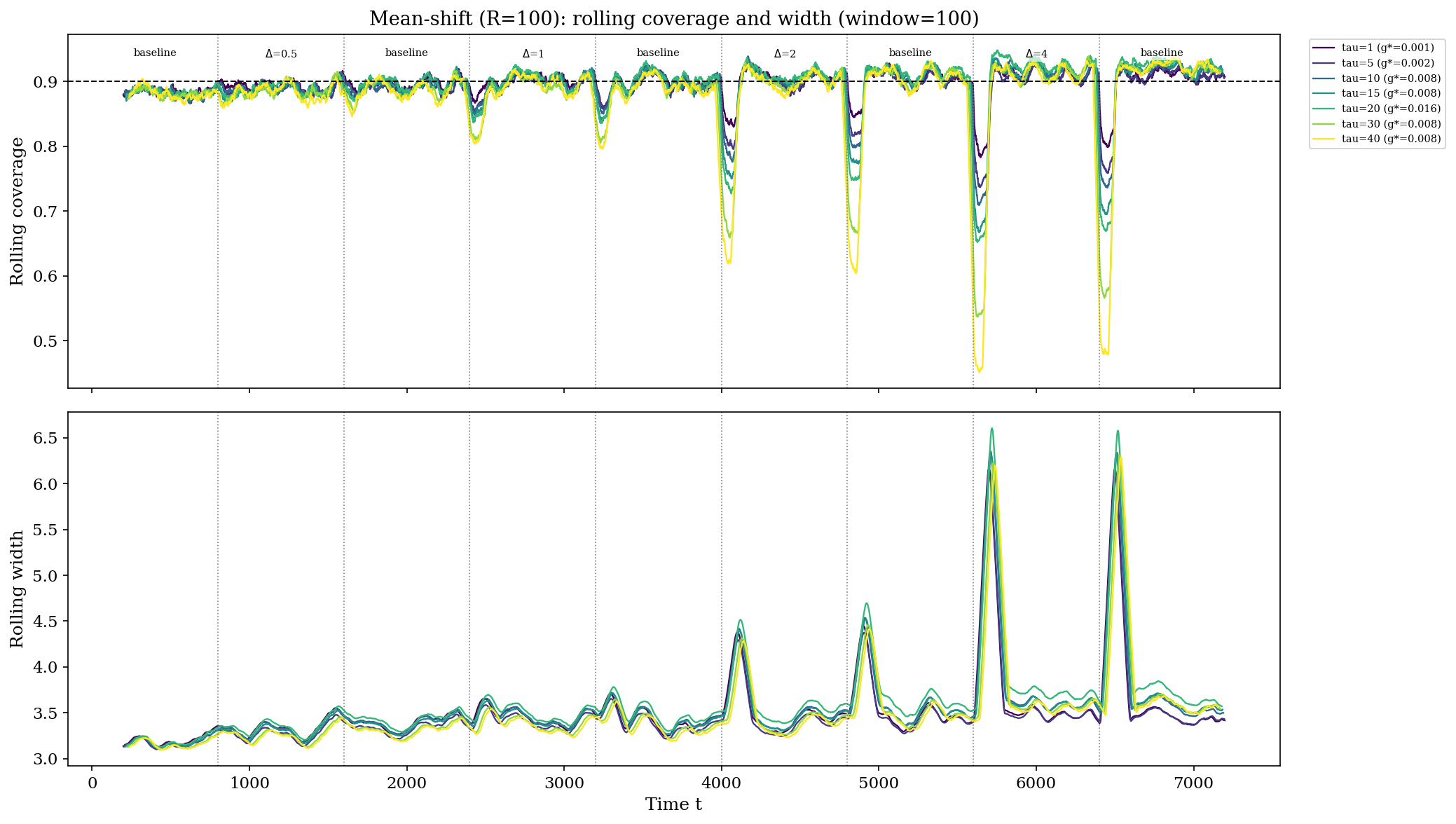}
    \caption{Rolling coverage (top) and interval width (bottom) along
    the combined mean-shift trajectory using the shortened calibration window
    $R=100$. Rolling quantities are computed over a window of $100$
    observations.}
    \label{fig:mean-shift-rolling-r100}
\end{figure}

\begin{figure}[H]
    \centering
    \includegraphics[width=\textwidth]{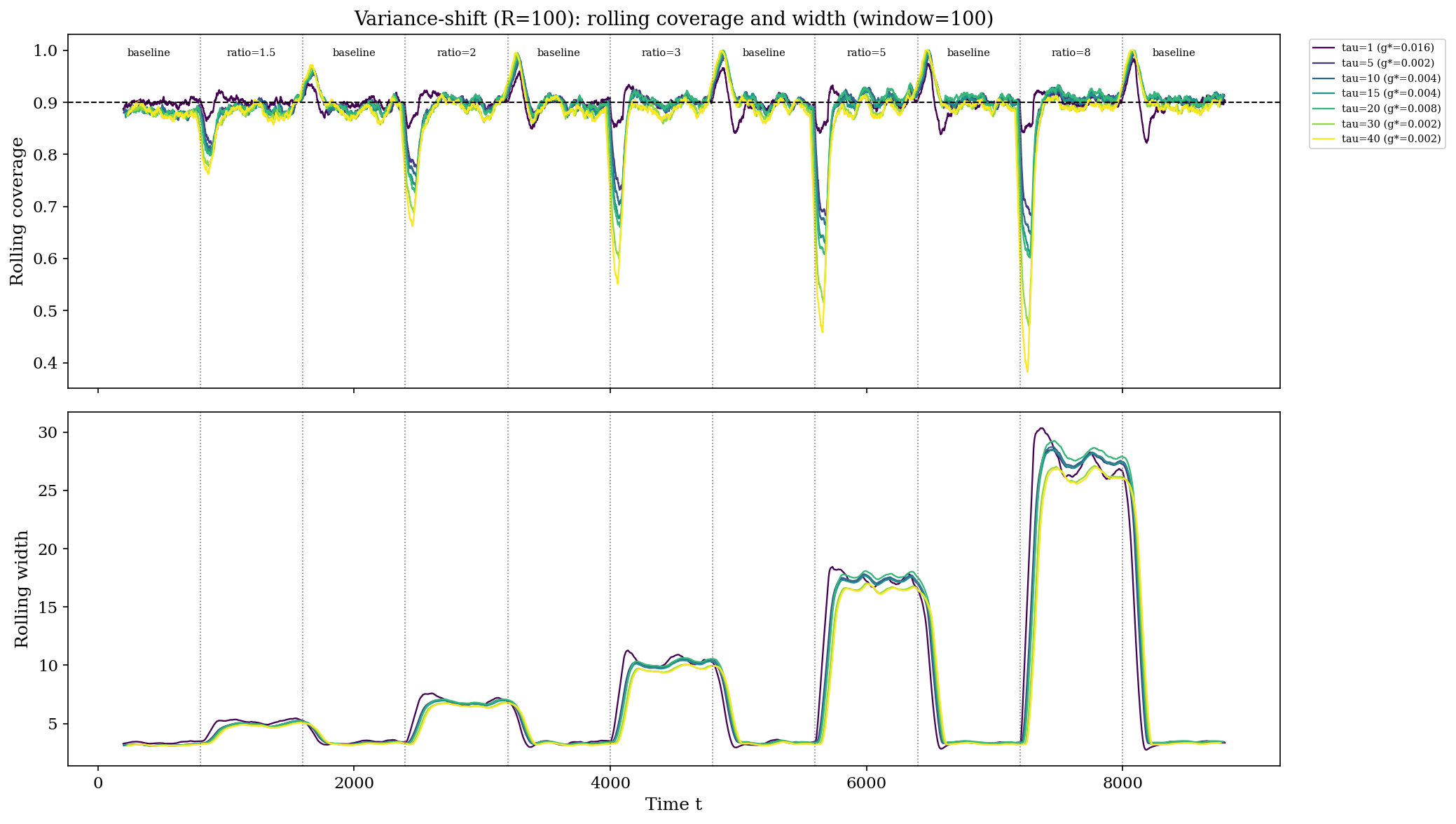}
    \caption{Rolling coverage (top) and interval width (bottom) along
    the combined variance-shift trajectory using the shortened calibration
    window $R=100$. Rolling quantities are computed over a window of $100$
    observations.}
    \label{fig:variance-shift-rolling-r100}
\end{figure}

\begin{figure}[H]
    \centering
    \includegraphics[width=\textwidth]{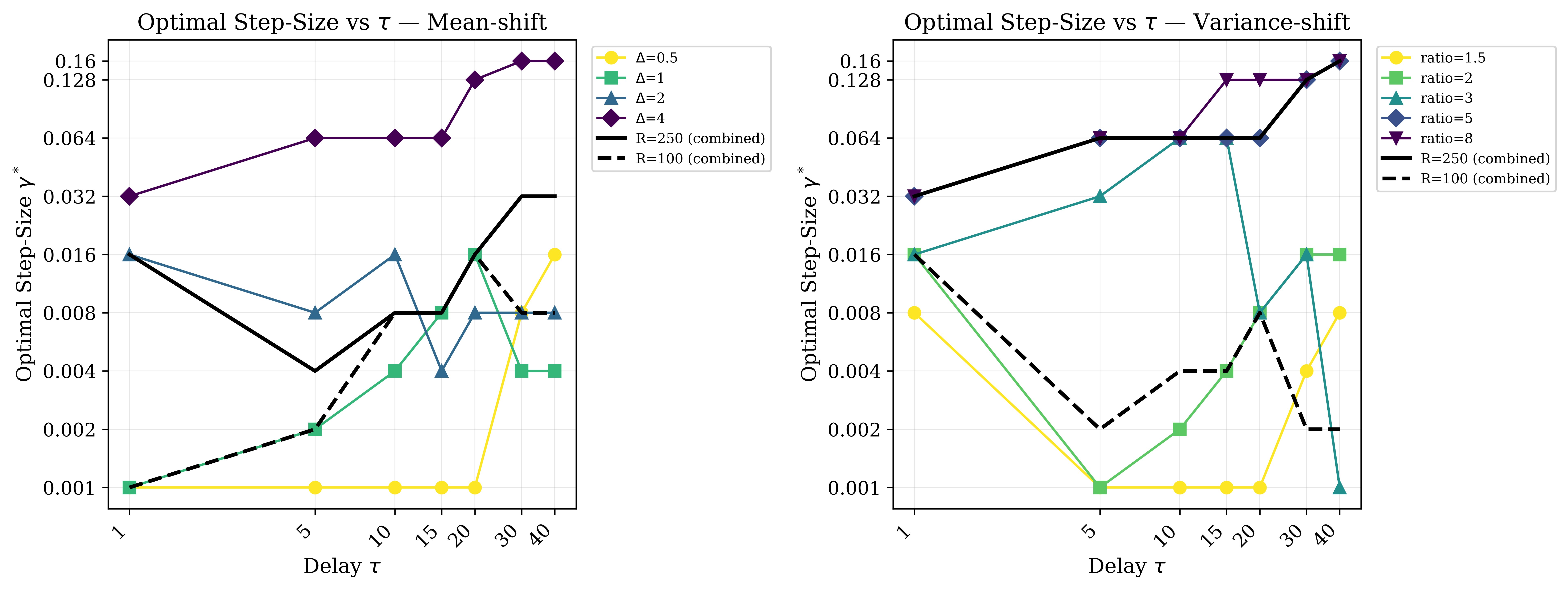}
    \caption{Selected step size $\gamma^\star$ as a function of feedback delay
    for the mean-shift family (left) and variance-shift family (right). Colored
    curves show the step sizes selected for the isolated-shift experiments using
    $R=250$. The solid and dashed black curves show the selections for the
    combined-shift experiments using $R=250$ and $R=100$, respectively.
    Step sizes are selected separately for each configuration using the
    coverage-aware interval-score criterion.}
    \label{fig:shift-optimal-step-size}
\end{figure}

\begin{figure}[H]
    \centering
    \includegraphics[width=\textwidth]{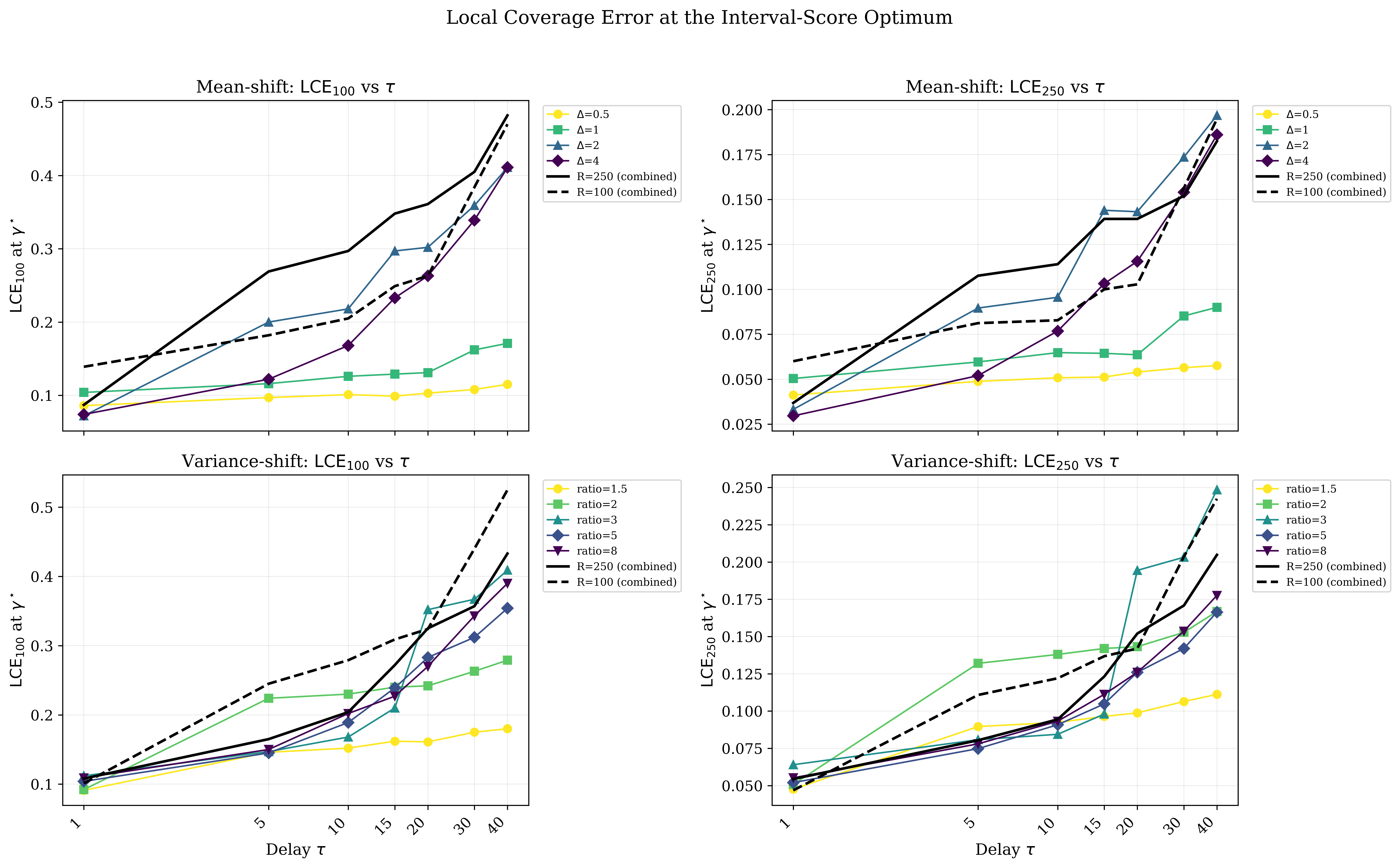}
    \caption{Local coverage error at $\gamma^\star$ as a
    function of feedback delay. The top row presents the mean-shift family, and
    the bottom row presents the variance-shift family; the left and right columns
    use local evaluation windows of $100$ and $250$ observations, respectively.
    Colored curves show the isolated-shift experiments using $R=250$. The solid
    and dashed black curves show the combined-shift experiments using $R=250$
    and $R=100$, respectively. Lower values indicate better
    local calibration.}
    \label{fig:shift-local-coverage-error}
\end{figure}

\subsection{Markov Switching}
\label{app:markov-results}

\begin{figure}[H]
    \centering
    \includegraphics[width=\textwidth]{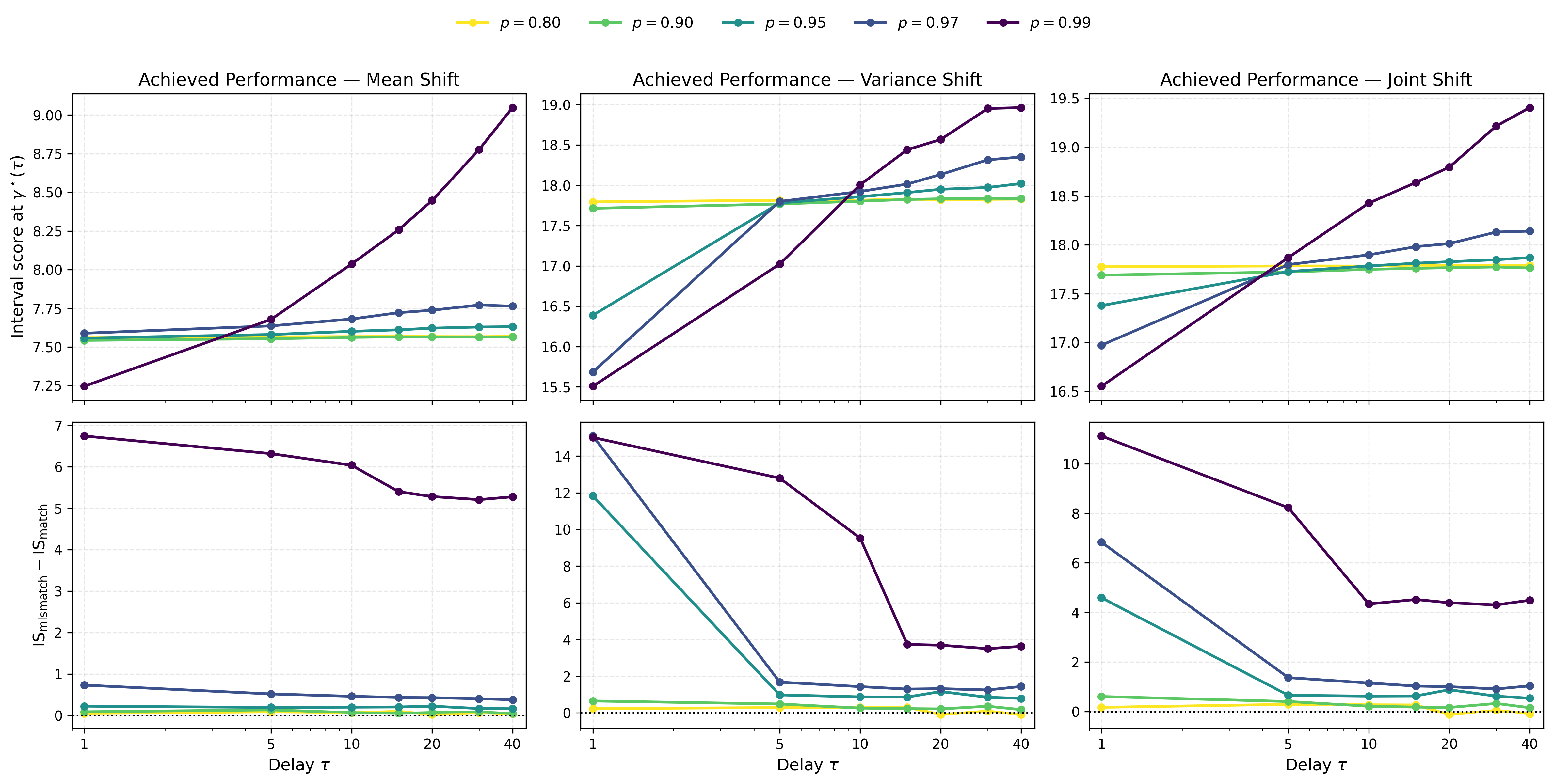}
    \caption{Achieved interval score (top) and mismatch-minus-match conditional
    interval score (bottom) under symmetric Markov switching with $R=250$,
    shown as functions of the raw delay $\tau$.}
    \label{fig:markov-symmetric-performance-mismatch-tau}
\end{figure}

\begin{figure}[H]
    \centering
    \includegraphics[width=\textwidth]{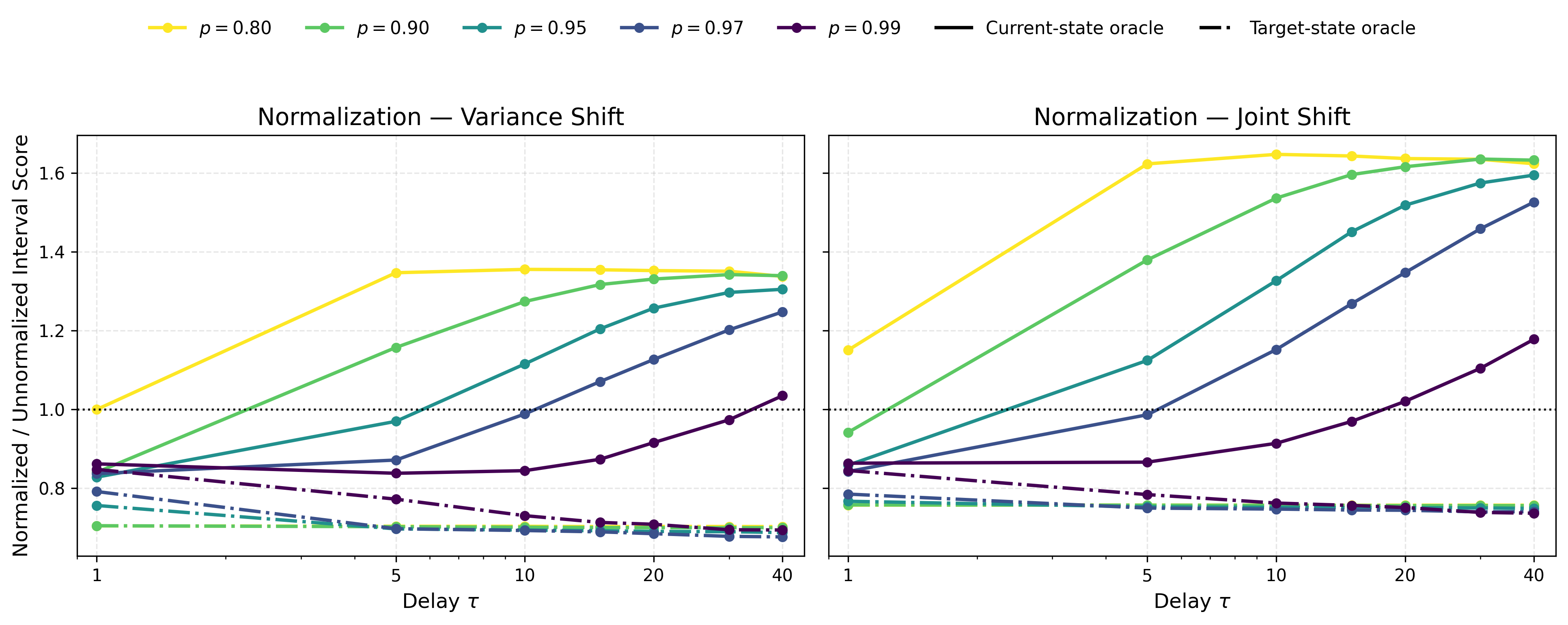}
    \caption{Normalized-to-unnormalized interval-score ratio for the current-state
    and target-state oracle constructions under symmetric Markov switching with
    $R=250$, shown as a function of the raw delay $\tau$.}
    \label{fig:markov-symmetric-normalization-tau}
\end{figure}

\bibliography{ref}

\end{document}